\documentclass[pdflatex,sn-mathphys-num]{sn-jnl}

\usepackage{graphicx}%
\usepackage{multirow}%
\usepackage{amsmath,amssymb,amsfonts}%
\usepackage{amsthm}%
\usepackage{mathrsfs}%
\usepackage[title]{appendix}%
\usepackage{xcolor}%
\usepackage{textcomp}%
\usepackage{manyfoot}%
\usepackage{booktabs}%
\usepackage{algorithm}%
\usepackage{algorithmicx}%
\usepackage{algpseudocode}%
\usepackage{listings}%
\usepackage{multirow} 
\usepackage{adjustbox} 
\usepackage{threeparttable} 
\usepackage{balance}  
\theoremstyle{thmstyleone}%
\theoremstyle{thmstyletwo}%
\theoremstyle{thmstylethree}%
\usepackage{svg} 
\usepackage{placeins}
\usepackage{comment}

\usepackage{caption} 
\newcounter{imagetable}

\begin{document}

\title[Article Title]{Modeling Day-Long ECG Signals to Predict Heart Failure Risk with Explainable AI}

\author*[1]{\fnm{Eran} \sur{Zvuloni}}\email{eranzvuloni@gmail.com}

\author[1,2]{\fnm{Ronit} \sur{Almog}}

\author[3]{\fnm{Michael} \sur{Glikson}}

\author[1]{\fnm{Shany} \sur{Brimer Biton}}

\author[4]{\fnm{Ilan} \sur{Green}}

\author[4]{\fnm{Izhar} \sur{Laufer}}

\author[5]{\fnm{Offer} \sur{Amir}}

\author*[1]{\fnm{Joachim} \spfx{A.} \sur{Behar}}\email{jbehar@technion.ac.il}

\affil[1]{\orgdiv{Faculty of Biomedical Engineering}, \orgname{Technion-IIT}, \city{Haifa}, \country{Israel}}

\affil[2]{\orgdiv{Epidemiology Unit}, \orgname{Rambam Health Care Campus}, \city{Haifa}, \country{Israel}}

\affil[3]{\orgdiv{Jesselson Integrated Heart Center, The Eisenberg R\&D Authority}, \orgname{Shaare Zedek Medical Center, Faculty of Medicine, Hebrew University}, \city{Jerusalem}, \country{Israel}}

\affil[4]{\orgname{Leumit Health Services}, \city{Tel Aviv-Yafo}, \country{Israel}}

\affil[5]{\orgdiv{Heart Institute, Hadassah Medical Center, Faculty of Medicine}, \orgname{Hebrew University}, \city{Jerusalem}, \country{Israel}}


\abstract{Heart failure (HF) affects 11.8\% of adults aged 65 and older, reducing quality of life and longevity. Preventing HF can reduce morbidity and mortality. We hypothesized that artificial intelligence (AI) applied to 24-hour single-lead electrocardiogram (ECG) data could predict the risk of HF within five years. To research this, the Technion-Leumit Holter ECG (TLHE) dataset, including 69,663 recordings from 47,729 patients, collected over 20 years was used. Our deep learning model, DeepHHF, trained on 24-hour ECG recordings, achieved an area under the receiver operating characteristic curve of 0.80 that outperformed a model using 30-second segments and a clinical score. High-risk individuals identified by DeepHHF had a two-fold chance of hospitalization or death incidents. Explainability analysis showed DeepHHF focused on arrhythmias and heart abnormalities. This study highlights the feasibility of deep learning to model 24-hour continuous ECG data, capturing paroxysmal events essential for reliable risk prediction. Artificial intelligence applied to single-lead Holter ECG is non-invasive, inexpensive, and widely accessible, making it a promising tool for HF risk prediction.}


\keywords{Heart failure, deep learning, Holter, electrophysiology, raw physiological time series}

\maketitle

\section{Introduction}\label{sec1}
Heart failure (HF) is a complex, generally chronic syndrome, characterized by a wide range of symptoms that can interfere with daily activities and carries significant morbidity and mortality implications~\cite{Groenewegen2020EpidemiologyFailure, Rossignol2019HeartTreatment, Heidenreich20222022Guidelines}. HF is marked by a continuous abnormal cardiac function and/or structure~\cite{Groenewegen2020EpidemiologyFailure}, along with clinical manifestations of elevated filling pressures and/or reduced cardiac output. 
Risk factors include among others coronary artery disease, arrhythmias, hypertension, cardiac toxins such as alcohol, valvular heart disease, infiltrative  and inflammatory diseases, and specific genetic mutations~\cite{Heidenreich20222022Guidelines}.
Its clinical severity may be classified based on functional status, such as the New York Heart Association (NYHA) classification, six-minute walk tests and cardiopulmonary exercise tests~\cite{Bennett2002ValidityDisease, Heidenreich20222022Guidelines}.  
HF is traditionally categorized according to the ejection fraction (EF), which is usually measured by echocardiography, as preserved (HFpEF), mildly reduced (HFmrEF) or reduced (HFrEF). A more recent four-stage (A-D) classification, introduced by the American College of Cardiology/American Heart Association (ACC/AHA), focuses on the continuum of disease, from risk factors to advanced HF~\cite{Bozkurt2021UniversalFailure}.

 Globally, HF affects approximately 64 million people~\cite{Groenewegen2020EpidemiologyFailure}, with a prevalence of 1-2\% among adults in developed countries and approximately 11.8\% in adults over 65~\cite{Groenewegen2020EpidemiologyFailure}. The increasing average age of the population and longer life expectancy have contributed to the increasing incidence of HF~\cite{Benjamin2017HeartAssociation}, which is now a leading cause of hospitalization in adults over 60 years~\cite{Rossignol2019HeartTreatment}. HF progresses over several years, resulting in a gradual decline in quality of life and increased mortality~\cite{Heidenreich20222022Guidelines}. The differential diagnosis of HF can be challenging due to the overlap of symptoms of comorbidities such as renal failure, chronic obstructive pulmonary disease and liver disease. The AHA emphasizes that early management of risk factors can prevent or slow HF progression, and advocates timely detection and prevention strategies~\cite{Heidenreich20222022Guidelines, Wang2023ImportanceFraction}. Indeed, early detection of HF has been associated with reduced mortality, improved quality of life and decreased economic burden on the health system~\cite{Heidenreich20222022Guidelines}. 

Artificial intelligence (AI), particularly deep learning (DL) algorithms, has become a powerful tool for uncovering complex patterns in data, and surpasses the limitations of traditional engineered features designed based on the data, which are constrained by human assumptions. DL has already been successfully applied for electrocardiogram (ECG) analysis, with various applications, such as arrhythmia diagnosis~\cite{Ribeiro2020AutomaticNetwork}, atrial fibrillation (AF) risk prediction~\cite{Biton2021AtrialLearning}, occlusion myocardial infarction detection~\cite{Al-Zaiti2023MachineInfarction}, and identifying chronic diseases such as kidney disease~\cite{Holmstrom2023DeepDisease}. DL-based ECG research has focused mainly on 12-lead ECG analysis, which provides a snapshot of the patient’s physiological state over a short period, typically around 10 seconds. Although standardized and commonly used in medical practice, the 12-lead ECG can miss paroxysmal events that could be captured by Holter ECG, which records ECG for 24 hours or more. In the context of HF, DL based on 12-lead ECG has been used to identify individuals with an EF~$\leq35$\%~\cite{Attia2019ScreeningElectrocardiogram, Attia2019ProspectiveDysfunction}, a condition often associated with HF. Despite this seminal study, the approach of identifying low EF is inherently focused on patients with HFrEF, and therefore would miss approximately half of cases of HF that are HFmrEF or HFpEF~\cite{Dunlay2017EpidemiologyFraction, Khan2024GlobalFailure}.

\begin{figure}[t]
\centering 
\includegraphics[width=\textwidth]{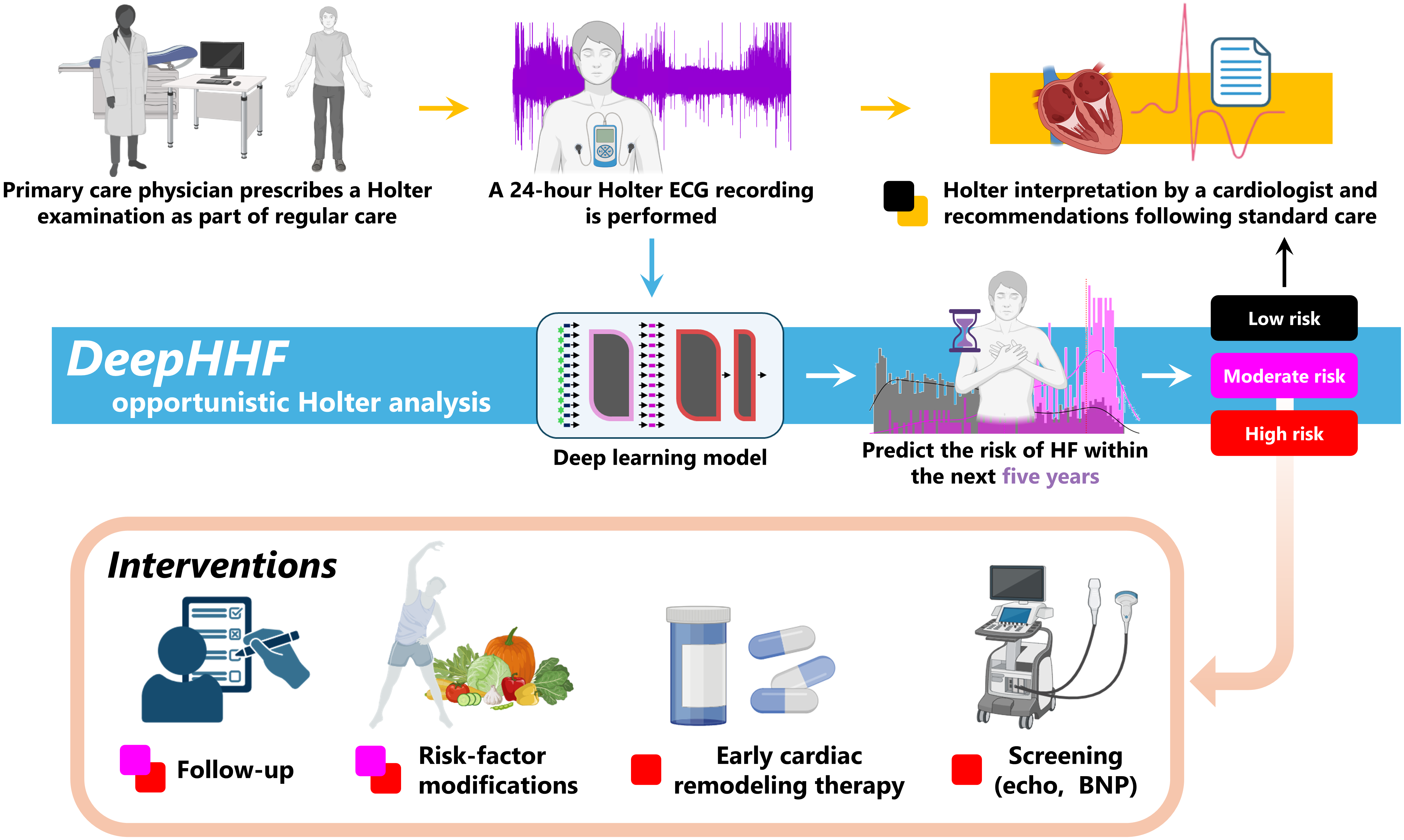}
\caption{\textbf{\textcolor{black}{Perspective clinical scenario for using the DeepHHF score.}} The patient undergoes a Holter ECG examination for regular indications, such as suspected arrhythmia or syncope. The ECG recording is then processed as an opportunistic analysis by the DeepHHF model and outputs a heart failure (HF) risk score. \textcolor{black}{Patients identified as being at moderate or high risk could be directed toward preventive actions, such as additional \textcolor{black}{predictive evaluations} with brain-type natriuretic peptide (BNP) testing or echocardiography.} The icons were created with BioRender.com.}\label{fig_overall}
\end{figure}

Since cardiovascular risk factors for HF are associated with cardiovascular conditions that can affect the heart's electrical conduction system, leading to changes in the ECG~\cite{Prinzen2022ElectricalTreatment}, identifying these patterns may enable the early detection of individuals at risk of future HF. Consequently, we hypothesized that AI applied to single-lead 24-hour ECG data can predict the risk of developing HF within five years. To test this hypothesis, we developed a large database of Holter ECG and associated 20-year-long clinical data. This work presents Deep Holter Heart Failure (DeepHHF), a DL model for 5-year HF risk prediction. Its output score can be obtained during primary care cardiac examinations and is expected to opportunistically aid in identifying high-risk individuals, thereby enabling their prioritization for periodic follow-up and interventions (Figure~\ref{fig_overall}).

\section*{Results}\label{sec2}
The main objective of this work was to develop a model capable of predicting HF risk from a single-lead ECG Holter examination performed in a primary care setting. To achieve this, the Technion-Leumit Holter ECG (TLHE) dataset was developed. The dataset included Holter ECG recordings coupled with comprehensive time-stamped demographic data, diagnoses, admissions, visits, lab results, measurements, treatments, and medications. The HF endpoint was established by extracting the first documented HF diagnosis per patient. Holter recordings within five years prior to the endpoint were considered to belong to the positive class (HF). All other recordings constituted the negative class (non-HF). \textcolor{black}{A total of 6,042 Holter examinations with a documented prior diagnosis of HF were excluded from the study cohort (Figure~\ref{fig_data})}. A DL model, denoted DeepHHF, was trained in two steps. First, an encoder was trained to extract features from independent 30-second windows of single-lead 24-hour Holter recordings. Second, features of the pre-trained encoder were fed into a sequential head that accounted for the sequential nature of the 24-hour long recording. In this way, following analysis of the entire 24-hour ECG recording, DeepHHF produced a score for probability of developing HF within 5 years.

\subsection*{Study cohort characteristics}
\begin{figure}[t]
\centering
\includegraphics[width=\textwidth]{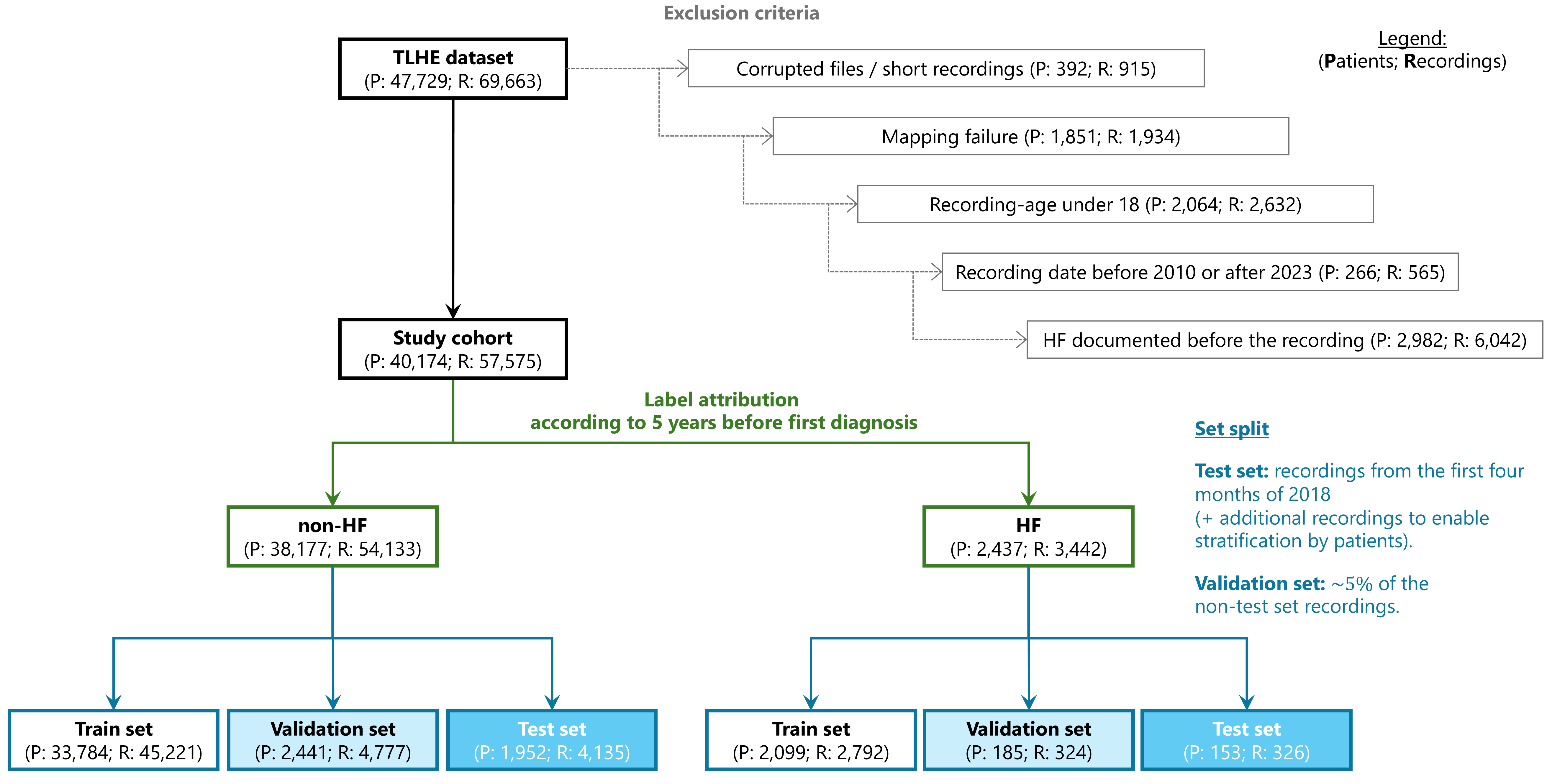}
\caption{\textbf{Cohort definition and experimental settings.} The flow diagram shows patient inclusion and exclusion criteria, the definition of the two classes with respect to heart failure diagnosis (HF) or lack thereof (non-HF) within five years of the Holter ECG recording, and the partition between training and test sets.}\label{fig_data}
\end{figure}

The TLHE dataset includes 69,663 Holter recordings from 47,729 unique patients collected from 20 primary care facilities in Israel, along with 20 years of clinical data. After applying exclusion criteria (Figure~\ref{fig_data}), and excluding recordings that had HF diagnosis documented prior to Holter examination, the study cohort consisted of 57,575 Holter ECG recordings from 40,174 unique patients, with $10,037$ patients (25.0\%) having more than one Holter recording. The median age at the recording time was 63 years (interquartile: 48–72) and 6.0\% of the recordings had an incident HF endpoint documented within 5 years of Holter examination. A total of 59.8\% of the patients were women. Furthermore, 26.2\% of the patients had a documented cardiac dysrhythmia before the recording time, 23.3\% had diabetes, 49.9\% had hypertension, and 25.4\% had ischemic heart disease (including myocardial infarction and acute coronary events). To prepare the study cohort data for the DL pipeline, the recordings were divided into a training set and a test set. \textcolor{black}{Specifically, the test set was defined to include all consecutive Holter recordings from the first four months of 2018, ensuring that all test set patients had a complete 5-year follow-up period.} A patient-specific stratification was applied to prevent possible information leakage between sets. The characteristics of the cohort are presented in Table~\ref{tab1}.

\subsection*{Deep learning}
Figure~\ref{fig_results_model}a shows the results obtained for the DL encoder model trained on 30-second window and DeepHHF which was trained on the entire 24-hour-long recording. When training on individual 30-second windows, the DL model yielded an area under the receiver operating characteristic curve (AUROC) of 0.77. \textcolor{black}{In this encoder-only setting, to establish a baseline representing a standard short-term clinical ECG spot-check, the model's performance was evaluated using a single 30-second segment extracted exactly one hour into the recording for each patient, ensuring a recording-level comparison.} When adding the sequential transformer head to the trained encoder in order to train the model over the full 24-hour recording, DeepHHF achieved AUROC=0.80. When separating between recordings belonging to male and female patients, AUROC remained 0.80 for both groups. The performance of the DeepHHF model was compared with that of the PCP-HF score~\cite{Khan201910-YearPopulation} for HF risk prediction, as recommended by the 2022 AHA/ACC/HFSA guidelines~\cite{Heidenreich20222022Guidelines}. The PCP-HF score classification performance had an AUROC=0.74 (Figure~\ref{fig_results_model}b), which was significantly (\textit{p}-value $<0.05$) lower than the DeepHHF AUROC. All performance reported are for the test set. Moreover, precision-recall curves for the different reported models are shown in Supplementary Fig. 8. 

\textcolor{black}{To investigate whether comorbidities were relied upon as shortcut features, a stratified subgroup analysis was performed. AF was selected as the most prominent potential shortcut feature due to its distinct ECG patterns. After patients with a prior AF or flutter diagnosis were excluded, DeepHHF was evaluated on an AF-absent subgroup (n=3,139). Strong discriminative performance was maintained (AUROC 0.82; 95\% CI: 0.79-0.84; Supplementary Fig. 11), confirming that predictive value is not strictly dependent on this prominent shortcut feature.}

\textcolor{black}{To evaluate whether incorporating demographic and comorbidity history variables from the TLHE dataset could improve model performance, these were combined with the DeepHHF score as input to a logistic regression classifier. Incorporating the patient variables listed in Table~\ref{tab1} resulted in an AUROC of 0.82 (Supplementary Fig. 7a). Similarly, augmenting the DeepHHF score with the variables used to compute the PCP-HF score also yielded an AUROC of 0.82 (Supplementary Fig. 7b).}

\subsection*{Error analysis}
\textcolor{black}{When evaluating DeepHHF performance according to discrete subgroups defined by the time interval between Holter examination and the first documented HF diagnosis (Figure~\ref{fig_results_error}a), the model demonstrated consistent performance with an AUROC=0.81 for examinations performed within the first two years prior to diagnosis. Performance remained strong for examinations conducted 2-4 years prior (AUROC=0.79) and 4-5 years prior (AUROC=0.80). To further confirm the model's ability to predict future events rather than merely detect pre-existing or subclinical disease, a lagged sensitivity analysis was performed. When progressively excluding patients diagnosed with HF in the early years following their Holter recording (e.g., excluding events within 1 to 3 years), overall performance experienced only a minimal reduction (AUROC shifting from 0.80 to 0.79). This underscores the model's robust long-term prognostic value.} In a second analysis, Kaplan-Meier survival curves were produced for both all-cause mortality and either cardiac/internal department hospitalization or all-cause mortality (Figure~\ref{fig_results_error}b). An 90\% specificity threshold, defined as the decision boundary separating between the high-risk group versus the low/moderate-risk groups in Figure~\ref{fig_results_model}c, was used to generate the curves for the false positive (FP), true negative (TN), true positive (TP), and false negative (FN) groups in the test set. The TP, FP and FN groups exhibited significantly lower (\textit{p}-value $<0.05$) survival rates than the TN group. Furthermore, the TP, FP and FN groups exhibited a higher proportion of comorbidities associated with HF compared to the TN group (Figure~\ref{fig_results_error}c). Of particular mention, the findings highlighted that while the FP group had no documented HF within five years of Holter examinations, the patients were in poor cardiovascular health. This is reflected by the low survival, similar to the TP group (Figure~\ref{fig_results_error}b), as well as the cardiovascular comorbidities (Figure~\ref{fig_results_error}c) with prevalence very similar to those in the TP group. In contrast, patients in the TN group had a much lower proportion of these comorbidities and had better survival rate than patients in the TP and FP groups.

\subsection*{Clinical value}
Figure~\ref{fig_results_model}c presents the distribution of the class (HF or non-HF) prediction by the DeepHHF model. Low, moderate, and high-risk groups were defined based on probability thresholds. A specificity of 70\% was used to distinguish low-risk from non-low-risk groups, while a specificity of 90\% was applied to separate non-high-risk from high-risk groups. The lower panels in Figure~\ref{fig_results_error}b show the Kaplan-Meier curves for the low, moderate and high risk groups. The TP and FP groups had the same survival trajectory, which emphasized that all individuals identified by the model as high-risk indeed had low survival. A comparison between the encoder part of the model based on 30-second recording inputs (dashed lines), versus the full DeepHHF model based on the whole 24-hour recording as input (solid lines) was made. The DeepHHF separated the different risk groups better, distinguishing more between patient who are prone to death and hospitalization incidents. 
Additionally, odds ratios were calculated to compare between the two groups, yielding an odds ratio of four for mortality and two for hospitalization or death. This indicates that individuals in the high-risk group are four times more likely to experience death and twice more likely to experience either hospitalization or death compared to those in the low/moderate-risk group.

\textcolor{black}{Following post-hoc calibration, the DeepHHF model demonstrated strong alignment between the predicted 5-year HF risk and the observed HF incidence in the test set, \textcolor{black}{achieving an expected calibration error (ECE) of 0.0166} (Supplementary Fig. 10). To understand the clinical implications of our previously defined risk strata within this calibrated space, we mapped our key operating points to the newly calculated absolute probabilities. The thresholds corresponding to 70\% and 90\% specificity translated to calibrated 5-year HF probabilities of approximately 7\% and 18\%, respectively. These operating points successfully separate the patient population into distinct groups with progressively higher observed HF incidence, confirming that the model's risk stratification remains robust and clinically meaningful even when evaluated in terms of absolute probability.}

\textcolor{black}{In terms of computational considerations, DeepHHF is a compact model by contemporary standards ($\approx 13$ million parameters), and end-to-end inference for a complete 24-hour Holter recording requires only a few milliseconds on modern hardware. Consequently, computational cost is negligible relative to currently available clinical and research AI infrastructure and is unlikely to represent a practical limitation for large-scale deployment.}

\subsection*{\textcolor{black}{Model explainability}}
Explainability analysis was conducted in order to provide insights on how DeepHHF makes its decisions. Figure~\ref{fig_results_explain}a shows an example of a 24-hour ECG recording from the positive class. The green curve highlights time intervals with high model attention. When using unsupervised learning to cluster heartbeats with a high level of attention, four clusters were discovered. Figure~\ref{fig_results_explain}b shows averaged beats reflecting these four clusters. The morphology of these averaged beats resembled: 1)~premature ventricular contraction with two clusters displayed in yellow and gray; 2)~normal sinus rhythm with one cluster displayed in green; and 3)~supraventricular ectopy with a narrow QRS and no P-wave displayed in pink.

\subsection*{External Validation}
DeepHHF was further evaluated on an external cohort from Rambam Health Care Campus (Haifa, Israel), consisting of 150 Holter examinations from unique patients without a HF diagnosis at baseline. Of these, 29 patients were diagnosed with HF within the following three years \textcolor{black}{Supplementary Table A3 summarize the external cohort characteristics)}. DeepHHF was applied to these recordings in a zero-shot setting (i.e., no fine-tuning), achieving an AUROC of 0.81 (95\% CI: 0.68-0.92). This analysis supports the validity of the model in a broader, more generalizable context.
\begin{figure}[t]
\centering
\includegraphics[width=0.8\textwidth]{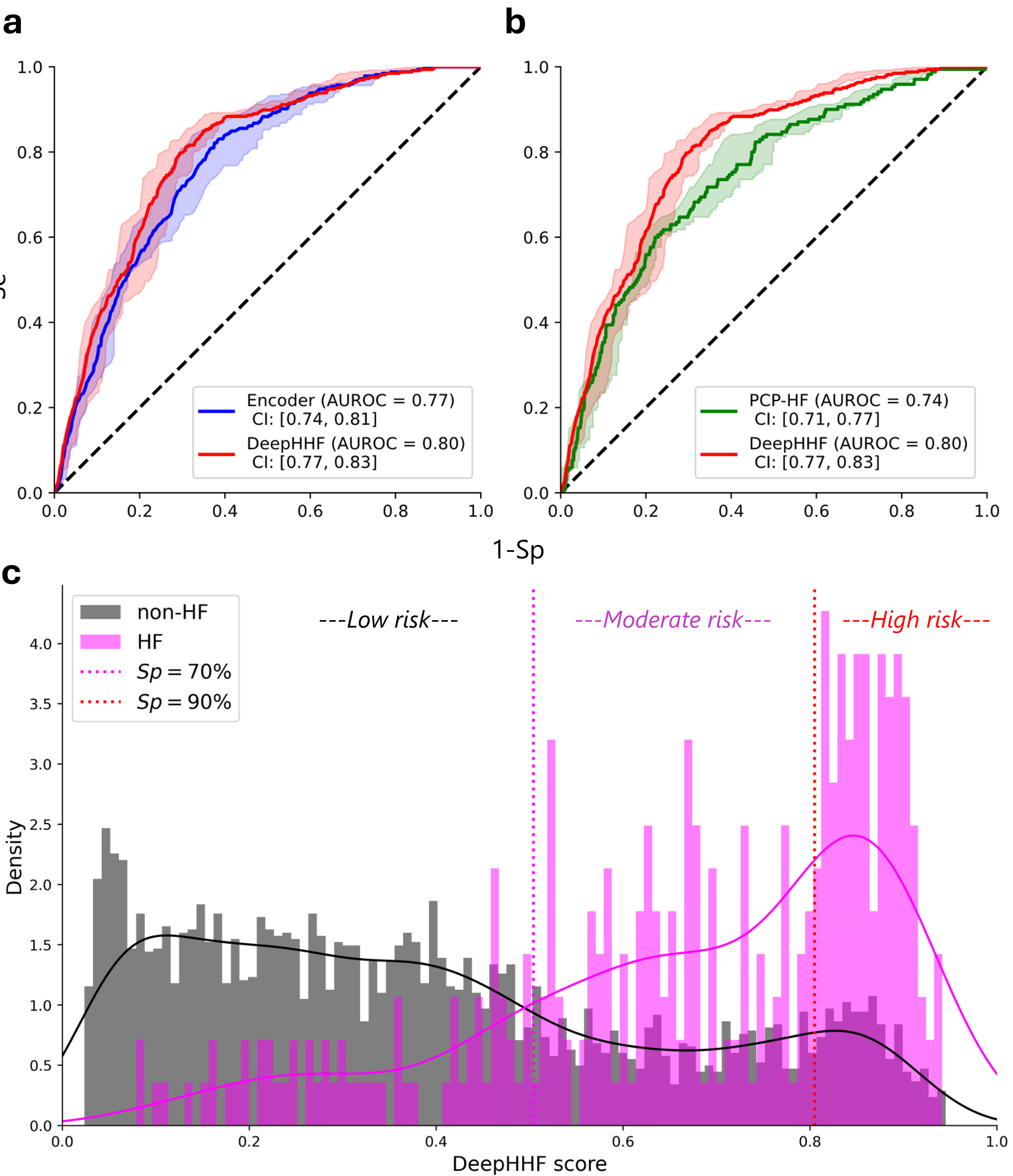}
\caption{\textbf{Models' performance}. Receiver operating characteristic curves (ROC) were produced for the test set. The area under ROC (AUROC) scores are provided with 95\% confidence intervals (CI), shown as shaded area and evaluated by bootstrapping the test set. \textbf{a,}~DeepHHF taking as input the 24-hour single ECG recording (red) vs. the encoder model taking as input a single 30-second ECG window (blue). \textcolor{black}{For the encoder-only analysis, performance was evaluated using a single 30-second segment extracted exactly one hour into the recording per patient, simulating a standard clinical short-term ECG spot-check and ensuring a recording-level comparison (e.g., 1 Holter = 1 Outcome).} \textbf{b,}~DeepHHF vs. the PCP-HF clinical score~\cite{Khan201910-YearPopulation} (green). The PCP-HF ROC curve is produced based on 1,917 test set examinations for which all the PCP-HF necessary variables were available (see Methods section). \textbf{c,}~DeepHHF probability output distribution, separating the non-HF (gray) and HF (purple) classes. Sub-groups were divided into low-, moderate- and- high-risk for HF according to the model probability risk scores. The moderate- and high-risk groups were defined by a probability decision threshold corresponding to Sp of 70\% and 90\%, respectively.}\label{fig_results_model}
\end{figure}


\begin{figure}[t]
\centering
\includegraphics[width=0.9\textwidth]{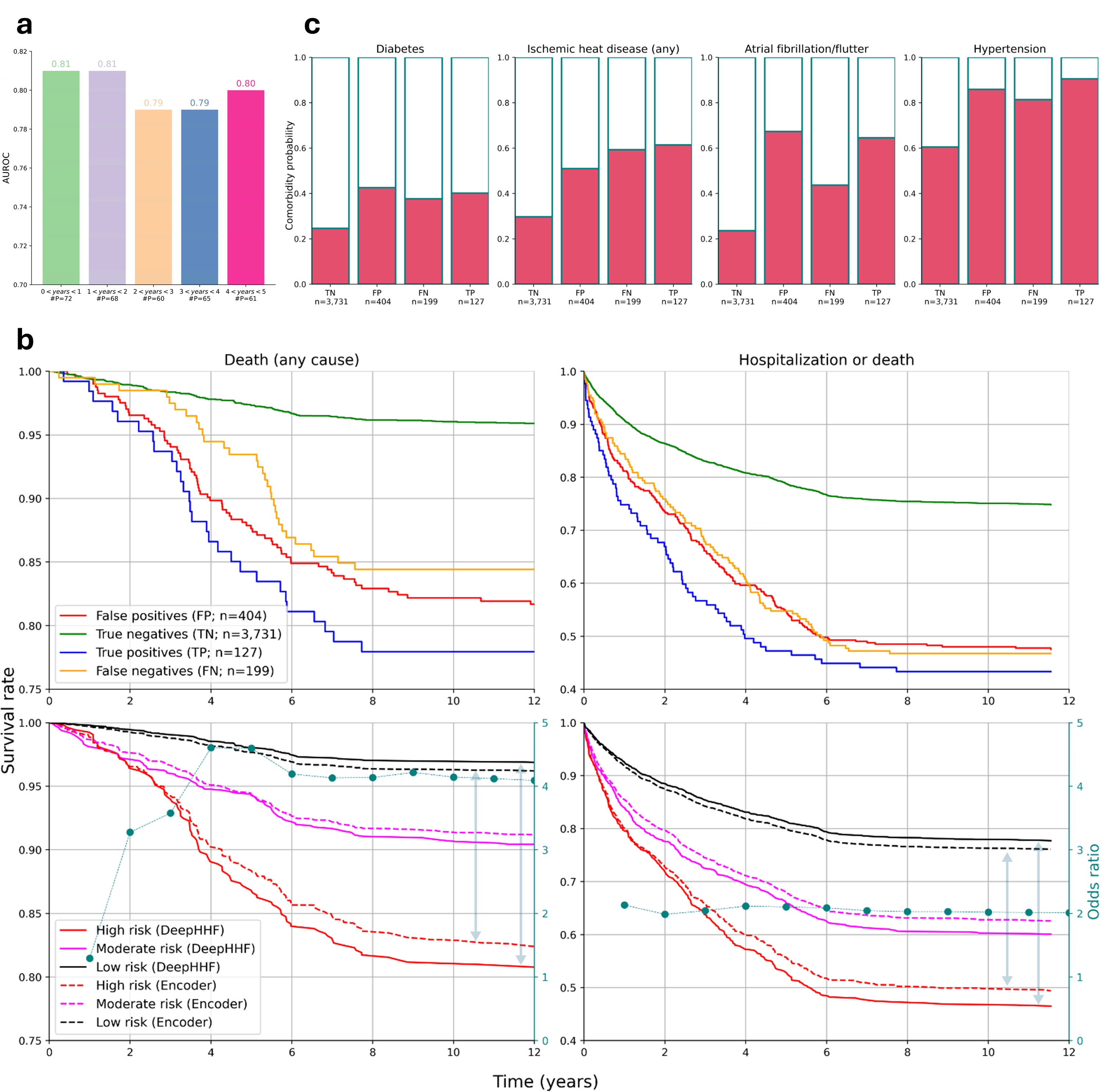}
\caption{\textbf{Error analysis and clinical value of the DeepHHF model.} The analysis is reported for the test set. \textbf{a,}~Model performance for subgroups according to the time interval between the recording date and the first documented diagnosis of HF. Area under receiver operating characteristic curves (AUROC) are shown for each subgroup. $\#P$ is the number of positive examples contained in each interval. \textbf{b,}~Kaplan-Meier analysis for death of any cause (left panels) and cardiac/internal department hospitalization or death of any cause (right panels). Upper panels show the Kaplan-Meier curves for the FP, TN, TP and FN groups. Lower panels show the Kaplan-Meier curves for the low, moderate and high risk groups as identified by the encoder (30-second window input) and by the DeepHHF model (full 24-hour recording input). The arrows illustrate the performance gap difference between using the short and long recordings. The odds ratio between the high and low/moderate survival curves are shown in the right y-axis (teal). \textcolor{black}{\textbf{c,}~Prevalence of known HF risk factors in the false positive (FP), true negative (TN), true positive (TP), and false negative (FN) groups, based on data available prior to the first HF diagnosis.}}\label{fig_results_error}
\end{figure}


\begin{figure}[t]
\centering
\includegraphics[width=\textwidth]{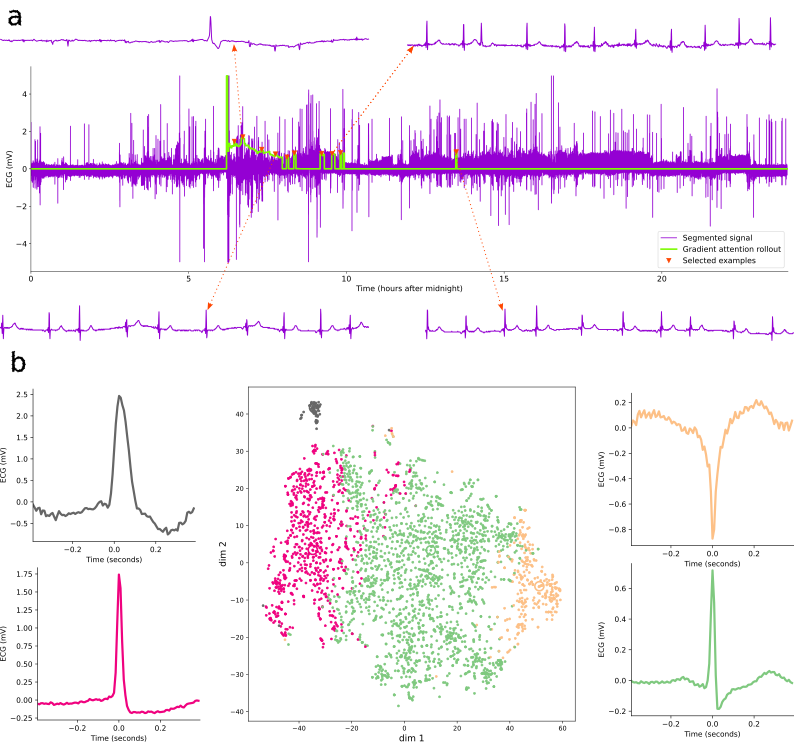}
\caption{\textcolor{black}{\textbf{DeepHHF model explainability analysis.}} The analysis was performed with the gradient attention rollout method~\cite{Abnar2020QuantifyingTransformers, GildenblatExploringTransformers}. \textbf{a,}~A single 24-hour recording example (purple) and associated gradient attention rollout map (green). The intervals with high attention were sampled (red triangles) for characteristic beat extraction.
Four examples of 10-second ECG segments in regions of high attention are shown. \textbf{b,}~\textbf{Middle panel:}~t-SNE visualization for the k-means clustering, applied on the beats extracted from HF-class recording. Each point in the graph corresponds to a single beat, with reduced dimensions by t-SNE from 100 ECG samples ($\sim0.78 ms$) into 2 dimensions (dim~1, dim~2). The colors represent the four clusters found.~\textbf{Side panels:} The beats from each cluster were averaged, obtaining characteristic beats for each cluster.}\label{fig_results_explain}
\end{figure}

\section*{Discussion}\label{sec3}

The main contribution of this research lies in the modeling of continuous 24-hour ECG recordings for HF risk prediction. To our knowledge, no previous work has explored the feasibility of predicting 5-year HF risk directly from raw ECG data. Although earlier studies have developed DL models using short 12-lead ECG recordings for cardiovascular risk prediction tasks~\cite{Biton2021AtrialLearning, Khurshid2022ECG-BasedFibrillation, Prifti2021DeepSyndrome, Hughes2023ADisease}, none have leveraged continuous 24-hour raw ECG recordings for this purpose. By using the full 24-hour signal, it is possible to capture temporal patterns that may be missed in shorter recordings, potentially improving prediction performance.

The 2022 AHA/ACC/HFSA guidelines~\cite{Heidenreich20222022Guidelines} recommended the use of risk scores to screen patients at high risk of developing HF. Among these, PCP-HF~\cite{Khan201910-YearPopulation} (2019) was highlighted as the most recent score, allowing evaluation of patient risk for HF over a 10-year period, with performance ranging from $0.74$ to $0.87$~\cite{Khan201910-YearPopulation} and $0.69$ to $0.84$~\cite{Khan2022ValidationOrganization} in race- and sex-specific models. Although DeepHHF was found to perform significantly (\textit{p}-value $<0.05$) better (AUROC=0.80) than PCP-HF (AUROC=0.74) on this research test set, it is important to note that the PCP-HF score was originally developed to predict the risk of HF among people aged 30-79 years without a history of cardiovascular disease. Alternative risk prediction scores for HF exist for groups with a cardiovascular history~\cite{Sinha2021Risk-BasedFailure}, but have shown modest performance.

DL analysis of ECG records has previously shown promise for HF detection. Attia et al.~\cite{Attia2019ScreeningElectrocardiogram, Attia2019ProspectiveDysfunction} developed a model using 12-lead ECG data to identify systolic cardiac dysfunction ($EF\leqslant35\%$). Their model achieved AUROC scores of 0.93 on a retrospective cohort~\cite{Attia2019ScreeningElectrocardiogram} and 0.91 on a prospective cohort~\cite{Attia2019ProspectiveDysfunction}. Despite this seminal research, the model was intended to detect HF prevalence rather than to predict the risk of new HF occurrences. Furthermore, the approach, which focuses on EF estimation and identifies only individuals with low EF, defined in their work as less than 35\%, inherently would overlooks nearly half of HF patients with either mildly reduced or preserved EF; HFmrEF and HFpEF respectively. On the contrary, our study addresses this unmet clinical need by identifying patients at risk for HF across the entire spectrum of EF subtypes. \textcolor{black}{Additionally, several recently published AI-ECG models have demonstrated the ability of 12-lead short-term ECG recordings to predict future HF occurrences. Some of these models relied on raw voltage time-series data, while others used image-based representations of ECG printouts. When extracting information solely from the ECG data, Dhingra et al~\cite{dhingra_heart_2025} reported C-statistics ranging from 0.72 to 0.81 in three multinational cohorts. Similarly, Butler et al.~\cite{butler_generalizable_2023} achieved an AUROC of 0.77 for predicting 10-year HF risk, and Akbilgic et al.~\cite{akbilgic_ecg-ai_2021} reported an AUROC of 0.76 in the ARIC study cohort. In contrast, DeepHHF leverages long-term single-lead Holter recordings and achieves an AUROC of 0.80, which is comparable or exceeds previously reported performances. However, those studies~\cite{butler_generalizable_2023, dhingra_heart_2025, akbilgic_ecg-ai_2021} were conducted in cohorts where HF was defined primarily by hospitalization events, whereas DeepHHF was evaluated using a broader definition of HF onset, largely captured through primary care diagnoses.}

A number of studies have focused on ECG-based risk prediction for cardiovascular and systemic diseases using short 12-lead recordings, typically around 10 seconds~\cite{Biton2021AtrialLearning, Khurshid2022ECG-BasedFibrillation, Prifti2021DeepSyndrome, Hughes2023ADisease}. Because ECG segments of approximately 30 seconds are commonly used in clinical practice, we adopted this duration as both the basic feature-extraction unit of the model and a reference baseline to assess the added value of long-term ECG modeling for HF risk prediction. \textcolor{black}{By evaluating the encoder on a single 30-second snapshot per patient, we confirmed that while a short-term spot-check provides a solid baseline, modeling the full 24-hour sequence to capture circadian dynamics and paroxysmal events yields a significantly improved predictive performance.} From an engineering perspective, modeling 24-hour raw ECG data is substantially more complex due to the large data volume and the need to capture long-range temporal dependencies. Nevertheless, using the full 24-hour ECG signal compared with a 30-second window resulted in a significant improvement in AUROC, from 0.77 to 0.80 (Figure~\ref{fig_results_model}a), indicating clear benefit from long-term modeling. This improvement is consistent with the paroxysmal nature of certain arrhythmias, such as AF or ectopic beats, which may require prolonged monitoring to be reliably detected and characterized.

\textcolor{black}{The performance of the model was highest for people diagnosed with HF within two years of Holter examination (AUROC of 0.81, Figure~\ref{fig_results_error}a), but remained remarkably robust for those diagnosed 3-5 years later (AUROC between 0.79 and 0.80). While it is expected that imminent HF is associated with a greater burden of cardiac abnormalities, our lagged sensitivity analysis showed only a minimal performance drop to an AUROC of 0.79 when excluding events occurring within 1 to 3 years post-recording. This confirms that DeepHHF's predictive power is not overly dependent on the detection of existing subclinical disease. Instead, the model is capable of capturing early electrophysiological patterns that reliably predict long-term incident HF.} In contrast, HF prediction several years before onset is more challenging, as it requires the model to detect more subtle changes in the ECG, while also adjusting for variability introduced by factors such as lifestyle changes and medical interventions. In addition, patients newly diagnosed with HF may be recorded in the electronic medical record (EMR) of the health maintenance organization (HMO) several months after their initial diagnosis. Consequently, undiagnosed prevalent HF cases with active symptoms and treatments at the time of Holter recording could influence the ECG data and potentially inflate model performance.
For this reason, it was important to observe that the performance of the model remained consistent within the first two years, as shown in Figure~\ref{fig_results_error}a. This shows the ability of the model to predict the risk of HF even in patients whose diagnosis was documented more than a year after their Holter examination.
\textcolor{black}{In addition, the TLHE dataset includes three ECG leads per Holter recording. DeepHHF was developed with a focus on single-lead ECG analysis to align with the modern development of single-lead ECG wearables, such as Zio by iRhythm Technologies, OmegaSnap by Bittium, and ePatch by Philips. \textcolor{black}{During the initial prototyping phase of this study, we conducted preliminary experiments testing various architectures using all three leads as a three-channel input on a small development subset. These early proof-of-concept tests yielded comparable predictive performance between the single-lead and multi-lead setups. Given the substantial computational complexity required to model long-range temporal dependencies in 24-hour continuous raw ECG data, and driven by our goal to maximize practical clinical applicability for single-lead devices, we elected to keep the input restricted and focus exclusively on optimizing the single-lead architecture. Future research may better leverage multi-lead information through alternative modeling strategies applied to fully optimized architectures.}}

The error analysis showed that specific comorbidities are associated with the FP group (Figure~\ref{fig_results_error}c). Specifically, comorbidites that are known risk factors for the development of HF~\cite{Heidenreich20222022Guidelines}, including hypertension, diabetes, ischemic heart disease and cardiac dysrhythmia, were observed to be highly prevalent in the FP group and in a proportion similar to that of the TP group, while having higher rates compared to the TN group. This observation suggests that DeepHHF is capable of learning subtle patterns in the ECG that are associated with cardiac electrical abnormalities as well as metabolic disorders and kidney failure (Supplementary Fig. 1). The similarity between the survival curve trajectories of the FP and TP groups (Figure~\ref{fig_results_error}b) highlights some model's additional non HF-specific clinical value as well.

DeepHHF has significant clinical value in identifying patients at high risk for HF. Our findings indicate that patients identified as high-risk had a forthcoming fourfold increased likelihood of all-cause mortality compared to low/moderate-risk patients. Similarly, high-risk patients had a forthcoming twofold increased likelihood of either all-cause death or cardiac/internal department hospitalization (Figure~\ref{fig_results_error}b). A recent randomized clinical trial~\cite{lin_ai-enabled_2024} demonstrated that using AI-ECG to identify patients with a high-risk of mortality can enable timely clinical interventions and reduce mortality. DeepHHF can also be applied to select high-risk populations in clinical trials focused on HF prevention. Since DeepHHF requires only a single-lead ECG without additional clinical variables, it can easily be integrated into standard Holter analysis software or utilized in modern ECG devices such as single-lead ECG patches.


\textcolor{black}{The explainability analysis was based on the gradient attention rollout method~\cite{Abnar2020QuantifyingTransformers, GildenblatExploringTransformers}. While traditional raw attention weights have been criticized for lacking causal grounding~\cite{jain_attention_2019}, this limitation was addressed by combining attention weights with gradient information to more accurately trace the influence of inputs on the model’s output. This approach has been shown to achieve better alignment with attribution methods based on input gradients and to provide more reliable explanations~\cite{chefer_transformer_2021, jo_gmar_2025}. By incorporating gradients directly into the explainability maps, concerns associated with purely visualizing raw attention are mitigated.}

Using this approach, the averaged beat analysis (Figure~\ref{fig_results_explain}b), based on the HF recordings, identified four types of characteristic ECG beats associated with the development of HF. \textcolor{black}{In particular, abnormal beat morphologies were identified that resemble premature ventricular contractions (two clusters in yellow and gray) and supraventricular ectopic beats (in pink). These morphologies indicate the presence of ectopic beats, which are known to be associated with HF development~\cite{Dukes2015VentricularDeath}. Furthermore, absent P waves were observed, a feature known to be associated with AF, which is a common comorbidity of HF~\cite{Maisel2003AtrialTherapy,Bergau2022AtrialFailure}.}

To evaluate the potential value of DeepHHF in improving clinical outcomes, we sought quantitative evidence on the effectiveness of preventive strategies in individuals at high-risk for developing HF. The STOP-HF randomized control  trial~\cite{ledwidge_natriuretic_2013} assessing the impact of brain-type natriuretic peptide (BNP) screening followed by HF prevention strategies demonstrated that the interventions implemented in individuals with elevated BNP levels ($\geq$ 50 pg/mL) effectively reduced emergency hospitalizations for major cardiovascular events (MACE), from an incidence rate ratio (IRR) of 0.60. However, in an editorial accompanying the trial results, Hernandez et al.~\cite{hernandez_preventing_2013} raised concerns about the feasibility of widespread BNP screening, and highlighted the need for targeted approaches to identify high-risk patients. Our findings show that DeepHHF improves the identification of moderate- and high-risk groups, thereby enabling reductions in both the number of patients requiring BNP \textcolor{black}{evaluation} and the number needed to screen (NNS) in order to prevent MACE hospitalizations. Specifically, when applying the settings from the STOP-HF trial~\cite{ledwidge_natriuretic_2013}, we observed MACE hospitalization incidence rates of 41.5 (compared with 40.4 reported by STOP-HF), 85.0, and 124.0 per 1,000 patient-years for the entire test set, moderate-risk subgroup, and high-risk subgroup, respectively. The corresponding NNSs were 61, 30, and 21, suggesting that DeepHHF significantly reduces the NNS. These results highlight the model’s potential to prioritize high-risk patients for preventive HF interventions. Additionally, when comparing the performance of DeepHHF using 24-hour Holter recordings with the encoder model using only 30-second ECG inputs, it was found that DeepHHF improved risk \textcolor{black}{prediction}. The encoder model yielded higher NNSs of 32 and 23 for moderate- and high-risk subgroups, respectively, compared to NNSs of 30 and 21 for DeepHHF. \textcolor{black}{When implemented at a population scale, this reduction in NNS translates to a clinically meaningful decrease in FPs, thereby averting unnecessary downstream testing and reducing healthcare burden.} These findings suggest that DeepHHF has the potential to optimize resource utilization and enhance clinical outcomes by focusing preventive care on patients most likely to benefit from it.

\textcolor{black}{In this context, DeepHHF is intended for risk prediction rather than calibrated absolute risk estimation. As shown in Figure~\ref{fig_results_model}c, the 90\% specificity threshold defines a high-risk group to limit unnecessary referrals, while the 70\% threshold identifies a moderate-risk group warranting increased clinical attention. The clinical relevance of these thresholds is supported by distinct survival trajectories (Figure~\ref{fig_results_error}b) and associated reductions in NNS. Future work should focus on refining threshold selection, assessing calibration where absolute risk communication is required, and validating these findings in larger external cohorts.}

This research had several limitations. First, despite the large dataset used for the experiments and the fact that the ECG data was collected in 20 different primary care facilities, the model should be evaluated in other centers around the world to cover a wider range of ethnicity, comorbidities, and medical center practices. \textcolor{black}{In this work, an additional dataset from Rambam Health Care Campus (Haifa, Israel) was obtained to support the external validation of our model. However, this dataset is of moderate size and may have limited variability, potentially reducing its ability to represent populations that differ substantially from the internal cohort.} Furthermore, the large cohort used in this research was not defined based on systematic clinical or demographic information, and therefore other Holter ECG samples may be different with varying clinical practice and policy of the healthcare system. This further emphasizes the importance of using external test sets in future work.  \textcolor{black}{An additional limitation is that all recordings were acquired using a single Holter device and a fixed single-lead configuration. While this ensured consistency across the dataset, generalizability to other vendors, patch-based devices, alternative acquisition settings, or variability in electrode placement and motion-related artifacts has not been established. Future work should evaluate performance across diverse devices, acquisition protocols, and signal quality conditions.}
 \textcolor{black}{Moreover, this study includes an inherent selection bias due to the use of Holter monitoring, which is typically performed for specific clinical indications. \textcolor{black}{Accordingly, DeepHHF is designed for risk prediction among patients referred for a Holter examination based on clinical indications in routine practice, and is not intended for evaluation in an unselected population-wide setting.} While this reflects real-world primary care practice, it may limit the generalizability of the findings to specific cardiovascular subgroups.}

An additional limitation of this study was that the predicted HF outcome was restricted to specific clinically documented stages of the disease, as recorded by physicians, rather than covering the full spectrum of possible HF presentations, including asymptomatic or atypical cases. \textcolor{black}{In addition, the outcome was defined as incident HF without stratification by EF phenotype (HFrEF, HFmrEF, HFpEF), which was not available to us from the HMO EMR. This limits the ability of the model to make distinctions between HF subtypes and to inform phenotype-specific clinical management or treatment considerations. Accordingly, the reported results should be interpreted as reflecting overall incident HF risk rather than risk associated with individual HF phenotypes.} This narrow focus may limit the ability of the model to detect earlier or less clearly defined stages of HF that are often missing from clinical records. However, these milder cases are generally associated with a lower clinical burden. \textcolor{black}{One additional important limitation of our study is the absence of a direct comparison between the predictive value of Holter ECG and 12-lead ECG for HF risk prediction. Indeed, while this research provided valuable empirical evidence supporting the value of long-term (24-hour) versus short-term (30-second) ECG recordings, these comparisons were based on single-lead recordings. Moreover, the use of trimmed or zero-padded recordings to match a 24-hour input length was intended to simplify the model design. \textcolor{black}{Given the tight distribution of recording durations (mean = 24.67h, median = 23.79h, IQR = 1.44h), the overall proportion of zero-padded data is minimal. Consequently, it is highly unlikely that prominent features from the padding itself were inadvertently learned by the model's attention mechanisms, and it is assumed that any potential bias was mitigated by the large size of the dataset.}} 
\textcolor{black}{Finally, regarding model explainability, this study focused on the morphological characteristics of the beats identified by the model. The temporal dynamics of the model's attention throughout the 24-hour recordings, and any potential relationship to circadian physiological effects, require further methodological alignment and investigation in future work.}
Additional improvements may be achieved through recently developed open ECG foundation models~\cite{McKeen2024ECG-FM:Model, Davies2024InterpretableData, Li2024AnDomains, Guo2023SiamAF:Detection, Abbaspourazad2023Large-scaleBiosignals} which can be exploited and fine-tuned for potentially improved performance of the downstream HF risk prediction task.

\section*{Methods}\label{sec5}

\subsection*{Technion-Leumit Holter ECG dataset elaboration}\label{subsec5_1}
\begin{figure}[t]
    \refstepcounter{imagetable} 
    \centering
    \captionof{table}{Study cohort characteristics. Holter examinations are divided into train and test sets and by whether they belong to the positive class (HF) or negative class (non-HF).}
    \includegraphics[width=0.9\textwidth]{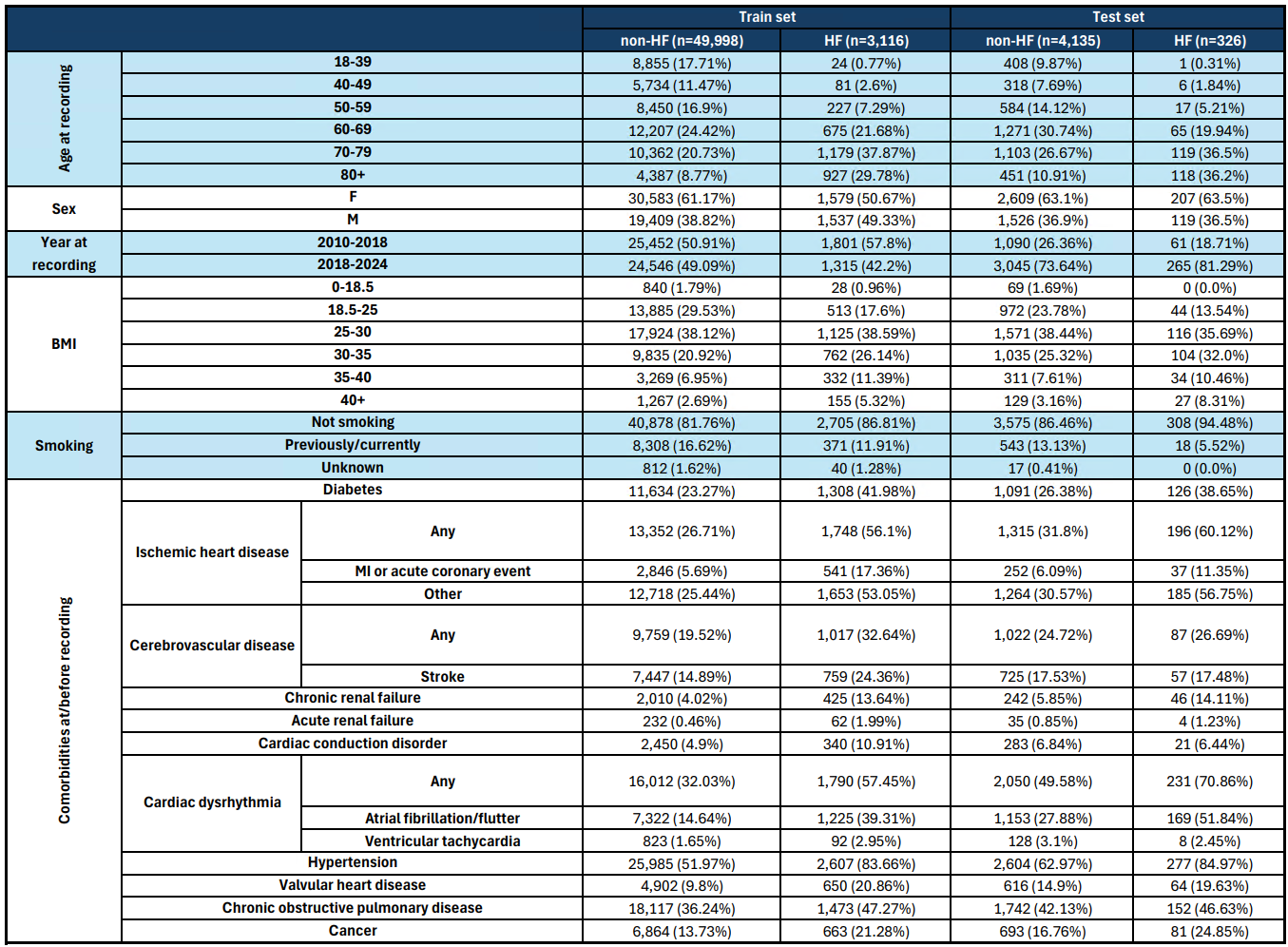}
    \label{tab1}
\end{figure}

The TLHE dataset was developed in partnership with the Israeli HMO Leumit. Leumit Health Services is one of the four HMOs active in Israel and provides medical services to more than 725,000 active members. The HMO has a comprehensive online computerized EMR database of all of its current and past members. Medical data from 2003 and onward are and have been collected from over 1.2 million individuals, allowing for a follow-up of up to 20 years. Leumit's EMR database includes data from primary and specialist community care providers, hospital admissions, laboratory test results, and pharmacies. Continuously updated mortality information was available from the national population registry. In addition to the EMR database, Leumit has access to raw Holter ECG data from corresponding patient examinations. This research included all Leumit patients with available Holter ECG recordings between 2010 and June 2023. The demographic and clinical information for the cohort is summarized in Table~\ref{tab1}.

The EMR data used was last updated on April 2024. Exclusion criteria (Figure~\ref{fig_data}) included corrupted files that could not be decoded from their binary format, recording duration below 20 hours, failure in mapping between the Holter recording and the patient EMR data, recordings from individuals under 18 years old. In addition, Holters timestamped with an inconsistent date, considered as outliers or wrongly dated, were excluded, i.e., Holters dated before 2010, which is when digitally storage of Holter data only began, or after 2023, that is after the inclusion period (Supplementary Fig. 6). Since our research interest is in risk prediction, all Holter recordings that had a diagnosis for HF documented prior to the Holter examination were discarded.  

The following variables were extracted from the EMR: patient demographics, body mass index (BMI), smoking status, medical diagnoses from both the clinic and hospital, hospital admissions, emergency department visits, clinic visits, lab results, measures (e.g., blood pressure), procedures, medication prescriptions and consumption, and death date. Data on hospital diagnoses were available starting from 2018. All diagnosis data were documented with ICD-9 codes. Prescribed medications were coded according to the Anatomical Therapeutic Chemical (ATC) Classification system. All events were dated with an absolute date. 

All Holter examinations, including patient preparation, placement of the device, and returning the device, were performed in one of 20 Leumit primary care facilities by Leumit nurses. Leumit patients may be referred for Holter examination by primary care physicians or cardiologists, with primary indications including syncope evaluation, assessment of palpitations or irregular heart rhythm, and the evaluation and follow-up of patients with known arrhythmias. The ECGs were remotely interpreted by certified cardiologists in a centralized private clinic. The Holter ECGs were recorded using the Lifecard CF device (Spacelabs Healthcare) and included 3 leads. On average, Holter recordings lasted for 24-hour. ECGs were originally sampled at $1024 Hz$ and downsampled to $128 Hz$ by the device with a dynamic range of $\pm 5 mV$ and a resolution of $2.5 \mu V$. The electrode positions for the single lead used in this research followed manufacturer's instructions: right sternal border at the level of the 2nd rib (-) and left anterior axillary lines and on the 6th ribs (+).

Information connecting between the recordings and the EMR data, in addition to complementary metadata, such as the recording date and the daytime hour of the recording start, was extracted from the binary files before de-identification. Respectively, the identifying data for the recordings was encoded in the same manner as the EMR data. Patient data were de-identified according to HMO standards and in accordance with the Israel Ministry of Health guidelines. In particular, the patient's ID and address as well as the physician's ID were encrypted and the date of birth was removed, leaving only the year of birth.

\subsection*{Class definition}\label{subsec5_2}
The endpoint was defined as the first documented diagnosis for HF in the EMR, as reported by the physician during a medical visit (clinic, emergency room or hospital). Accordingly, the endpoint date was defined as the first date out of any of the HF diagnoses documented for the patient. The ICD-9 codes that were used to identify HF diagnosis in the EMR are listed in Supplementary Table A1. These were selected after consulting with the Leumit director of family medicine (co-author IG), to ensure inclusion of all possible codes that appear as ``Heart failure" in the doctors' electronic system interface, or codes that could have been given during hospitalization. To verify the validity of the HF diagnosis endpoints, several methods also used in previous research~\cite{Khan2022ValidationOrganization} were used, as described in Supplementary section A1 and in Supplementary Fig. 2, 3, 4, 5. Each Holter recording was labeled as HF if an HF diagnosis was documented within 5-year of the examination. In contrast, Holter recordings that did not have a HF diagnosis documented within 5 years from the recording were labeled as non-HF (Figure~\ref{fig_data}).

\subsection*{Dataset split and preprocessing}\label{subsec5_4}
After applying the exclusion criteria and defining the labels, the recordings were divided into train, validation and test sets. The test set was comprised of all recordings from the first four months (January-April) of 2018. This period was selected because it marked the earliest availability of hospital diagnosis data, ensuring the reference labels for HF were as comprehensive and reliable as possible. Additionally, it allowed for at least a 5-year follow-up, as clinical data Supplementary through April 2024. To avoid information leakage, a stratification step was applied, in which all patient recordings of those who had a recording performed within this time frame (January-April 2018) were also included in the test set. Subsequently, 5\% of the remaining recordings were randomly selected to form the validation set and the same patient stratification step was applied. The remaining recordings were allocated to the train set. This process resulted in $48,013$, $5,101$, and $4,461$ recordings from $35,884$, $2,626$, and $2,105$ patients for the train, validation and test sets, respectively. \textcolor{black}{The first lead of the three available channels was selected as a fixed input for all experiments to ensure anatomical consistency and prevent dynamic selection bias. Across all Leumit Holter recordings, this specific lead corresponds to a modified Lead II vector (from the right sternal border to the left anterior axillary line). This lead was utilized because it provides a clinically robust signal for ambulatory monitoring and directly matches the primary vector used by modern single-lead ECG wearables, thereby demonstrating the model's reliability in a real-world setting.} To standardize the dataset, the length of the Holter recordings was adjusted to 24 hours\textcolor{black}{; specifically, any excess data was trimmed from the end of the recording, and shorter recordings were zero-padded at the end.}

\subsection*{Deep learning model}\label{subsec5_5}
\begin{figure}[h]
\centering
\includegraphics[width=\textwidth]{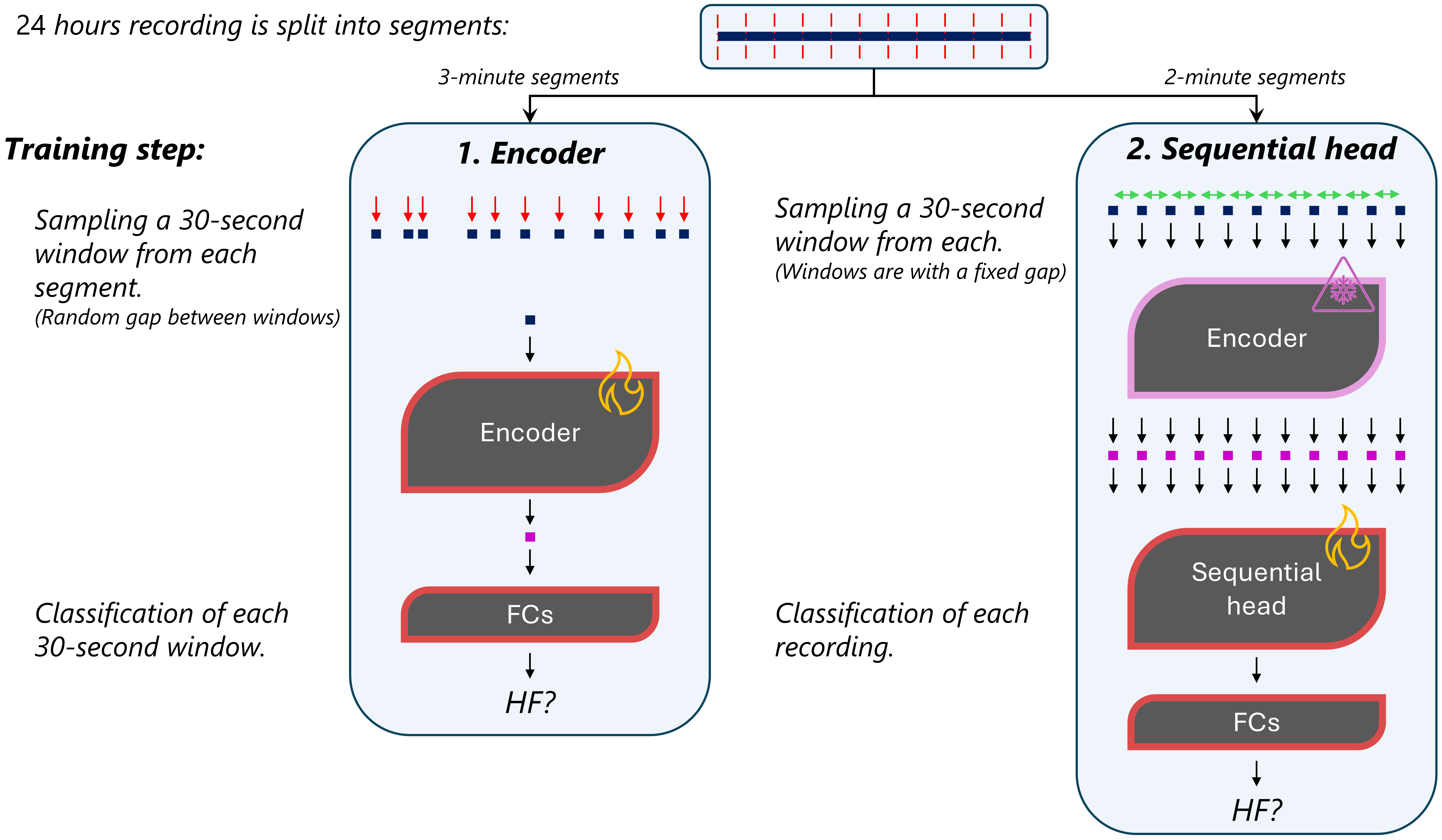}
\caption{\textbf{The two training steps and the window sampling technique in each.} \textbf{Step~1~- training the encoder:} The 24-hour recording is split into 3-minute segments, and a 30-second window is sampled from each segment. The gap between the windows is random. The fully-connected layers (FCs) output a HF score for each individual window. \textbf{Step~2~- training the sequential head:} The 24-hour recording is split into 2-minute segments and the gap between the 30-second windows is fixed, forming a sequence. The encoder's weights are frozen and it is used for feature extraction. The FCs (different set) output a HF score for the entire recording.}\label{fig_training}
\end{figure}

The DL model was trained in two steps (Figure~\ref{fig_training}). In the first step, in each epoch, 30-second windows (3,840 samples) were randomly sampled once every 3-minute segment in the Holter recordings. This resulted in 480 windows per recording. These windows were used to train an encoder for the binary classification task of interest. Correspondingly, fully connected layers (FCs, Figure~\ref{fig_training}) were used to classify the extracted features of each 30-second window independently. \textcolor{black}{While the encoder was trained on multiple sampled windows per recording, its baseline classification performance was evaluated on a strictly patient-level basis to ensure a fair comparison with the full sequence model. Specifically, for the baseline encoder evaluation, a single 30-second window extracted exactly one hour into the recording (specifically, the 21st chronological 30-second window) was used per patient to simulate a short-term clinical spot-check, avoiding initial placement noise and maintaining a '1 Holter = 1 Outcome' methodology.}
In the second step, the 30-second windows were sampled for every 2-minute segment, resulting in 720 windows per recording. In this step, the trained encoder was frozen and used as a feature generator for each 30-second windows, to train the sequential head. The extracted features were connected from the different windows to form a sequential representation of the entire 24-hour ECG recording, and to be processed by the sequential head. Then the processed sequence went through FCs (different weights than the encoder FCs) for the final whole Holter classification. Thus, in summary, the DeepHHF model generates a score for the entire Holter recording based on the extracted features.

The architecture of both parts of the model is summarized in Supplementary Fig. 9. The encoder basic building block was adapted from the EnCodec architecture~\cite{Defossez2022HighCompression}. The motivation behind this choice was to create a low-dimensional compact representation in the latent space. These were passed to the second training step, enabling a large number of encoded 30-second windows to be included while considering the entire 24-hour recording. In summary, this building block consisted of a residual block, following a strided convolutional layer, thereby substantially reducing the dimensions after each block. Overall, the encoder consisted of four building blocks, followed by two FC layers. The sequential head was built from a transformer encoder followed by two FC layers. For both steps, the loss function used was a class-weighted binary cross-entropy, with class weights calculated based on the proportion of positive to negative labels in the train set. The loss function was integrated with a sigmoid function to obtain probabilities at the FCs output. A learning rate of $10^{-3}$ was used in the first step, as attempts with different rates did not yield better convergence, whereas in the second step a learning rate of $5 \times 10^{-5}$ was used after hyperparameters search. An unlimited number of epochs was used, with early stopping applied after eight consecutive epochs with no improvement. Improvement was measured based on the AUROC over the validation set. The DL pipeline was implemented via PyTorch version 2.2.2 with Python version 3.8.19. All computations were executed on Microsoft Azure using the NC48ads A100 v4 instance (48 vCPUs, 440 GiB memory). A batch size of 32 was used in both training steps, which was tuned to maximize the memory allocation in the GPU during training.

Hyperparameters were optimized using Bayesian search via Optuna~\cite{Akiba2019Optuna:Framework} version 3.6.1. The optimization attempted to maximize the validation set AUROC score. For the encoder, the filter size of first convolution layer, size of strides and corresponding convolution filter size of the final layer inside the building block, the number of filters (same for all convolutional layers), and the dropout rate $p$ were tuned. For the sequential head, the segment size to sample the 30-second windows from (was tuned to 2 minutes, see Figure~\ref{fig_training}), and the learning rate, transformer $d_{model}$, number of heads in the transformer attention layers, number of transformer encoder layers, and the feedforward dimension (transformer linear layer size) were tuned.

\subsection*{Rambam Holter databank cohort for external validation}
\textcolor{black}{To evaluate the generalization performance of the model on an independent external test set, data were obtained from the Holter databank (IRB number: 0057-23-RMB-D) of Rambam Health Care Campus, a secondary care medical center in Haifa, Israel. Holter recordings collected between 2017 and 2018 were extracted from individuals over 30 years of age who did not have a pacemaker and were not receiving cardiac pharmacological treatment at the time of examination. Clinical follow-up data were available through November 2021. In total, 150 Holter examinations were included, comprising 29 from patients who were subsequently diagnosed with HF and 121 from patients without documented future HF (Non-HF).}

\subsection*{PCP-HF score computation}\label{subsec5_6}
The PCP-HF clinical score was calculated using the equation parameters reported by Khan et al.~\cite{Khan201910-YearPopulation} (in their Table 2) for the estimation of the 10-year risk of HF. The score was also validated for 5-year risk prediction based on other Israeli HMO EMR extracted data~\cite{Khan2022ValidationOrganization}. Since individuals with black skin make up less than 2\% of the total population in Israel~\cite{2023The2023}, only the white male and female coefficients were used, similarly to Khan et al~\cite{Khan2022ValidationOrganization}. Patient age at the time of Holter examination and smoking history were obtained from the Leumit EMR patient's demographic file. BMI and systolic blood pressure were extracted from the EMR measures file. Glucose, cholesterol, and high-density lipoprotein cholesterol (HDL-C) levels were extracted from the patient's laboratory results file. The measurement and laboratory results closest to the Holter index date, which were documented within a window of two years before to two months after this date, were used. Treatments for diabetes and hypertension were defined by at least one dispensed prescription of a relevant medication (based on ATC codes as per Khan et al.~\cite{Khan2022ValidationOrganization}) within the same time frame as the measurements and laboratory results. To measure the duration of the QRS complex, the QRS durations from the first five minutes of the ECG of the twelfth recording hour, that is, about the middle of the recording, were averaged. This process involved filtering the signal, detecting R-peaks and R-on and R-off fiducials using the NeuroKit2~\cite{Makowski2021NeuroKit2:Processing} version 0.2.9, and calculating the difference between R-off and R-on for each detected R-peak. For systolic blood pressure and glucose, the untreated coefficients were used. If any variable was missing for a given recording, then it was excluded from the ROC curve computation. Consequently, 1,917 out of 4,461 cases in the test were considered.

\subsection*{Post-hoc calibration}\label{subsec5_11}
\textcolor{black}{The model's calibration was evaluated to ensure the outputs could be interpreted as absolute risk estimates. To obtain calibrated 5-year probabilities, a post-hoc logistic regression was applied over the validation set. This regression utilized the uncalibrated DeepHHF model score as the single independent variable and the presence of incident HF within 5 years from the recording as the binary label. The fitted logistic calibrator was subsequently applied to the test set to map the raw risk scores to absolute 5-year HF probabilities. \textcolor{black}{To quantitatively and visually assess this calibration, the calibration curve and the ECE were computed using decile-based binning.}}

\subsection*{Model explainability analysis}\label{subsec5_10}
The model explainability analysis was performed using gradient attention rollout~\cite{Abnar2020QuantifyingTransformers, GildenblatExploringTransformers}. The technique involved modification of the original vision transformer code and its adaptation to 1D time-series signals. Accordingly, the multi-head attention from each transformer encoder layer (Supplementary Fig. 9) was extracted and weighted by its gradient. These were accumulated to a single attention across all the encoder layers. The different heads were averaged, and a discard ratio of $0.9$ was applied, as suggested in the original technique, simplifying visualization and analysis.

Unsupervised learning was employed to evaluate the characteristics of beats identified by DeepHHF. Time intervals with high DeepHHF attention were identified for all test set recordings classified as HF. For each of these intervals, a 10-second segment was retained, and R-peaks were detected using the jqrs peak detector~\cite{Gendelman2021PhysioZooConduction, behar_physiozoo_2023}. Beats were then segmented with 50 samples on either side of the R-peak, resulting in beat lengths of $0.78 ms$. K-means clustering was then applied to the segmented beats and silhouette analysis was used to determine the optimal number of clusters. Only clusters with a minimum of 30 beats were retained. Finally, t-SNE was employed for visualization. The K-means, silhouette and t-SNE algorithms were applied using scikit-learn~\cite{Pedregosa2011Scikit-learn:Python} version 1.3.0.


\subsection*{Kaplan-Meier curves}\label{subsec5_7}
Three HF risk groups were defined: low, moderate, and high risk. Moderate risk was defined by the probability threshold that yields a specificity of 70\% and high risk by the probability threshold that yields a specificity of 90\%. Due to the low proportion of HF patients in the cohort, high specificity is needed to limit the number of FP. Figure~\ref{fig_results_model}c shows the probability decision threshold associated with this definition.

The clinical value of DeepHHF was assessed by evaluating survival of the high risk group vs. low/moderate-risk group (Figure~\ref{fig_results_model}c). The Kaplan-Meier curves were plotted with respect to the Holter examination dates for the test set. Two survival analyses were performed: one with respect to death incidents, and the second with respect to either documented hospitalization admissions in cardiac and internal departments or death incidents. For an Holter examination, the time interval between the examination date and a follow-up hospitalization and death event were calculated. The date of the latest documented event for all individuals in the test set was used for censorship. Survival curves were produced both for the TP, FP, TN and FN groups as well as for the different risk groups. For the latest, the odds ratios between the high and moderate/low groups were computed. The encoder model survival curves were produced as well for the risk groups the same manner. The Kaplan-Meier curves were generated using the lifelines~\cite{Davidson-Pilon2019Lifelines:Python} version 0.27.8.


\subsection*{Performance measures and statistical analysis}\label{subsec5_9}
The predictive performance of the obtained classifiers was measured by the AUROC determined with the test set. The AUROC CI was estimated by bootstrapping for 1,000 iterations over the test set. Specifically, in each iteration, an AUROC was calculated for a random subset comprised of 250 non-HF examples and 250 HF examples, resulting in a total of $500$ examples from the test set. \textcolor{black}{An analogous bootstrap procedure was applied to the external validation cohort as well.} The statistical significance between two classifiers was obtained using a Student's t-test between the bootstrapped test set distributions. To compare and evaluate the statistical significance between two Kaplan-Meier curves, the logrank test was applied using the lifelines~\cite{Davidson-Pilon2019Lifelines:Python} version 0.27.8. For all statistical tests, significance was set at \textit{p}-value $<0.05$. For all CIs, the confidence level was determined at $95\%$. \textcolor{black}{Additionally, an AF-absent subgroup analysis was conducted on the test set; all recordings from patients with a prior documented EMR diagnosis of AF or flutter were strictly excluded.}

\section*{Declarations}
\textbf{Data availability}: All requests for raw data can be addressed to Leumit Start (co-author IL). Any data and materials that can be shared will be released via the standard Data Transfer Agreement of the Leumit Health Maintenance Organization. Source Data are provided with this paper.

\textbf{Code availability}: The experiments can be reproduced by utilizing the details provided in the Methods section. For DeepHHF, this includes the model architecture (Supplementary Fig. 9). Our trained model is also made publicly available at (URL upon publication) and is provided for academic research purposes and under a GNU GPL license. Main Python package versions: Python - 3.8.19; PyTorch - 2.2.2; scikit-learn - 1.3.0; lifelines - 0.27.8; Optuna - 3.6.1; NeuroKit2 - 0.2.9.

\textbf{Acknowledgments:} We thank Nuriel Burak from Leumit Health Services for his invaluable support in exporting and maintaining the data from the HMO systems, as well as for his assistance in supporting the computing system throughout the project. EZ acknowledges The Miriam and Aaron Gutwirth Memorial Fellowship. EZ and JB acknowledge the Technion EVPR Fund: Hittman Family Fund. The research was supported by a cloud computing grant from the Israel Council of Higher Education, administered by the Israel Data Science Initiative. We acknowledge the assistance of ChatGPT, an AI-based language model developed by OpenAI, for its help in editing the English language of this manuscript. 

\textbf{Author contributions}: JB and EZ conceived and designed the research. EZ developed the dataset and algorithms and performed the analysis under the supervision of JB. RA guided the epidemiological methodological design and heart failure label verification. OA provided medical guidance on interpreting the data and the clinical outlook of the research in heart failure management. MG provided guidance on interpreting the electrophysiology data and the research design. IG provided medical guidance on interpreting the clinical variables from the health maintenance organization Leumit and the results of the research. IL supported the development of the dataset, data curation, data management and cloud computing platform. JB and EZ drafted the first version of the manuscript. All authors edited, have read and approved the manuscript

\textbf{Competing interests}: The authors declare no competing financial or non-financial interests.

\textbf{Ethics approval and consent to participate}: The study protocol was approved by the statutory Leumit Health Services Institutional Review Board (approval number 0025-22 LEU). Informed consent was waived because this large-scale retrospective study was performed on deidentified electronic health records.

\balance  
\bibliography{references_Zotero}


\begin{thebibliography}{51}
\ifx \bisbn   \undefined \def \bisbn  #1{ISBN #1}\fi
\ifx \binits  \undefined \def \binits#1{#1}\fi
\ifx \bauthor  \undefined \def \bauthor#1{#1}\fi
\ifx \batitle  \undefined \def \batitle#1{#1}\fi
\ifx \bjtitle  \undefined \def \bjtitle#1{#1}\fi
\ifx \bvolume  \undefined \def \bvolume#1{\textbf{#1}}\fi
\ifx \byear  \undefined \def \byear#1{#1}\fi
\ifx \bissue  \undefined \def \bissue#1{#1}\fi
\ifx \bfpage  \undefined \def \bfpage#1{#1}\fi
\ifx \blpage  \undefined \def \blpage #1{#1}\fi
\ifx \burl  \undefined \def \burl#1{\textsf{#1}}\fi
\ifx \doiurl  \undefined \def \doiurl#1{\url{https://doi.org/#1}}\fi
\ifx \betal  \undefined \def \betal{\textit{et al.}}\fi
\ifx \binstitute  \undefined \def \binstitute#1{#1}\fi
\ifx \binstitutionaled  \undefined \def \binstitutionaled#1{#1}\fi
\ifx \bctitle  \undefined \def \bctitle#1{#1}\fi
\ifx \beditor  \undefined \def \beditor#1{#1}\fi
\ifx \bpublisher  \undefined \def \bpublisher#1{#1}\fi
\ifx \bbtitle  \undefined \def \bbtitle#1{#1}\fi
\ifx \bedition  \undefined \def \bedition#1{#1}\fi
\ifx \bseriesno  \undefined \def \bseriesno#1{#1}\fi
\ifx \blocation  \undefined \def \blocation#1{#1}\fi
\ifx \bsertitle  \undefined \def \bsertitle#1{#1}\fi
\ifx \bsnm \undefined \def \bsnm#1{#1}\fi
\ifx \bsuffix \undefined \def \bsuffix#1{#1}\fi
\ifx \bparticle \undefined \def \bparticle#1{#1}\fi
\ifx \barticle \undefined \def \barticle#1{#1}\fi
\bibcommenthead
\ifx \bconfdate \undefined \def \bconfdate #1{#1}\fi
\ifx \botherref \undefined \def \botherref #1{#1}\fi
\ifx \url \undefined \def \url#1{\textsf{#1}}\fi
\ifx \bchapter \undefined \def \bchapter#1{#1}\fi
\ifx \bbook \undefined \def \bbook#1{#1}\fi
\ifx \bcomment \undefined \def \bcomment#1{#1}\fi
\ifx \oauthor \undefined \def \oauthor#1{#1}\fi
\ifx \citeauthoryear \undefined \def \citeauthoryear#1{#1}\fi
\ifx \endbibitem  \undefined \def \endbibitem {}\fi
\ifx \bconflocation  \undefined \def \bconflocation#1{#1}\fi
\ifx \arxivurl  \undefined \def \arxivurl#1{\textsf{#1}}\fi
\csname PreBibitemsHook\endcsname

\bibitem[\protect\citeauthoryear{Groenewegen et~al.}{2020}]{Groenewegen2020EpidemiologyFailure}
\begin{barticle}
\bauthor{\bsnm{Groenewegen}, \binits{A.}},
\bauthor{\bsnm{Rutten}, \binits{F.H.}},
\bauthor{\bsnm{Mosterd}, \binits{A.}},
\bauthor{\bsnm{Hoes}, \binits{A.W.}}:
\batitle{Epidemiology of heart failure}.
\bjtitle{European Journal of Heart Failure}
\bvolume{22}(\bissue{8}),
\bfpage{1342}--\blpage{1356}
(\byear{2020})
\doiurl{10.1002/ejhf.1858}
\end{barticle}
\endbibitem

\bibitem[\protect\citeauthoryear{Rossignol et~al.}{2019}]{Rossignol2019HeartTreatment}
\begin{barticle}
\bauthor{\bsnm{Rossignol}, \binits{P.}},
\bauthor{\bsnm{Hernandez}, \binits{A.F.}},
\bauthor{\bsnm{Solomon}, \binits{S.D.}},
\bauthor{\bsnm{Zannad}, \binits{F.}}:
\batitle{Heart failure drug treatment}.
\bjtitle{The Lancet}
\bvolume{393}(\bissue{10175}),
\bfpage{1034}--\blpage{1044}
(\byear{2019})
\doiurl{10.1016/S0140-6736(18)31808-7}
\end{barticle}
\endbibitem

\bibitem[\protect\citeauthoryear{Heidenreich et~al.}{2022}]{Heidenreich20222022Guidelines}
\begin{barticle}
\bauthor{\bsnm{Heidenreich}, \binits{P.A.}},
\bauthor{\bsnm{Bozkurt}, \binits{B.}},
\bauthor{\bsnm{Aguilar}, \binits{D.}},
\bauthor{\bsnm{Allen}, \binits{L.A.}},
\bauthor{\bsnm{Byun}, \binits{J.J.}},
\bauthor{\bsnm{Colvin}, \binits{M.M.}},
\bauthor{\bsnm{Deswal}, \binits{A.}},
\bauthor{\bsnm{Drazner}, \binits{M.H.}},
\bauthor{\bsnm{Dunlay}, \binits{S.M.}},
\bauthor{\bsnm{Evers}, \binits{L.R.}},
\bauthor{\bsnm{Fang}, \binits{J.C.}},
\bauthor{\bsnm{Fedson}, \binits{S.E.}},
\bauthor{\bsnm{Fonarow}, \binits{G.C.}},
\bauthor{\bsnm{Hayek}, \binits{S.S.}},
\bauthor{\bsnm{Hernandez}, \binits{A.F.}},
\bauthor{\bsnm{Khazanie}, \binits{P.}},
\bauthor{\bsnm{Kittleson}, \binits{M.M.}},
\bauthor{\bsnm{Lee}, \binits{C.S.}},
\bauthor{\bsnm{Link}, \binits{M.S.}},
\bauthor{\bsnm{Milano}, \binits{C.A.}},
\bauthor{\bsnm{Nnacheta}, \binits{L.C.}},
\bauthor{\bsnm{Sandhu}, \binits{A.T.}},
\bauthor{\bsnm{Stevenson}, \binits{L.W.}},
\bauthor{\bsnm{Vardeny}, \binits{O.}},
\bauthor{\bsnm{Vest}, \binits{A.R.}},
\bauthor{\bsnm{Yancy}, \binits{C.W.}}:
\batitle{2022 {AHA}/{ACC}/{HFSA} guideline for the management of heart failure: {A} report of the {American} {College} of {Cardiology}/{American} {Heart} {Association} {Joint} {Committee} on {Clinical} {Practice} {Guidelines}}.
\bjtitle{Circulation}
\bvolume{145}(\bissue{18}),
\bfpage{895}--\blpage{1032}
(\byear{2022})
\doiurl{10.1161/CIR.0000000000001063}
\end{barticle}
\endbibitem

\bibitem[\protect\citeauthoryear{Bennett et~al.}{2002}]{Bennett2002ValidityDisease}
\begin{barticle}
\bauthor{\bsnm{Bennett}, \binits{J.A.}},
\bauthor{\bsnm{Riegel}, \binits{B.}},
\bauthor{\bsnm{Bittner}, \binits{V.}},
\bauthor{\bsnm{Nichols}, \binits{J.}}:
\batitle{Validity and reliability of the {NYHA} classes for measuring research outcomes in patients with cardiac disease}.
\bjtitle{Heart \& Lung}
\bvolume{31}(\bissue{4}),
\bfpage{262}--\blpage{270}
(\byear{2002})
\doiurl{10.1067/mhl.2002.124554}
\end{barticle}
\endbibitem

\bibitem[\protect\citeauthoryear{Bozkurt et~al.}{2021}]{Bozkurt2021UniversalFailure}
\begin{barticle}
\bauthor{\bsnm{Bozkurt}, \binits{B.}},
\bauthor{\bsnm{Coats}, \binits{A.J.}},
\bauthor{\bsnm{Tsutsui}, \binits{H.}},
\bauthor{\bsnm{Abdelhamid}, \binits{M.}},
\bauthor{\bsnm{Adamopoulos}, \binits{S.}},
\bauthor{\bsnm{Albert}, \binits{N.}},
\bauthor{\bsnm{Anker}, \binits{S.D.}},
\bauthor{\bsnm{Atherton}, \binits{J.}},
\bauthor{\bsnm{Böhm}, \binits{M.}},
\bauthor{\bsnm{Butler}, \binits{J.}},
\bauthor{\bsnm{Drazner}, \binits{M.H.}},
\bauthor{\bsnm{Felker}, \binits{G.M.}},
\bauthor{\bsnm{Filippatos}, \binits{G.}},
\bauthor{\bsnm{Fonarow}, \binits{G.C.}},
\bauthor{\bsnm{Fiuzat}, \binits{M.}},
\bauthor{\bsnm{Gomez-Mesa}, \binits{J.E.}},
\bauthor{\bsnm{Heidenreich}, \binits{P.}},
\bauthor{\bsnm{Imamura}, \binits{T.}},
\bauthor{\bsnm{Januzzi}, \binits{J.}},
\bauthor{\bsnm{Jankowska}, \binits{E.A.}},
\bauthor{\bsnm{Khazanie}, \binits{P.}},
\bauthor{\bsnm{Kinugawa}, \binits{K.}},
\bauthor{\bsnm{Lam}, \binits{C.S.P.}},
\bauthor{\bsnm{Matsue}, \binits{Y.}},
\bauthor{\bsnm{Metra}, \binits{M.}},
\bauthor{\bsnm{Ohtani}, \binits{T.}},
\bauthor{\bsnm{Francesco~Piepoli}, \binits{M.}},
\bauthor{\bsnm{Ponikowski}, \binits{P.}},
\bauthor{\bsnm{Rosano}, \binits{G.M.C.}},
\bauthor{\bsnm{Sakata}, \binits{Y.}},
\bauthor{\bsnm{SeferoviĆ}, \binits{P.}},
\bauthor{\bsnm{Starling}, \binits{R.C.}},
\bauthor{\bsnm{Teerlink}, \binits{J.R.}},
\bauthor{\bsnm{Vardeny}, \binits{O.}},
\bauthor{\bsnm{Yamamoto}, \binits{K.}},
\bauthor{\bsnm{Yancy}, \binits{C.}},
\bauthor{\bsnm{Zhang}, \binits{J.}},
\bauthor{\bsnm{Zieroth}, \binits{S.}}:
\batitle{Universal definition and classification of heart failure: {A} report of the {Heart} {Failure} {Society} of {America}, {Heart} {Failure} {Association} of the {European} {Society} of {Cardiology}, {Japanese} {Heart} {Failure} {Society} and {Writing} {Committee} of the {Universal} {Definition} o}.
\bjtitle{Journal of Cardiac Failure}
\bvolume{27}(\bissue{4}),
\bfpage{387}--\blpage{413}
(\byear{2021})
\doiurl{10.1016/J.CARDFAIL.2021.01.022}
\end{barticle}
\endbibitem

\bibitem[\protect\citeauthoryear{Benjamin et~al.}{2017}]{Benjamin2017HeartAssociation}
\begin{barticle}
\bauthor{\bsnm{Benjamin}, \binits{E.J.}},
\bauthor{\bsnm{Blaha}, \binits{M.J.}},
\bauthor{\bsnm{Chiuve}, \binits{S.E.}},
\bauthor{\bsnm{Cushman}, \binits{M.}},
\bauthor{\bsnm{Das}, \binits{S.R.}},
\bauthor{\bsnm{Deo}, \binits{R.}},
\bauthor{\bsnm{De~Ferranti}, \binits{S.D.}},
\bauthor{\bsnm{Floyd}, \binits{J.}},
\bauthor{\bsnm{Fornage}, \binits{M.}},
\bauthor{\bsnm{Gillespie}, \binits{C.}},
\bauthor{\bsnm{Isasi}, \binits{C.R.}},
\bauthor{\bsnm{Jim'nez}, \binits{M.C.}},
\bauthor{\bsnm{Jordan}, \binits{L.C.}},
\bauthor{\bsnm{Judd}, \binits{S.E.}},
\bauthor{\bsnm{Lackland}, \binits{D.}},
\bauthor{\bsnm{Lichtman}, \binits{J.H.}},
\bauthor{\bsnm{Lisabeth}, \binits{L.}},
\bauthor{\bsnm{Liu}, \binits{S.}},
\bauthor{\bsnm{Longenecker}, \binits{C.T.}},
\bauthor{\bsnm{MacKey}, \binits{R.H.}},
\bauthor{\bsnm{Matsushita}, \binits{K.}},
\bauthor{\bsnm{Mozaffarian}, \binits{D.}},
\bauthor{\bsnm{Mussolino}, \binits{M.E.}},
\bauthor{\bsnm{Nasir}, \binits{K.}},
\bauthor{\bsnm{Neumar}, \binits{R.W.}},
\bauthor{\bsnm{Palaniappan}, \binits{L.}},
\bauthor{\bsnm{Pandey}, \binits{D.K.}},
\bauthor{\bsnm{Thiagarajan}, \binits{R.R.}},
\bauthor{\bsnm{Reeves}, \binits{M.J.}},
\bauthor{\bsnm{Ritchey}, \binits{M.}},
\bauthor{\bsnm{Rodriguez}, \binits{C.J.}},
\bauthor{\bsnm{Roth}, \binits{G.A.}},
\bauthor{\bsnm{Rosamond}, \binits{W.D.}},
\bauthor{\bsnm{Sasson}, \binits{C.}},
\bauthor{\bsnm{Towfghi}, \binits{A.}},
\bauthor{\bsnm{Tsao}, \binits{C.W.}},
\bauthor{\bsnm{Turner}, \binits{M.B.}},
\bauthor{\bsnm{Virani}, \binits{S.S.}},
\bauthor{\bsnm{Voeks}, \binits{J.H.}},
\bauthor{\bsnm{Willey}, \binits{J.Z.}},
\bauthor{\bsnm{Wilkins}, \binits{J.T.}},
\bauthor{\bsnm{Wu}, \binits{J.H.Y.}},
\bauthor{\bsnm{Alger}, \binits{H.M.}},
\bauthor{\bsnm{Wong}, \binits{S.S.}},
\bauthor{\bsnm{Muntner}, \binits{P.}}:
\batitle{Heart disease and stroke statistics'2017 update: {A} report from the {American} {Heart} {Association}}.
\bjtitle{Circulation}
\bvolume{135}(\bissue{10}),
\bfpage{146}--\blpage{603}
(\byear{2017})
\doiurl{10.1161/CIR.0000000000000485/ASSET/065F105B-352E-48D3-AC2B-2F0A69AE4BE0/ASSETS/CIR.0000000000000485.FP.PNG}
\end{barticle}
\endbibitem

\bibitem[\protect\citeauthoryear{Wang et~al.}{2023}]{Wang2023ImportanceFraction}
\begin{barticle}
\bauthor{\bsnm{Wang}, \binits{H.}},
\bauthor{\bsnm{Gao}, \binits{C.}},
\bauthor{\bsnm{Guignard-Duff}, \binits{M.}},
\bauthor{\bsnm{Cole}, \binits{C.}},
\bauthor{\bsnm{Hall}, \binits{C.}},
\bauthor{\bsnm{Larman}, \binits{M.}},
\bauthor{\bsnm{Baruah}, \binits{R.}},
\bauthor{\bsnm{Gao}, \binits{H.}},
\bauthor{\bsnm{Mamza}, \binits{J.B.}},
\bauthor{\bsnm{Lang}, \binits{C.C.}},
\bauthor{\bsnm{Mordi}, \binits{I.}}:
\batitle{Importance of early diagnosis and treatment of heart failure across the spectrum of ejection fraction}.
\bjtitle{European Heart Journal}
\bvolume{44}(\bissue{Supplement\_2}),
\bfpage{655}--\blpage{892}
(\byear{2023})
\doiurl{10.1093/eurheartj/ehad655.892}
\end{barticle}
\endbibitem

\bibitem[\protect\citeauthoryear{Ribeiro et~al.}{2020}]{Ribeiro2020AutomaticNetwork}
\begin{barticle}
\bauthor{\bsnm{Ribeiro}, \binits{A.H.}},
\bauthor{\bsnm{Ribeiro}, \binits{M.H.}},
\bauthor{\bsnm{Paixão}, \binits{G.M.M.}},
\bauthor{\bsnm{Oliveira}, \binits{D.M.}},
\bauthor{\bsnm{Gomes}, \binits{P.R.}},
\bauthor{\bsnm{Canazart}, \binits{J.A.}},
\bauthor{\bsnm{Ferreira}, \binits{M.P.S.}},
\bauthor{\bsnm{Andersson}, \binits{C.R.}},
\bauthor{\bsnm{Macfarlane}, \binits{P.W.}},
\bauthor{\bsnm{Meira}, \binits{W.}},
\bauthor{\bsnm{Schön}, \binits{T.B.}},
\bauthor{\bsnm{Ribeiro}, \binits{A.L.P.}}:
\batitle{Automatic diagnosis of the 12-lead {ECG} using a deep neural network}.
\bjtitle{Nature Communications}
\bvolume{11}(\bissue{1}),
\bfpage{1760}--\blpage{1760}
(\byear{2020})
\doiurl{10.1038/s41467-020-15432-4}
\end{barticle}
\endbibitem

\bibitem[\protect\citeauthoryear{Biton et~al.}{2021}]{Biton2021AtrialLearning}
\begin{barticle}
\bauthor{\bsnm{Biton}, \binits{S.}},
\bauthor{\bsnm{Gendelman}, \binits{S.}},
\bauthor{\bsnm{Ribeiro}, \binits{A.H.}},
\bauthor{\bsnm{Miana}, \binits{G.}},
\bauthor{\bsnm{Moreira}, \binits{C.}},
\bauthor{\bsnm{Ribeiro}, \binits{A.L.P.}},
\bauthor{\bsnm{Behar}, \binits{J.A.}}:
\batitle{Atrial fibrillation risk prediction from the 12-lead electrocardiogram using digital biomarkers and deep representation learning}.
\bjtitle{European Heart Journal - Digital Health}
\bvolume{2}(\bissue{4}),
\bfpage{576}--\blpage{585}
(\byear{2021})
\doiurl{10.1093/EHJDH/ZTAB071}
\end{barticle}
\endbibitem

\bibitem[\protect\citeauthoryear{Al-Zaiti et~al.}{2023}]{Al-Zaiti2023MachineInfarction}
\begin{barticle}
\bauthor{\bsnm{Al-Zaiti}, \binits{S.S.}},
\bauthor{\bsnm{Martin-Gill}, \binits{C.}},
\bauthor{\bsnm{Zègre-Hemsey}, \binits{J.K.}},
\bauthor{\bsnm{Bouzid}, \binits{Z.}},
\bauthor{\bsnm{Faramand}, \binits{Z.}},
\bauthor{\bsnm{Alrawashdeh}, \binits{M.O.}},
\bauthor{\bsnm{Gregg}, \binits{R.E.}},
\bauthor{\bsnm{Helman}, \binits{S.}},
\bauthor{\bsnm{Riek}, \binits{N.T.}},
\bauthor{\bsnm{Kraevsky-Phillips}, \binits{K.}},
\bauthor{\bsnm{Clermont}, \binits{G.}},
\bauthor{\bsnm{Akcakaya}, \binits{M.}},
\bauthor{\bsnm{Sereika}, \binits{S.M.}},
\bauthor{\bsnm{Van~Dam}, \binits{P.}},
\bauthor{\bsnm{Smith}, \binits{S.W.}},
\bauthor{\bsnm{Birnbaum}, \binits{Y.}},
\bauthor{\bsnm{Saba}, \binits{S.}},
\bauthor{\bsnm{Sejdic}, \binits{E.}},
\bauthor{\bsnm{Callaway}, \binits{C.W.}}:
\batitle{Machine learning for {ECG} diagnosis and risk stratification of occlusion myocardial infarction}.
\bjtitle{Nature Medicine}
\bvolume{29},
\bfpage{1804}--\blpage{1813}
(\byear{2023})
\doiurl{10.1038/s41591-023-02396-3}
\end{barticle}
\endbibitem

\bibitem[\protect\citeauthoryear{Holmstrom et~al.}{2023}]{Holmstrom2023DeepDisease}
\begin{barticle}
\bauthor{\bsnm{Holmstrom}, \binits{L.}},
\bauthor{\bsnm{Christensen}, \binits{M.}},
\bauthor{\bsnm{Yuan}, \binits{N.}},
\bauthor{\bsnm{Weston~Hughes}, \binits{J.}},
\bauthor{\bsnm{Theurer}, \binits{J.}},
\bauthor{\bsnm{Jujjavarapu}, \binits{M.}},
\bauthor{\bsnm{Fatehi}, \binits{P.}},
\bauthor{\bsnm{Kwan}, \binits{A.}},
\bauthor{\bsnm{Sandhu}, \binits{R.K.}},
\bauthor{\bsnm{Ebinger}, \binits{J.}},
\bauthor{\bsnm{Cheng}, \binits{S.}},
\bauthor{\bsnm{Zou}, \binits{J.}},
\bauthor{\bsnm{Chugh}, \binits{S.S.}},
\bauthor{\bsnm{Ouyang}, \binits{D.}}:
\batitle{Deep learning-based electrocardiographic screening for chronic kidney disease}.
\bjtitle{Communications Medicine}
\bvolume{3},
\bfpage{73}--\blpage{73}
(\byear{2023})
\doiurl{10.1038/s43856-023-00278-w}
\end{barticle}
\endbibitem

\bibitem[\protect\citeauthoryear{Attia et~al.}{2019a}]{Attia2019ScreeningElectrocardiogram}
\begin{barticle}
\bauthor{\bsnm{Attia}, \binits{Z.I.}},
\bauthor{\bsnm{Kapa}, \binits{S.}},
\bauthor{\bsnm{Lopez-Jimenez}, \binits{F.}},
\bauthor{\bsnm{McKie}, \binits{P.M.}},
\bauthor{\bsnm{Ladewig}, \binits{D.J.}},
\bauthor{\bsnm{Satam}, \binits{G.}},
\bauthor{\bsnm{Pellikka}, \binits{P.A.}},
\bauthor{\bsnm{Enriquez-Sarano}, \binits{M.}},
\bauthor{\bsnm{Noseworthy}, \binits{P.A.}},
\bauthor{\bsnm{Munger}, \binits{T.M.}},
\bauthor{\bsnm{Asirvatham}, \binits{S.J.}},
\bauthor{\bsnm{Scott}, \binits{C.G.}},
\bauthor{\bsnm{Carter}, \binits{R.E.}},
\bauthor{\bsnm{Friedman}, \binits{P.A.}}:
\batitle{Screening for cardiac contractile dysfunction using an artificial intelligence–enabled electrocardiogram}.
\bjtitle{Nature Medicine}
\bvolume{25}(\bissue{1}),
\bfpage{70}--\blpage{74}
(\byear{2019})
\doiurl{10.1038/s41591-018-0240-2}
\end{barticle}
\endbibitem

\bibitem[\protect\citeauthoryear{Attia et~al.}{2019b}]{Attia2019ProspectiveDysfunction}
\begin{barticle}
\bauthor{\bsnm{Attia}, \binits{Z.I.}},
\bauthor{\bsnm{Kapa}, \binits{S.}},
\bauthor{\bsnm{Yao}, \binits{X.}},
\bauthor{\bsnm{Lopez-Jimenez}, \binits{F.}},
\bauthor{\bsnm{Mohan}, \binits{T.L.}},
\bauthor{\bsnm{Pellikka}, \binits{P.A.}},
\bauthor{\bsnm{Carter}, \binits{R.E.}},
\bauthor{\bsnm{Shah}, \binits{N.D.}},
\bauthor{\bsnm{Friedman}, \binits{P.A.}},
\bauthor{\bsnm{Noseworthy}, \binits{P.A.}}:
\batitle{Prospective validation of a deep learning electrocardiogram algorithm for the detection of left ventricular systolic dysfunction}.
\bjtitle{Journal of Cardiovascular Electrophysiology}
\bvolume{30}(\bissue{5}),
\bfpage{668}--\blpage{674}
(\byear{2019})
\doiurl{10.1111/JCE.13889}
\end{barticle}
\endbibitem

\bibitem[\protect\citeauthoryear{Dunlay et~al.}{2017}]{Dunlay2017EpidemiologyFraction}
\begin{barticle}
\bauthor{\bsnm{Dunlay}, \binits{S.M.}},
\bauthor{\bsnm{Roger}, \binits{V.L.}},
\bauthor{\bsnm{Redfield}, \binits{M.M.}}:
\batitle{Epidemiology of heart failure with preserved ejection fraction}.
\bjtitle{Nature Reviews Cardiology}
\bvolume{14}(\bissue{10}),
\bfpage{591}--\blpage{602}
(\byear{2017})
\doiurl{10.1038/nrcardio.2017.65}
\end{barticle}
\endbibitem

\bibitem[\protect\citeauthoryear{Khan et~al.}{2024}]{Khan2024GlobalFailure}
\begin{barticle}
\bauthor{\bsnm{Khan}, \binits{M.S.}},
\bauthor{\bsnm{Shahid}, \binits{I.}},
\bauthor{\bsnm{Bennis}, \binits{A.}},
\bauthor{\bsnm{Rakisheva}, \binits{A.}},
\bauthor{\bsnm{Metra}, \binits{M.}},
\bauthor{\bsnm{Butler}, \binits{J.}}:
\batitle{Global epidemiology of heart failure}.
\bjtitle{Nature Reviews Cardiology}
\bvolume{21}(\bissue{10}),
\bfpage{717}--\blpage{734}
(\byear{2024})
\doiurl{10.1038/s41569-024-01046-6}
\end{barticle}
\endbibitem

\bibitem[\protect\citeauthoryear{Prinzen et~al.}{2022}]{Prinzen2022ElectricalTreatment}
\begin{barticle}
\bauthor{\bsnm{Prinzen}, \binits{F.W.}},
\bauthor{\bsnm{Auricchio}, \binits{A.}},
\bauthor{\bsnm{Mullens}, \binits{W.}},
\bauthor{\bsnm{Linde}, \binits{C.}},
\bauthor{\bsnm{Huizar}, \binits{J.F.}}:
\batitle{Electrical management of heart failure: from pathophysiology to treatment}.
\bjtitle{European Heart Journal}
\bvolume{43}(\bissue{20}),
\bfpage{1917}--\blpage{1927}
(\byear{2022})
\doiurl{10.1093/eurheartj/ehac088}
\end{barticle}
\endbibitem

\bibitem[\protect\citeauthoryear{Khan et~al.}{2019}]{Khan201910-YearPopulation}
\begin{barticle}
\bauthor{\bsnm{Khan}, \binits{S.S.}},
\bauthor{\bsnm{Ning}, \binits{H.}},
\bauthor{\bsnm{Shah}, \binits{S.J.}},
\bauthor{\bsnm{Yancy}, \binits{C.W.}},
\bauthor{\bsnm{Carnethon}, \binits{M.}},
\bauthor{\bsnm{Berry}, \binits{J.D.}},
\bauthor{\bsnm{Mentz}, \binits{R.J.}},
\bauthor{\bsnm{O’Brien}, \binits{E.}},
\bauthor{\bsnm{Correa}, \binits{A.}},
\bauthor{\bsnm{Suthahar}, \binits{N.}},
\bauthor{\bsnm{Boer}, \binits{R.A.}},
\bauthor{\bsnm{Wilkins}, \binits{J.T.}},
\bauthor{\bsnm{Lloyd-Jones}, \binits{D.M.}}:
\batitle{10-{Year} risk equations for incident heart failure in the general population}.
\bjtitle{Journal of the American College of Cardiology}
\bvolume{73}(\bissue{19}),
\bfpage{2388}--\blpage{2397}
(\byear{2019})
\doiurl{10.1016/j.jacc.2019.02.057}
\end{barticle}
\endbibitem

\bibitem[\protect\citeauthoryear{Abnar and Zuidema}{2020}]{Abnar2020QuantifyingTransformers}
\begin{bchapter}
\bauthor{\bsnm{Abnar}, \binits{S.}},
\bauthor{\bsnm{Zuidema}, \binits{W.}}:
\bctitle{Quantifying attention flow in transformers}.
In: \bbtitle{Proceedings of the 58th {Annual} {Meeting} of the {Association} for {Computational} {Linguistics}},
pp. \bfpage{4190}--\blpage{4197}.
\bpublisher{Association for Computational Linguistics},
\blocation{Online}
(\byear{2020}).
\doiurl{10.18653/v1/2020.acl-main.385}
\end{bchapter}
\endbibitem

\bibitem[\protect\citeauthoryear{Gildenblat}{2020}]{GildenblatExploringTransformers}
\begin{botherref}
\oauthor{\bsnm{Gildenblat}, \binits{J.}}:
Exploring {Explainability} for {Vision} {Transformers}
(2020).
\url{https://jacobgil.github.io/deeplearning/vision-transformer-explainability}
\end{botherref}
\endbibitem

\bibitem[\protect\citeauthoryear{Khurshid et~al.}{2022}]{Khurshid2022ECG-BasedFibrillation}
\begin{barticle}
\bauthor{\bsnm{Khurshid}, \binits{S.}},
\bauthor{\bsnm{Friedman}, \binits{S.}},
\bauthor{\bsnm{Reeder}, \binits{C.}},
\bauthor{\bsnm{Di~Achille}, \binits{P.}},
\bauthor{\bsnm{Diamant}, \binits{N.}},
\bauthor{\bsnm{Singh}, \binits{P.}},
\bauthor{\bsnm{Harrington}, \binits{L.X.}},
\bauthor{\bsnm{Wang}, \binits{X.}},
\bauthor{\bsnm{Al-Alusi}, \binits{M.A.}},
\bauthor{\bsnm{Sarma}, \binits{G.}},
\bauthor{\bsnm{Foulkes}, \binits{A.S.}},
\bauthor{\bsnm{Ellinor}, \binits{P.T.}},
\bauthor{\bsnm{Anderson}, \binits{C.D.}},
\bauthor{\bsnm{Ho}, \binits{J.E.}},
\bauthor{\bsnm{Philippakis}, \binits{A.A.}},
\bauthor{\bsnm{Batra}, \binits{P.}},
\bauthor{\bsnm{Lubitz}, \binits{S.A.}}:
\batitle{{ECG}-based deep learning and clinical risk factors to predict atrial fibrillation}.
\bjtitle{Circulation}
\bvolume{145}(\bissue{2}),
\bfpage{122}--\blpage{133}
(\byear{2022})
\doiurl{10.1161/CIRCULATIONAHA.121.057480}
\end{barticle}
\endbibitem

\bibitem[\protect\citeauthoryear{Prifti et~al.}{2021}]{Prifti2021DeepSyndrome}
\begin{barticle}
\bauthor{\bsnm{Prifti}, \binits{E.}},
\bauthor{\bsnm{Fall}, \binits{A.}},
\bauthor{\bsnm{Davogustto}, \binits{G.}},
\bauthor{\bsnm{Pulini}, \binits{A.}},
\bauthor{\bsnm{Denjoy}, \binits{I.}},
\bauthor{\bsnm{Funck-Brentano}, \binits{C.}},
\bauthor{\bsnm{Khan}, \binits{Y.}},
\bauthor{\bsnm{Durand-Salmon}, \binits{A.}},
\bauthor{\bsnm{Badilini}, \binits{F.}},
\bauthor{\bsnm{Wells}, \binits{Q.S.}},
\bauthor{\bsnm{Leenhardt}, \binits{A.}},
\bauthor{\bsnm{Zucker}, \binits{J.D.}},
\bauthor{\bsnm{Roden}, \binits{D.M.}},
\bauthor{\bsnm{Extramiana}, \binits{F.}},
\bauthor{\bsnm{Salem}, \binits{J.E.}}:
\batitle{Deep learning analysis of electrocardiogram for risk prediction of drug-induced arrhythmias and diagnosis of long {QT} syndrome}.
\bjtitle{European Heart Journal}
\bvolume{42}(\bissue{38}),
\bfpage{3948}--\blpage{3961}
(\byear{2021})
\doiurl{10.1093/EURHEARTJ/EHAB588}
\end{barticle}
\endbibitem

\bibitem[\protect\citeauthoryear{Hughes et~al.}{2023}]{Hughes2023ADisease}
\begin{barticle}
\bauthor{\bsnm{Hughes}, \binits{J.W.}},
\bauthor{\bsnm{Tooley}, \binits{J.}},
\bauthor{\bsnm{Torres~Soto}, \binits{J.}},
\bauthor{\bsnm{Ostropolets}, \binits{A.}},
\bauthor{\bsnm{Poterucha}, \binits{T.}},
\bauthor{\bsnm{Christensen}, \binits{M.K.}},
\bauthor{\bsnm{Yuan}, \binits{N.}},
\bauthor{\bsnm{Ehlert}, \binits{B.}},
\bauthor{\bsnm{Kaur}, \binits{D.}},
\bauthor{\bsnm{Kang}, \binits{G.}},
\bauthor{\bsnm{Rogers}, \binits{A.}},
\bauthor{\bsnm{Narayan}, \binits{S.}},
\bauthor{\bsnm{Elias}, \binits{P.}},
\bauthor{\bsnm{Ouyang}, \binits{D.}},
\bauthor{\bsnm{Ashley}, \binits{E.}},
\bauthor{\bsnm{Zou}, \binits{J.}},
\bauthor{\bsnm{Perez}, \binits{M.V.}}:
\batitle{A deep learning-based electrocardiogram risk score for long term cardiovascular death and disease}.
\bjtitle{npj Digital Medicine}
\bvolume{6},
\bfpage{169}--\blpage{169}
(\byear{2023})
\doiurl{10.1038/s41746-023-00916-6}
\end{barticle}
\endbibitem

\bibitem[\protect\citeauthoryear{Khan et~al.}{2022}]{Khan2022ValidationOrganization}
\begin{barticle}
\bauthor{\bsnm{Khan}, \binits{S.S.}},
\bauthor{\bsnm{Barda}, \binits{N.}},
\bauthor{\bsnm{Greenland}, \binits{P.}},
\bauthor{\bsnm{Dagan}, \binits{N.}},
\bauthor{\bsnm{Lloyd-Jones}, \binits{D.M.}},
\bauthor{\bsnm{Balicer}, \binits{R.}},
\bauthor{\bsnm{Rasmussen-Torvik}, \binits{L.J.}}:
\batitle{Validation of heart failure-specific risk equations in 1.3 million {Israeli} adults and usefulness of combining ambulatory and hospitalization data from a aarge integrated health care organization}.
\bjtitle{The American Journal of Cardiology}
\bvolume{168},
\bfpage{105}--\blpage{109}
(\byear{2022})
\doiurl{10.1016/j.amjcard.2021.12.017}
\end{barticle}
\endbibitem

\bibitem[\protect\citeauthoryear{Sinha et~al.}{2021}]{Sinha2021Risk-BasedFailure}
\begin{barticle}
\bauthor{\bsnm{Sinha}, \binits{A.}},
\bauthor{\bsnm{Gupta}, \binits{D.K.}},
\bauthor{\bsnm{Yancy}, \binits{C.W.}},
\bauthor{\bsnm{Shah}, \binits{S.J.}},
\bauthor{\bsnm{Rasmussen-Torvik}, \binits{L.J.}},
\bauthor{\bsnm{McNally}, \binits{E.M.}},
\bauthor{\bsnm{Greenland}, \binits{P.}},
\bauthor{\bsnm{Lloyd-Jones}, \binits{D.M.}},
\bauthor{\bsnm{Khan}, \binits{S.S.}}:
\batitle{Risk-based approach for the prediction and prevention of heart failure}.
\bjtitle{Circulation: Heart Failure}
\bvolume{14}(\bissue{2}),
\bfpage{007761}
(\byear{2021})
\doiurl{10.1161/CIRCHEARTFAILURE.120.007761}
\end{barticle}
\endbibitem

\bibitem[\protect\citeauthoryear{Dhingra et~al.}{2025}]{dhingra_heart_2025}
\begin{barticle}
\bauthor{\bsnm{Dhingra}, \binits{L.S.}},
\bauthor{\bsnm{Aminorroaya}, \binits{A.}},
\bauthor{\bsnm{Sangha}, \binits{V.}},
\bauthor{\bsnm{Pedroso}, \binits{A.F.}},
\bauthor{\bsnm{Asselbergs}, \binits{F.W.}},
\bauthor{\bsnm{Brant}, \binits{L.C.C.}},
\bauthor{\bsnm{Barreto}, \binits{S.M.}},
\bauthor{\bsnm{Ribeiro}, \binits{A.L.P.}},
\bauthor{\bsnm{Krumholz}, \binits{H.M.}},
\bauthor{\bsnm{Oikonomou}, \binits{E.K.}},
\bauthor{\bsnm{Khera}, \binits{R.}}:
\batitle{Heart failure risk stratification using artificial intelligence applied to electrocardiogram images: a multinational study}.
\bjtitle{European Heart Journal}
\bvolume{46}(\bissue{11}),
\bfpage{1044}--\blpage{1053}
(\byear{2025})
\doiurl{10.1093/eurheartj/ehae914} .
Accessed 2025-07-12
\end{barticle}
\endbibitem

\bibitem[\protect\citeauthoryear{Butler et~al.}{2023}]{butler_generalizable_2023}
\begin{barticle}
\bauthor{\bsnm{Butler}, \binits{L.}},
\bauthor{\bsnm{Karabayir}, \binits{I.}},
\bauthor{\bsnm{Kitzman}, \binits{D.W.}},
\bauthor{\bsnm{Alonso}, \binits{A.}},
\bauthor{\bsnm{Tison}, \binits{G.H.}},
\bauthor{\bsnm{Chen}, \binits{L.Y.}},
\bauthor{\bsnm{Chang}, \binits{P.P.}},
\bauthor{\bsnm{Clifford}, \binits{G.}},
\bauthor{\bsnm{Soliman}, \binits{E.Z.}},
\bauthor{\bsnm{Akbilgic}, \binits{O.}}:
\batitle{A generalizable electrocardiogram-based artificial intelligence model for 10-year heart failure risk prediction}.
\bjtitle{Cardiovascular Digital Health Journal}
\bvolume{4}(\bissue{6}),
\bfpage{183}--\blpage{190}
(\byear{2023})
\doiurl{10.1016/j.cvdhj.2023.11.003} .
Accessed 2025-07-12
\end{barticle}
\endbibitem

\bibitem[\protect\citeauthoryear{Akbilgic et~al.}{2021}]{akbilgic_ecg-ai_2021}
\begin{barticle}
\bauthor{\bsnm{Akbilgic}, \binits{O.}},
\bauthor{\bsnm{Butler}, \binits{L.}},
\bauthor{\bsnm{Karabayir}, \binits{I.}},
\bauthor{\bsnm{Chang}, \binits{P.P.}},
\bauthor{\bsnm{Kitzman}, \binits{D.W.}},
\bauthor{\bsnm{Alonso}, \binits{A.}},
\bauthor{\bsnm{Chen}, \binits{L.Y.}},
\bauthor{\bsnm{Soliman}, \binits{E.Z.}}:
\batitle{{ECG}-{AI}: electrocardiographic artificial intelligence model for prediction of heart failure}.
\bjtitle{European Heart Journal - Digital Health}
\bvolume{2}(\bissue{4}),
\bfpage{626}--\blpage{634}
(\byear{2021})
\doiurl{10.1093/ehjdh/ztab080} .
Accessed 2025-07-12
\end{barticle}
\endbibitem

\bibitem[\protect\citeauthoryear{Lin et~al.}{2024}]{lin_ai-enabled_2024}
\begin{barticle}
\bauthor{\bsnm{Lin}, \binits{C.-S.}},
\bauthor{\bsnm{Liu}, \binits{W.-T.}},
\bauthor{\bsnm{Tsai}, \binits{D.-J.}},
\bauthor{\bsnm{Lou}, \binits{Y.-S.}},
\bauthor{\bsnm{Chang}, \binits{C.-H.}},
\bauthor{\bsnm{Lee}, \binits{C.-C.}},
\bauthor{\bsnm{Fang}, \binits{W.-H.}},
\bauthor{\bsnm{Wang}, \binits{C.-C.}},
\bauthor{\bsnm{Chen}, \binits{Y.-Y.}},
\bauthor{\bsnm{Lin}, \binits{W.-S.}},
\bauthor{\bsnm{Cheng}, \binits{C.-C.}},
\bauthor{\bsnm{Lee}, \binits{C.-C.}},
\bauthor{\bsnm{Wang}, \binits{C.-H.}},
\bauthor{\bsnm{Tsai}, \binits{C.-S.}},
\bauthor{\bsnm{Lin}, \binits{S.-H.}},
\bauthor{\bsnm{Lin}, \binits{C.}}:
\batitle{{AI}-enabled electrocardiography alert intervention and all-cause mortality: a pragmatic randomized clinical trial}.
\bjtitle{Nature Medicine}
\bvolume{30}(\bissue{5}),
\bfpage{1461}--\blpage{1470}
(\byear{2024})
\doiurl{10.1038/s41591-024-02961-4} .
\bcomment{Publisher: Nature Publishing Group}.
Accessed 2024-11-01
\end{barticle}
\endbibitem

\bibitem[\protect\citeauthoryear{Jain and Wallace}{2019}]{jain_attention_2019}
\begin{botherref}
\oauthor{\bsnm{Jain}, \binits{S.}},
\oauthor{\bsnm{Wallace}, \binits{B.C.}}:
Attention is not {Explanation}.
arXiv.
arXiv:1902.10186 [cs]
(2019).
\doiurl{10.48550/arXiv.1902.10186} .
\url{http://arxiv.org/abs/1902.10186}
Accessed 2025-07-04
\end{botherref}
\endbibitem

\bibitem[\protect\citeauthoryear{Chefer et~al.}{2021}]{chefer_transformer_2021}
\begin{bchapter}
\bauthor{\bsnm{Chefer}, \binits{H.}},
\bauthor{\bsnm{Gur}, \binits{S.}},
\bauthor{\bsnm{Wolf}, \binits{L.}}:
\bctitle{Transformer {Interpretability} {Beyond} {Attention} {Visualization}},
pp. \bfpage{782}--\blpage{791}
(\byear{2021})
\end{bchapter}
\endbibitem

\bibitem[\protect\citeauthoryear{Jo et~al.}{2025}]{jo_gmar_2025}
\begin{botherref}
\oauthor{\bsnm{Jo}, \binits{S.}},
\oauthor{\bsnm{Jang}, \binits{G.}},
\oauthor{\bsnm{Park}, \binits{H.}}:
{GMAR}: {Gradient}-{Driven} {Multi}-{Head} {Attention} {Rollout} for {Vision} {Transformer} {Interpretability}.
arXiv.
arXiv:2504.19414 [cs]
(2025).
\doiurl{10.48550/arXiv.2504.19414} .
\url{http://arxiv.org/abs/2504.19414}
Accessed 2025-07-04
\end{botherref}
\endbibitem

\bibitem[\protect\citeauthoryear{Dukes et~al.}{2015}]{Dukes2015VentricularDeath}
\begin{barticle}
\bauthor{\bsnm{Dukes}, \binits{J.W.}},
\bauthor{\bsnm{Dewland}, \binits{T.A.}},
\bauthor{\bsnm{Vittinghoff}, \binits{E.}},
\bauthor{\bsnm{Mandyam}, \binits{M.C.}},
\bauthor{\bsnm{Heckbert}, \binits{S.R.}},
\bauthor{\bsnm{Siscovick}, \binits{D.S.}},
\bauthor{\bsnm{Stein}, \binits{P.K.}},
\bauthor{\bsnm{Psaty}, \binits{B.M.}},
\bauthor{\bsnm{Sotoodehnia}, \binits{N.}},
\bauthor{\bsnm{Gottdiener}, \binits{J.S.}},
\bauthor{\bsnm{Marcus}, \binits{G.M.}}:
\batitle{Ventricular ectopy as a predictor of heart failure and death}.
\bjtitle{Journal of the American College of Cardiology}
\bvolume{66}(\bissue{2}),
\bfpage{101}--\blpage{109}
(\byear{2015})
\doiurl{10.1016/J.JACC.2015.04.062}
\end{barticle}
\endbibitem

\bibitem[\protect\citeauthoryear{Maisel and Stevenson}{2003}]{Maisel2003AtrialTherapy}
\begin{barticle}
\bauthor{\bsnm{Maisel}, \binits{W.H.}},
\bauthor{\bsnm{Stevenson}, \binits{L.W.}}:
\batitle{Atrial fibrillation in heart failure: epidemiology, pathophysiology, and rationale for therapy}.
\bjtitle{The American Journal of Cardiology}
\bvolume{91}(\bissue{6}),
\bfpage{2}--\blpage{8}
(\byear{2003})
\doiurl{10.1016/S0002-9149(02)03373-8}
\end{barticle}
\endbibitem

\bibitem[\protect\citeauthoryear{Bergau et~al.}{2022}]{Bergau2022AtrialFailure}
\begin{barticle}
\bauthor{\bsnm{Bergau}, \binits{L.}},
\bauthor{\bsnm{Bengel}, \binits{P.}},
\bauthor{\bsnm{Sciacca}, \binits{V.}},
\bauthor{\bsnm{Fink}, \binits{T.}},
\bauthor{\bsnm{Sohns}, \binits{C.}},
\bauthor{\bsnm{Sommer}, \binits{P.}}:
\batitle{Atrial fibrillation and heart failure}.
\bjtitle{Journal of Clinical Medicine}
\bvolume{11}(\bissue{9}),
\bfpage{2510}--\blpage{2510}
(\byear{2022})
\doiurl{10.3390/JCM11092510}
\end{barticle}
\endbibitem

\bibitem[\protect\citeauthoryear{Ledwidge et~al.}{2013}]{ledwidge_natriuretic_2013}
\begin{barticle}
\bauthor{\bsnm{Ledwidge}, \binits{M.}},
\bauthor{\bsnm{Gallagher}, \binits{J.}},
\bauthor{\bsnm{Conlon}, \binits{C.}},
\bauthor{\bsnm{Tallon}, \binits{E.}},
\bauthor{\bsnm{O’Connell}, \binits{E.}},
\bauthor{\bsnm{Dawkins}, \binits{I.}},
\bauthor{\bsnm{Watson}, \binits{C.}},
\bauthor{\bsnm{O’Hanlon}, \binits{R.}},
\bauthor{\bsnm{Bermingham}, \binits{M.}},
\bauthor{\bsnm{Patle}, \binits{A.}},
\bauthor{\bsnm{Badabhagni}, \binits{M.R.}},
\bauthor{\bsnm{Murtagh}, \binits{G.}},
\bauthor{\bsnm{Voon}, \binits{V.}},
\bauthor{\bsnm{Tilson}, \binits{L.}},
\bauthor{\bsnm{Barry}, \binits{M.}},
\bauthor{\bsnm{McDonald}, \binits{L.}},
\bauthor{\bsnm{Maurer}, \binits{B.}},
\bauthor{\bsnm{McDonald}, \binits{K.}}:
\batitle{Natriuretic {Peptide}–{Based} {Screening} and {Collaborative} {Care} for {Heart} {Failure}: {The} {STOP}-{HF} {Randomized} {Trial}}.
\bjtitle{JAMA}
\bvolume{310}(\bissue{1}),
\bfpage{66}--\blpage{74}
(\byear{2013})
\doiurl{10.1001/jama.2013.7588} .
Accessed 2024-12-12
\end{barticle}
\endbibitem

\bibitem[\protect\citeauthoryear{Hernandez}{2013}]{hernandez_preventing_2013}
\begin{barticle}
\bauthor{\bsnm{Hernandez}, \binits{A.F.}}:
\batitle{Preventing {Heart} {Failure}}.
\bjtitle{JAMA}
\bvolume{310}(\bissue{1}),
\bfpage{44}--\blpage{45}
(\byear{2013})
\doiurl{10.1001/jama.2013.7589} .
Accessed 2024-12-12
\end{barticle}
\endbibitem

\bibitem[\protect\citeauthoryear{McKeen et~al.}{2024}]{McKeen2024ECG-FM:Model}
\begin{botherref}
\oauthor{\bsnm{McKeen}, \binits{K.}},
\oauthor{\bsnm{Oliva}, \binits{L.}},
\oauthor{\bsnm{Masood}, \binits{S.}},
\oauthor{\bsnm{Toma}, \binits{A.}},
\oauthor{\bsnm{Rubin}, \binits{B.}},
\oauthor{\bsnm{Wang}, \binits{B.}}:
{ECG}-{FM}: {An} {Open} {Electrocardiogram} {Foundation} {Model}.
arXiv
(2024).
\doiurl{10.48550/ARXIV.2408.05178} .
\url{https://arxiv.org/abs/2408.05178}
\end{botherref}
\endbibitem

\bibitem[\protect\citeauthoryear{Davies et~al.}{2024}]{Davies2024InterpretableData}
\begin{botherref}
\oauthor{\bsnm{Davies}, \binits{H.J.}},
\oauthor{\bsnm{Monsen}, \binits{J.}},
\oauthor{\bsnm{Mandic}, \binits{D.P.}}:
Interpretable {Pre}-{Trained} {Transformers} for {Heart} {Time}-{Series} {Data}.
arXiv
(2024).
\doiurl{10.48550/ARXIV.2407.20775} .
\url{https://arxiv.org/abs/2407.20775}
\end{botherref}
\endbibitem

\bibitem[\protect\citeauthoryear{Li et~al.}{2024}]{Li2024AnDomains}
\begin{botherref}
\oauthor{\bsnm{Li}, \binits{J.}},
\oauthor{\bsnm{Aguirre}, \binits{A.}},
\oauthor{\bsnm{Moura}, \binits{J.}},
\oauthor{\bsnm{Liu}, \binits{C.}},
\oauthor{\bsnm{Zhong}, \binits{L.}},
\oauthor{\bsnm{Sun}, \binits{C.}},
\oauthor{\bsnm{Clifford}, \binits{G.}},
\oauthor{\bsnm{Westover}, \binits{B.}},
\oauthor{\bsnm{Hong}, \binits{S.}}:
An electrocardiogram foundation model built on over 10 million recordings with external evaluation across multiple domains.
arXiv
(2024).
\doiurl{10.48550/arXiv.2410.04133} .
\url{http://arxiv.org/abs/2410.04133}
\end{botherref}
\endbibitem

\bibitem[\protect\citeauthoryear{Guo et~al.}{2024}]{Guo2023SiamAF:Detection}
\begin{botherref}
\oauthor{\bsnm{Guo}, \binits{Z.}},
\oauthor{\bsnm{Ding}, \binits{C.}},
\oauthor{\bsnm{Do}, \binits{D.H.}},
\oauthor{\bsnm{Shah}, \binits{A.}},
\oauthor{\bsnm{Lee}, \binits{R.J.}},
\oauthor{\bsnm{Hu}, \binits{X.}},
\oauthor{\bsnm{Rudin}, \binits{C.}}:
{SiamAF}: {Learning} {Shared} {Information} from {ECG} and {PPG} {Signals} for {Robust} {Atrial} {Fibrillation} {Detection}.
arXiv
(2024).
\doiurl{10.48550/ARXIV.2310.09203} .
\url{https://arxiv.org/abs/2310.09203}
\end{botherref}
\endbibitem

\bibitem[\protect\citeauthoryear{Abbaspourazad et~al.}{2024}]{Abbaspourazad2023Large-scaleBiosignals}
\begin{bchapter}
\bauthor{\bsnm{Abbaspourazad}, \binits{S.}},
\bauthor{\bsnm{Elachqar}, \binits{O.}},
\bauthor{\bsnm{Miller}, \binits{A.C.}},
\bauthor{\bsnm{Emrani}, \binits{S.}},
\bauthor{\bsnm{Nallasamy}, \binits{U.}},
\bauthor{\bsnm{Shapiro}, \binits{I.}}:
\bctitle{Large-scale {Training} of {Foundation} {Models} for {Wearable} {Biosignals}}.
In: \bbtitle{The {Twelfth} {International} {Conference} on {Learning} {Representations} - {ICLR} 2024}
(\byear{2024}).
\burl{https://openreview.net/forum?id=pC3WJHf51j}
\end{bchapter}
\endbibitem

\bibitem[\protect\citeauthoryear{Défossez et~al.}{2023}]{Defossez2022HighCompression}
\begin{botherref}
\oauthor{\bsnm{Défossez}, \binits{A.}},
\oauthor{\bsnm{Copet}, \binits{J.}},
\oauthor{\bsnm{Synnaeve}, \binits{G.}},
\oauthor{\bsnm{Adi}, \binits{Y.}}:
High {Fidelity} {Neural} {Audio} {Compression}.
Transactions on Machine Learning Research
(2023)
\end{botherref}
\endbibitem

\bibitem[\protect\citeauthoryear{Akiba et~al.}{2019}]{Akiba2019Optuna:Framework}
\begin{bchapter}
\bauthor{\bsnm{Akiba}, \binits{T.}},
\bauthor{\bsnm{Sano}, \binits{S.}},
\bauthor{\bsnm{Yanase}, \binits{T.}},
\bauthor{\bsnm{Ohta}, \binits{T.}},
\bauthor{\bsnm{Koyama}, \binits{M.}}:
\bctitle{Optuna: {A} {Next}-generation {Hyperparameter} {Optimization} {Framework}}.
In: \bbtitle{Proceedings of the 25th {ACM} {SIGKDD} {International} {Conference} on {Knowledge} {Discovery} \& {Data} {Mining}}.
\bsertitle{{KDD} '19},
pp. \bfpage{2623}--\blpage{2631}.
\bpublisher{Association for Computing Machinery},
\blocation{New York, NY, USA}
(\byear{2019}).
\doiurl{10.1145/3292500.3330701}
\end{bchapter}
\endbibitem

\bibitem[\protect\citeauthoryear{{Israel Central Bureau of Statistics}}{2023}]{2023The2023}
\begin{botherref}
\oauthor{\bsnm{{Israel Central Bureau of Statistics}}}:
The {Population} of {Ethiopian} {Origin} in {Israel}: {Selected} {Data} {Published} on the {Occasion} of the {Sigd} {Festival} 2023.
Technical report,
Jerusalem
(2023).
\url{https://www.cbs.gov.il/en/mediarelease/Pages/2023/The-Ethiopian-Population-in-Israel-2023.aspx}
\end{botherref}
\endbibitem

\bibitem[\protect\citeauthoryear{Makowski et~al.}{2021}]{Makowski2021NeuroKit2:Processing}
\begin{barticle}
\bauthor{\bsnm{Makowski}, \binits{D.}},
\bauthor{\bsnm{Pham}, \binits{T.}},
\bauthor{\bsnm{Lau}, \binits{Z.J.}},
\bauthor{\bsnm{Brammer}, \binits{J.C.}},
\bauthor{\bsnm{Lespinasse}, \binits{F.}},
\bauthor{\bsnm{Pham}, \binits{H.}},
\bauthor{\bsnm{Scholzel}, \binits{C.}},
\bauthor{\bsnm{Chen}, \binits{S.H.A.}}:
\batitle{{NeuroKit2}: {A} {Python} toolbox for neurophysiological signal processing}.
\bjtitle{Behavior Research Methods}
\bvolume{53}(\bissue{4}),
\bfpage{1689}--\blpage{1696}
(\byear{2021})
\doiurl{10.3758/S13428-020-01516-Y/TABLES/3}
\end{barticle}
\endbibitem

\bibitem[\protect\citeauthoryear{Gendelman et~al.}{2021}]{Gendelman2021PhysioZooConduction}
\begin{bchapter}
\bauthor{\bsnm{Gendelman}, \binits{S.}},
\bauthor{\bsnm{Biton}, \binits{S.}},
\bauthor{\bsnm{Derman}, \binits{R.}},
\bauthor{\bsnm{Zvuloni}, \binits{E.}},
\bauthor{\bsnm{Levy}, \binits{J.}},
\bauthor{\bsnm{Lugassy}, \binits{S.}},
\bauthor{\bsnm{Alexandrovich}, \binits{A.}},
\bauthor{\bsnm{Behar}, \binits{J.A.}}:
\bctitle{{PhysioZoo} {ECG}: {Digital} electrocardiography biomarkers to assess cardiac conduction}.
In: \bbtitle{2021 {Computing} in {Cardiology} ({CinC})},
vol. \bseriesno{2021-September},
pp. \bfpage{1}--\blpage{4}.
\bpublisher{IEEE},
\blocation{Brno, Czech Republic}
(\byear{2021}).
\doiurl{10.23919/CinC53138.2021.9662857}
\end{bchapter}
\endbibitem

\bibitem[\protect\citeauthoryear{Behar et~al.}{2023}]{behar_physiozoo_2023}
\begin{bchapter}
\bauthor{\bsnm{Behar}, \binits{J.A.}},
\bauthor{\bsnm{Levy}, \binits{J.}},
\bauthor{\bsnm{Zvuloni}, \binits{E.}},
\bauthor{\bsnm{Gendelman}, \binits{S.}},
\bauthor{\bsnm{Rosenberg}, \binits{A.}},
\bauthor{\bsnm{Biton}, \binits{S.}},
\bauthor{\bsnm{Derman}, \binits{R.}},
\bauthor{\bsnm{A.~Sobel}, \binits{J.}},
\bauthor{\bsnm{Alexandrovich}, \binits{A.}},
\bauthor{\bsnm{Charlton}, \binits{P.}},
\bauthor{\bsnm{Marton Aron~Goda}, \binits{D.}}:
\bctitle{{PhysioZoo}: {The} {Open} {Physiological} {Biomarkers} {Resource}}.
In: \bbtitle{Computing in {Cardiology}}.
\bpublisher{IEEE Computer Society},
\blocation{Atlanta, Georgia, USA}
(\byear{2023}).
\doiurl{10.22489/CinC.2023.190}
\end{bchapter}
\endbibitem

\bibitem[\protect\citeauthoryear{Pedregosa et~al.}{2011}]{Pedregosa2011Scikit-learn:Python}
\begin{barticle}
\bauthor{\bsnm{Pedregosa}, \binits{F.}},
\bauthor{\bsnm{Varoquaux}, \binits{G.}},
\bauthor{\bsnm{Gramfort}, \binits{A.}},
\bauthor{\bsnm{Michel}, \binits{V.}},
\bauthor{\bsnm{Thirion}, \binits{B.}},
\bauthor{\bsnm{Grisel}, \binits{O.}},
\bauthor{\bsnm{Blondel}, \binits{M.}},
\bauthor{\bsnm{Prettenhofer}, \binits{P.}},
\bauthor{\bsnm{Weiss}, \binits{R.}},
\bauthor{\bsnm{Dubourg}, \binits{V.}},
\bauthor{\bsnm{Vanderplas}, \binits{J.}},
\bauthor{\bsnm{Passos}, \binits{A.}},
\bauthor{\bsnm{Cournapeau}, \binits{D.}},
\bauthor{\bsnm{Brucher}, \binits{M.}},
\bauthor{\bsnm{Perrot}, \binits{M.}},
\bauthor{\bsnm{Duchesnay}, \binits{E.}}:
\batitle{Scikit-learn: {Machine} {Learning} in {Python}}.
\bjtitle{Journal of Machine Learning Research}
\bvolume{12}(\bissue{85}),
\bfpage{2825}--\blpage{2830}
(\byear{2011})
\end{barticle}
\endbibitem

\bibitem[\protect\citeauthoryear{Davidson-Pilon}{2019}]{Davidson-Pilon2019Lifelines:Python}
\begin{barticle}
\bauthor{\bsnm{Davidson-Pilon}, \binits{C.}}:
\batitle{lifelines: survival analysis in {Python}}.
\bjtitle{Journal of Open Source Software}
\bvolume{4}(\bissue{40}),
\bfpage{1317}
(\byear{2019})
\doiurl{10.21105/joss.01317}
\end{barticle}
\endbibitem

\bibitem[\protect\citeauthoryear{Goyal et~al.}{2020}]{goyal_assembling_2020}
\begin{barticle}
\bauthor{\bsnm{Goyal}, \binits{P.}},
\bauthor{\bsnm{Mefford}, \binits{M.T.}},
\bauthor{\bsnm{Chen}, \binits{L.}},
\bauthor{\bsnm{Sterling}, \binits{M.R.}},
\bauthor{\bsnm{Durant}, \binits{R.W.}},
\bauthor{\bsnm{Safford}, \binits{M.M.}},
\bauthor{\bsnm{Levitan}, \binits{E.B.}}:
\batitle{Assembling and validating a heart failure-free cohort from the {Reasons} for {Geographic} and {Racial} {Differences} in {Stroke} ({REGARDS}) study}.
\bjtitle{BMC Medical Research Methodology}
\bvolume{20}(\bissue{1}),
\bfpage{53}
(\byear{2020})
\doiurl{10.1186/s12874-019-0890-x} .
Accessed 2024-11-01
\end{barticle}
\endbibitem

\bibitem[\protect\citeauthoryear{McDonagh et~al.}{2021}]{mcdonagh_2021_2021}
\begin{barticle}
\bauthor{\bsnm{McDonagh}, \binits{T.A.}},
\bauthor{\bsnm{Metra}, \binits{M.}},
\bauthor{\bsnm{Adamo}, \binits{M.}},
\bauthor{\bsnm{Gardner}, \binits{R.S.}},
\bauthor{\bsnm{Baumbach}, \binits{A.}},
\bauthor{\bsnm{Böhm}, \binits{M.}},
\bauthor{\bsnm{Burri}, \binits{H.}},
\bauthor{\bsnm{Butler}, \binits{J.}},
\bauthor{\bsnm{Čelutkienė}, \binits{J.}},
\bauthor{\bsnm{Chioncel}, \binits{O.}},
\bauthor{\bsnm{Cleland}, \binits{J.G.F.}},
\bauthor{\bsnm{Coats}, \binits{A.J.S.}},
\bauthor{\bsnm{Crespo-Leiro}, \binits{M.G.}},
\bauthor{\bsnm{Farmakis}, \binits{D.}},
\bauthor{\bsnm{Gilard}, \binits{M.}},
\bauthor{\bsnm{Heymans}, \binits{S.}},
\bauthor{\bsnm{Hoes}, \binits{A.W.}},
\bauthor{\bsnm{Jaarsma}, \binits{T.}},
\bauthor{\bsnm{Jankowska}, \binits{E.A.}},
\bauthor{\bsnm{Lainscak}, \binits{M.}},
\bauthor{\bsnm{Lam}, \binits{C.S.P.}},
\bauthor{\bsnm{Lyon}, \binits{A.R.}},
\bauthor{\bsnm{McMurray}, \binits{J.J.V.}},
\bauthor{\bsnm{Mebazaa}, \binits{A.}},
\bauthor{\bsnm{Mindham}, \binits{R.}},
\bauthor{\bsnm{Muneretto}, \binits{C.}},
\bauthor{\bsnm{Francesco~Piepoli}, \binits{M.}},
\bauthor{\bsnm{Price}, \binits{S.}},
\bauthor{\bsnm{Rosano}, \binits{G.M.C.}},
\bauthor{\bsnm{Ruschitzka}, \binits{F.}},
\bauthor{\bsnm{Kathrine~Skibelund}, \binits{A.}},
\bauthor{\bsnm{{ESC Scientific Document Group}}}:
\batitle{2021 {ESC} {Guidelines} for the diagnosis and treatment of acute and chronic heart failure: {Developed} by the {Task} {Force} for the diagnosis and treatment of acute and chronic heart failure of the {European} {Society} of {Cardiology} ({ESC}) {With} the special contribution of the {Heart} {Failure} {Association} ({HFA}) of the {ESC}}.
\bjtitle{European Heart Journal}
\bvolume{42}(\bissue{36}),
\bfpage{3599}--\blpage{3726}
(\byear{2021})
\doiurl{10.1093/eurheartj/ehab368} .
Accessed 2024-11-01
\end{barticle}
\endbibitem

\end{thebibliography}

\FloatBarrier
\newpage


\renewcommand{\appendixname}{Supplement}
\renewcommand{\appendixpagename}{Supplement}
\renewcommand{\appendixtocname}{Supplement}
\renewcommand{\thesection}{S\arabic{section}}
\captionsetup[figure]{name=Supplementary Fig.}
\captionsetup[table]{name=Supplementary Table}
\renewcommand{\figurename}{Supplementary Fig.}
\renewcommand{\tablename}{Supplementary Table}
\begin{appendices}

\section{Supplementary Information}\label{secA1}

\subsection{Heart failure label verification}\label{subsecA1_1}

To verify the validity of the HF label, several investigations were conducted. The primary goal was to establish the label specificity and to prevent the mislabeling of non-HF cases as positive HF diagnoses. The validity analyses were performed for all the HF diagnoses in the database, that is, before applying the ``recording date before 2010 or after 2023" and ``HF before recording" exclusion criteria (main Figure~\ref{fig_data}).

\textbf{(1) Prescribed medication analysis:} This method was considered as the most informative, following validation approaches applied in prior studies~\cite{Khan2022ValidationOrganization, goyal_assembling_2020}. Medication prescriptions were analyzed from five guideline-directed medical therapy groups recommended for HF patients~\cite{Heidenreich20222022Guidelines, mcdonagh_2021_2021}. These groups included MRA, SGLT2i, agents targeting the renin-angiotensin system (including ACE inhibitors, ARBs, and ARNIs), diuretics (both low- and high-ceiling), and beta-blockers. The complete list of medications, along with their corresponding ATC codes, is provided in Supplementary Table~\ref{tab_medications}. The proportion of patients receiving these drugs was determined based on prescription dates relative to the time of HF diagnosis.

In Supplementary Figure~\ref{ext_fig_drugs_pos}, the prescription patterns are presented by dividing patients into categories according to their prescription starting time. The categories included the entire data collection period. A ``new or renewed" prescription was defined as one started during the year before and the first year after diagnosis. Renewed prescription in the period was defined as if it was used previously but not during the second year before diagnosis. A ``future prescription" referred to those started after the first year following diagnosis. Finally, a ``history of prescription" indicated prescriptions already given before the two years preceding diagnosis. For patients without any prescriptions recorded in the available data period, the category ``not documented" was assigned.

A “minimum 1” group was included to capture whether patients had received at least one drug from the specified medication groups. This assignment was made by following a priority order, where the ``new or renewed" category took precedence over ``future prescription", which in turn took precedence over ``history of prescription", followed finally by ``not documented". It was found that 71.6\% of HF patients had been prescribed at least one medication group during the ``new or renewed" period, and another 10.4\% after this period. In addition, more than 99\% were treated with at least one drug from those groups. For some patients with a new HF diagnosis, differences in pharmacological treatment included dose adjustments and changes in specific drugs within the therapeutic class. Although not explicitly detailed, these adjustments are reflected in the overall usage proportions.

\textbf{(2) Repeated HF diagnosis analysis:} The repeated documentation of HF diagnoses across multiple patient visits is expected to improve diagnostic specificity. It was confirmed that diagnoses were not automatically re-documented during each new visit but were recorded only when deemed relevant to the visit’s content by the physician. Since multiple visits for HF-related reasons may reflect greater disease severity, excluding patients with only a single or few documented diagnoses could introduce selection bias by disproportionately excluding those with milder forms of the disease. This exclusion could also reduce both the sensitivity of the study and the sample size available for analysis. To mitigate these concerns and to decide whether repeated HF documentations should be used as an inclusion criteria, a comparison of demographic characteristics and comorbidity prevalence was conducted among patients with one, two, and three documented HF diagnoses, as presented in Supplementary Table~\ref{ext_fig_tab_diag}. No significant differences were identified between these groups.

\textbf{(3) Echocardiogram examinations analysis:} A corresponding analysis to the medication prescription evaluation was conducted for variables related to transthoracic or transesophageal echocardiogram (TTE or TEE) examinations, as shown in Supplementary Figure~\ref{ext_fig_echo_pos}. It was expected that patients diagnosed with HF would have a documented TTE or TEE examination around the time of diagnosis. The analysis classified examinations into three categories. The ``inside range" category included examinations performed from one year before to two years after the initial diagnosis. Examinations documented outside of this period were assigned to the ``outside range" category, while cases with no documented examination during the available data period were classified as ``not documented." It was observed that 82.5\% of HF patients had undergone at least one echocardiogram examination within the defined ``inside range" period. Since the TLHE dataset excludes echocardiograms performed during hospitalization, this result aligns with expectations.

\textbf{(4) Comparing between populations with/without hospital diagnosis data:} Patients were additionally compared based on whether their first HF diagnosis occurred before or after 2018, the year when hospital diagnoses were incorporated into the dataset, potentially enhancing the sensitivity of the diagnostic labels. As shown in Supplementary Table~\ref{ext_fig_tab_diag}, the different population subsets exhibited similar characteristics.

Furthermore, the analyses of prescribed medications and echocardiogram examinations were expanded to include additional groups: patients without an HF diagnosis ($\#HF=0$), those with a low number of HF diagnoses ($0 < \#HF < 3$), and those with a high number of diagnoses ($\#HF \geq 3$), as shown in Supplementary Figure~\ref{ext_fig_echo_drugs}. ``New or renewed" prescriptions for at least one medication group were identified in 58.5\% and 76.4\% of patients with low and high numbers of HF diagnoses, respectively. In patients of $\#HF \geq 3$, there was a higher proportion of new high-ceiling diuretic prescriptions following diagnosis with 63.5\% for $\#HF \geq 3$ versus 38.2\% for $0 < \#HF < 3$, with an overall usage rate of 86.6\% in this group. This finding supports the hypothesis that the frequency of HF documentation correlates with disease severity and may serve as an additional indicator. Similarly, echocardiogram examinations within the ``inside range" period were documented for 75.4\% and 85.0\% of patients with low and high numbers of HF diagnoses, respectively.

\newpage
\subsection{Window selection strategy for training steps}\label{subsecA1_3}
As described in Main Figure~\ref{fig_training}, in the first encoder training step, a window of 30 seconds was selected randomly from every consecutive 3 minute segment of the recording. This ended up with 480 windows of 30-second from each recording, i.e., a total of 4 hours was processed. In the second sequential step, a 30-second window was selected randomly within a 2-minute segment only once, and then the specific position inside the segment was applied for all the consecutive 2-minute segments. Thus, the gap between each 30-second window was a constant of 2 minutes. In this second step, the number of sequence elements was search as a hyper-parameter. The optimal number according to the validation set was 720, thus a total of 6 hours sampling the whole recording. This strategy allowed to reduce the computational load used. The randomization of the 30 seconds sampled made it in practice to be used as data augmentation technique, since in each epoch a different form of the same signal was used.
\end{appendices}


\renewcommand{\figurename}{Supplementary Fig.} 
\renewcommand{\thefigure}{\arabic{figure}}
\setcounter{figure}{0} 

\newpage
\subsection*{Supplementary figures}

\begin{figure*}[h]
\centering
\includegraphics[width=\textwidth]{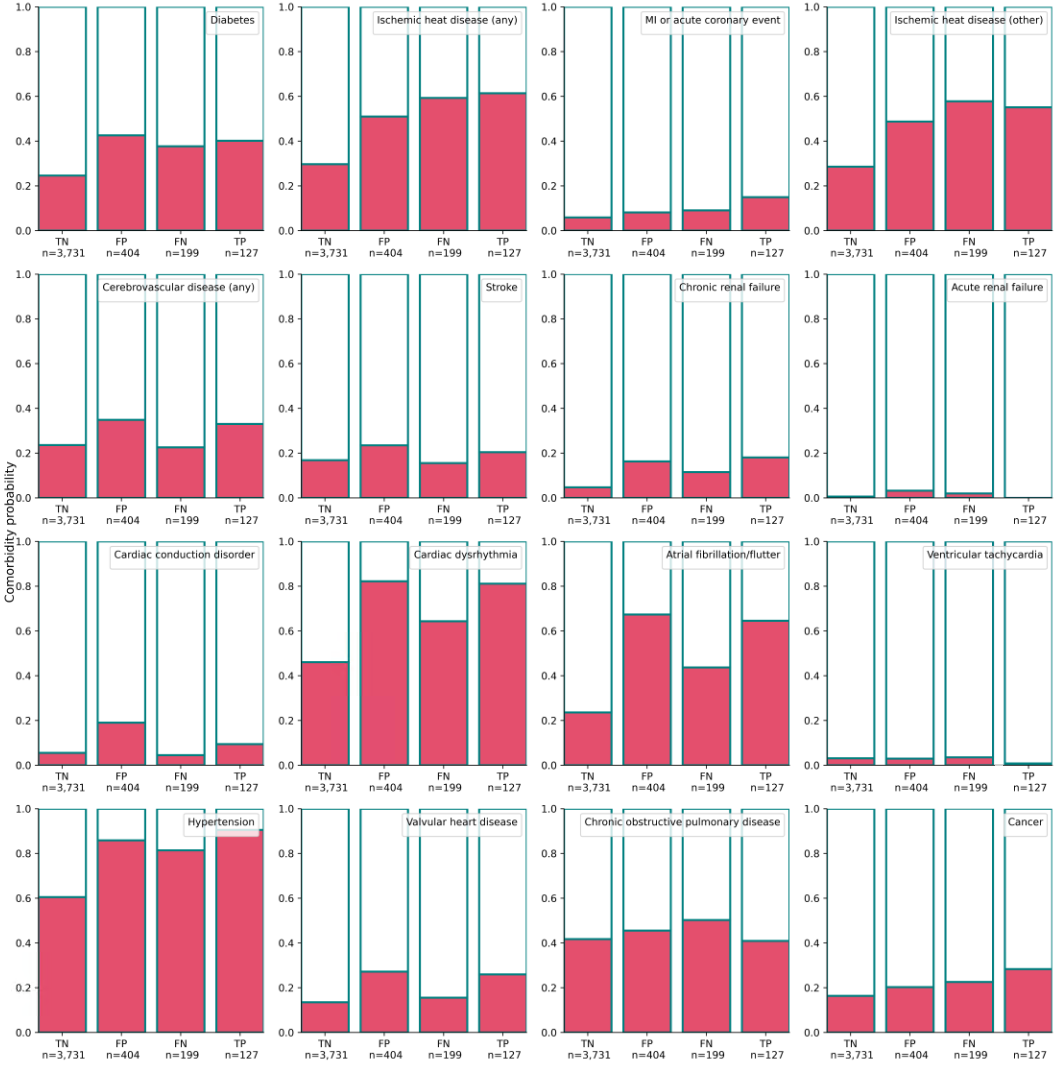}
\caption{\textcolor{black}{\textbf{Probability of comorbidities described in Main Table~\ref{tab1} stratified by DeepHHF classification groups.} Comorbidities were extracted from electronic medical records documented prior to the Holter examination. Classification groups correspond to true positive (TP), false positive (FP), true negative (TN), and false negative (FN) predictions, defined using a decision threshold yielding 90\% specificity, corresponding to the high-risk threshold shown in Main Figure~\ref{fig_results_model}c.}}
\label{ext_fig_com_errors}
\end{figure*}

\newpage
\begin{figure*}[h]
\centering
\includegraphics[width=\textwidth]{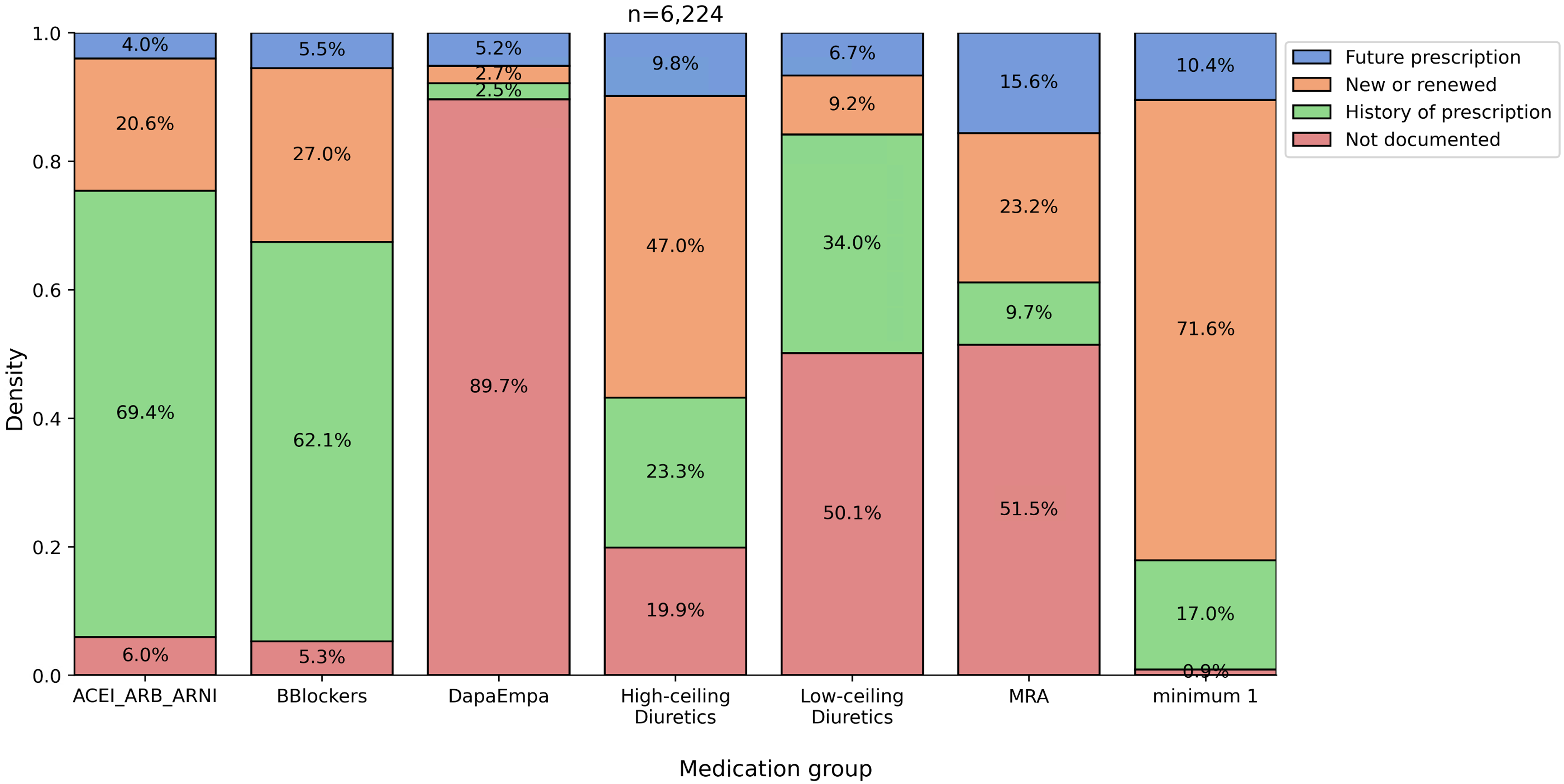}
\caption{\textbf{HF label verification by examining variables of prescribed medications associated with HF treatment.} Comparison is between four periods with respect to interval between prescription and diagnosis. The medication groups are detailed in Supplementary Table~\ref{tab_medications}: Mineralocorticoid receptor antagonists (MRA); Sodium-glucose co-transporter-2 (SGLT2) inhibitors (DapaEmpa);
Angiotensin-converting enzyme inhibitors (ACE-I);
Angiotensin II receptor blockers (ARB);
Angiotensin receptor-neprilysin inhibitors (ARNI);
Diuretics (exc. MRA); and
Beta-adrenergic blocking agents (BBlockers).}
\label{ext_fig_drugs_pos}
\end{figure*}

\newpage
\begin{figure*}[h]
\centering
\includegraphics[width=0.5\textwidth]{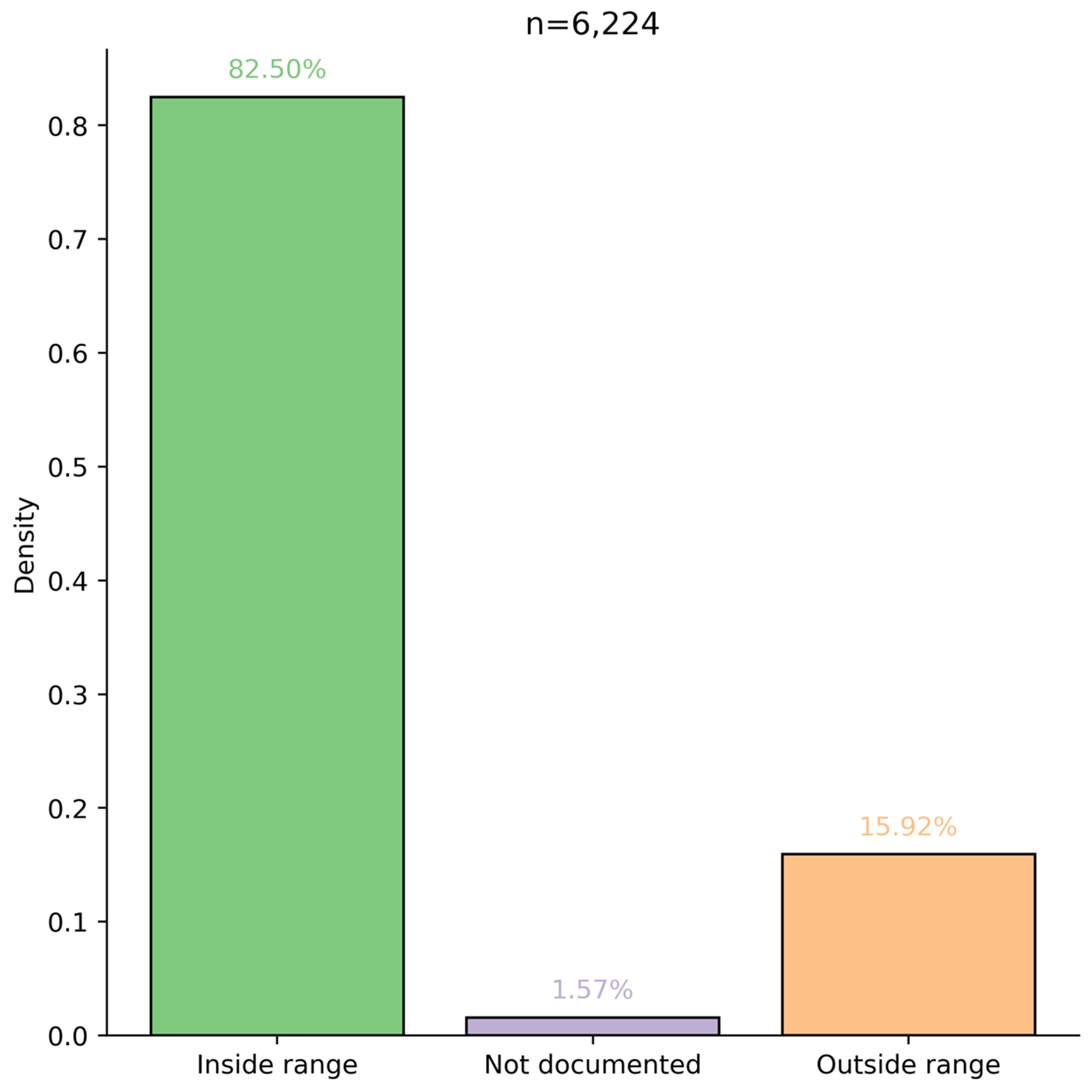}
\caption{\textcolor{black}{\textbf{HF label verification using transthoracic and transesophageal echocardiography (TTE/TEE) examination records.} Echocardiography-derived variables were examined to support verification of HF diagnosis labels. For each patient, TTE/TEE examinations were categorized based on their timing relative to the first documented HF diagnosis: within a defined window (“inside range”), outside this window (“outside range”), or not documented. The verification window was defined as one year prior to and two years following the first documented HF diagnosis.}}
\label{ext_fig_echo_pos}
\end{figure*}

\newpage
\begin{figure*}[h]
\centering
\includegraphics[width=\textwidth]{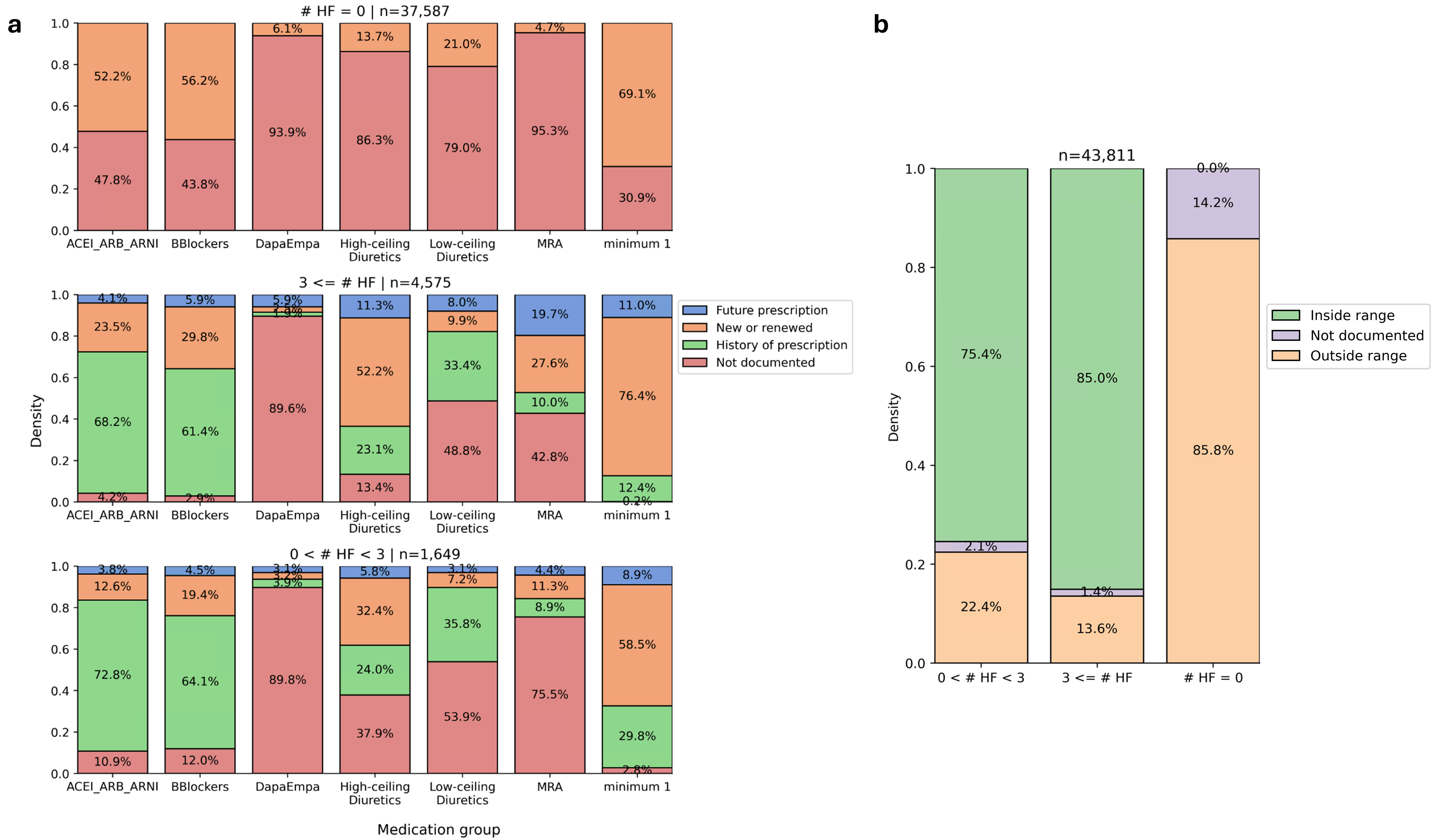}
\caption{\textcolor{black}{\textbf{Supplementary analysis for HF label verification using HF-related treatments and repeated HF diagnosis entries in the EMR.}
Patients were grouped based on the number of documented HF diagnosis entries in the EMR: $3<=HF$, $0 < \#HF < 3$, and $HF=0$ (non-HF). Panel a extends the analysis presented in Supplementary Figure~\ref{ext_fig_drugs_pos} by examining HF-related treatments, and panel b extends Supplementary Figure~\ref{ext_fig_echo_pos} by examining echocardiography-associated variables. This analysis was used to support the reliability of HF outcome labeling.}}
\label{ext_fig_echo_drugs}
\end{figure*}

\newpage
\begin{figure*}[h]
\centering
\includegraphics[width=\textwidth]{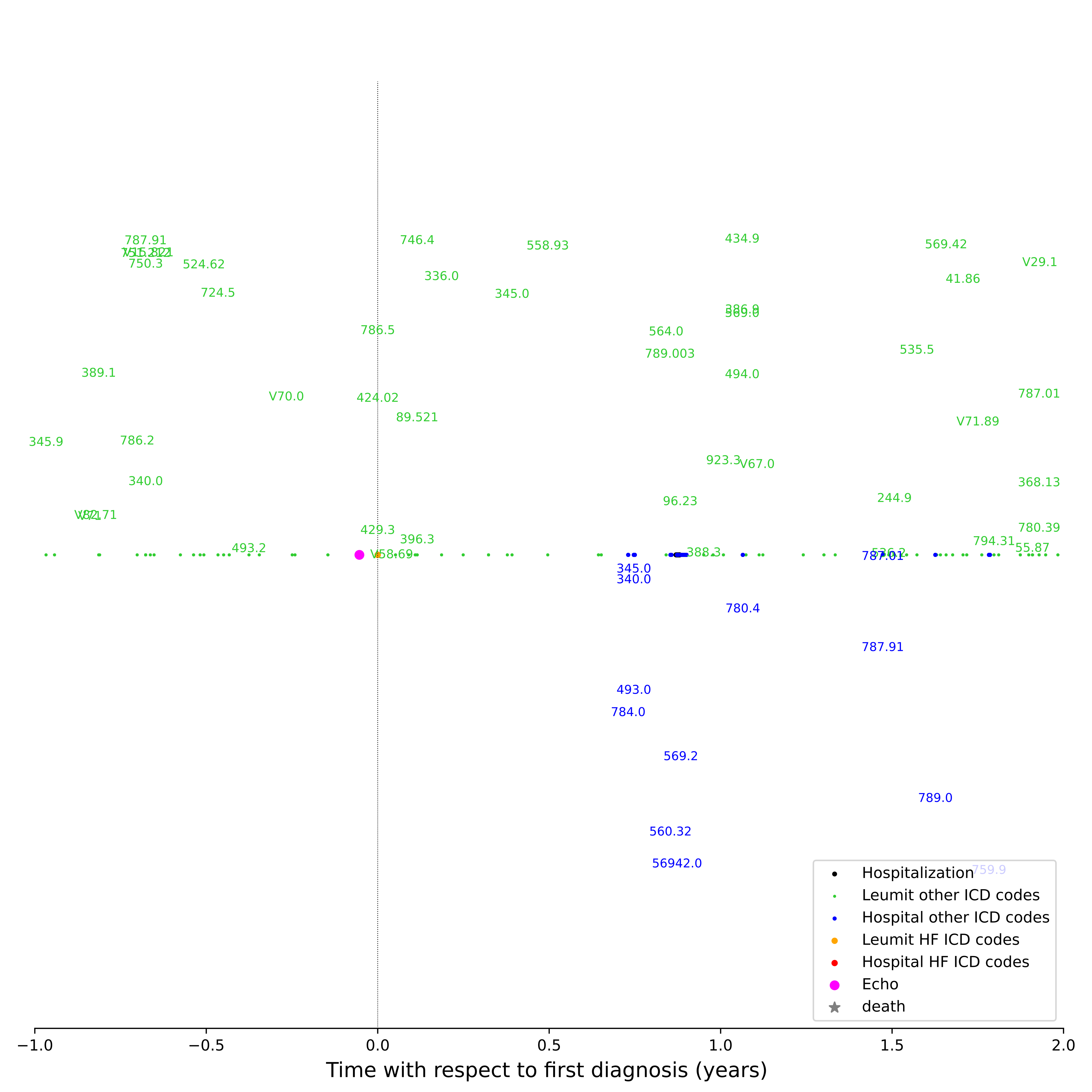}
\caption{\textcolor{black}{\textbf{Example of an individual analysis for an outlier patient case with a positive HF diagnosis.} The figure illustrates the detailed validation approach used to review the patient’s medical records around their first HF occurrence. This visualization includes the full set of ICD-9 codes recorded for the patient both in hospital and in primary care (Leumit), alongside other relevant events such as echocardiography examinations, hospitalizations, or death. In this case, the echocardiography performed shortly before the HF event strengthened the confidence in the HF diagnosis.}}
\label{ext_patient_timeline}
\end{figure*}

\newpage
\begin{figure*}[h]
\centering
\includegraphics[width=\textwidth]{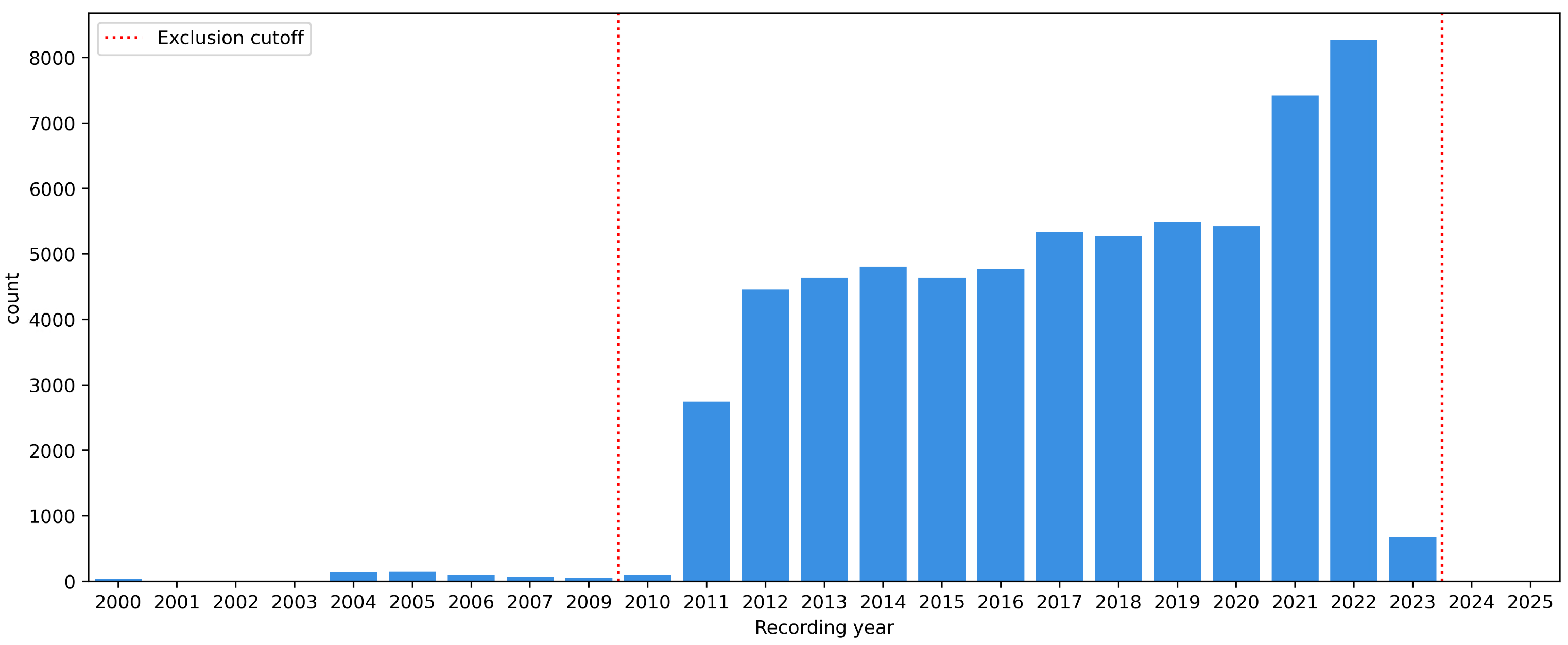}
\caption{\textbf{Histogram of the Holter recordings dates.} Holter recordings with inconsistent timestamps were considered outliers or misdated and were therefore excluded. Systematic recording began in 2010, with the last recording collected in 2023. Consequently, the data inclusion interval for the study cohort spans from 2010 to 2023, as marked by the red dotted lines.}
\label{ext_fig_recording_date}
\end{figure*}

\newpage
\begin{figure*}[h]
\centering
\includegraphics[width=\textwidth]{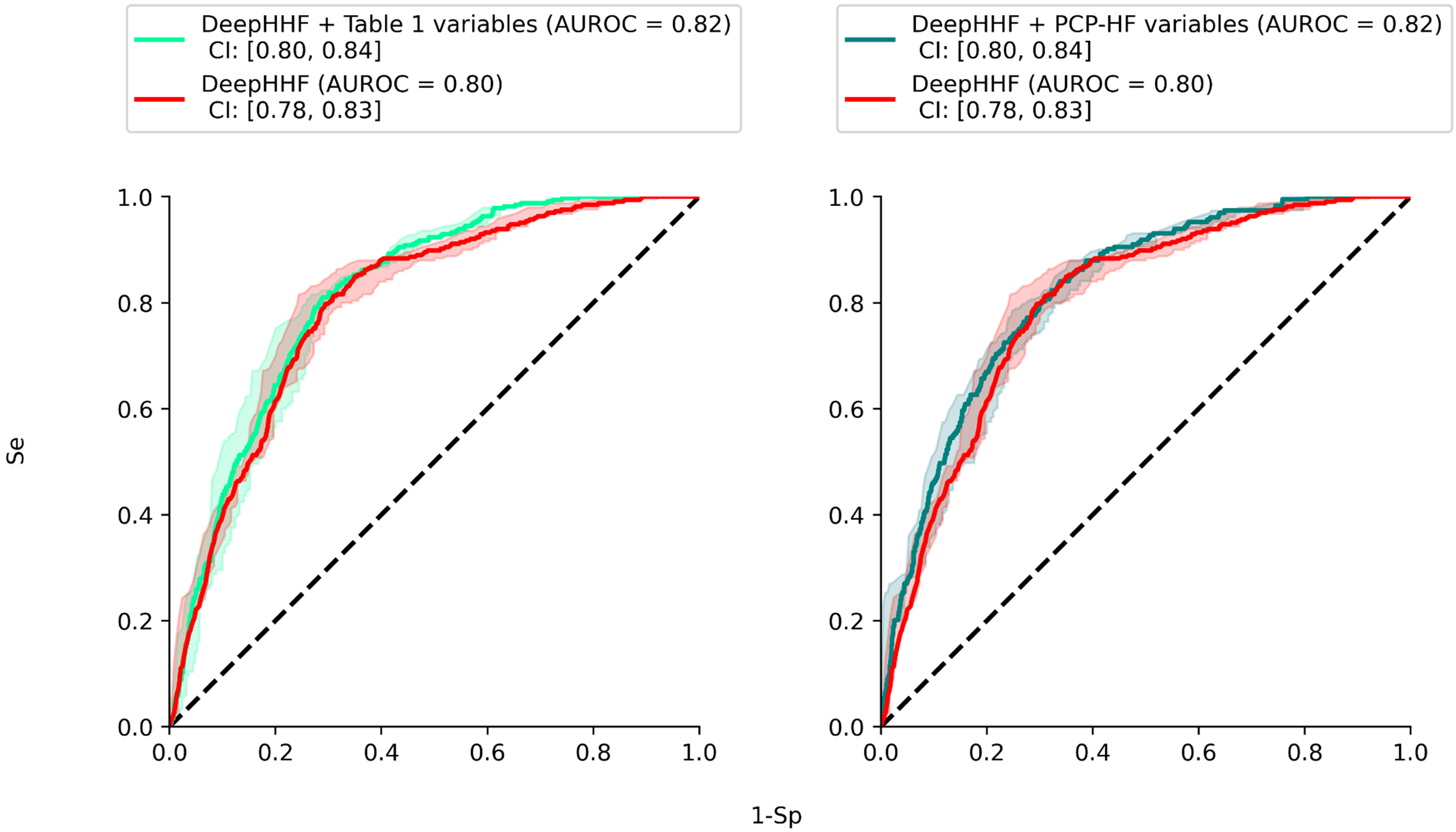}
\caption{\textcolor{black}{textbf{Combining variables with the DeepHHF score}. A logistic regression classifier was trained to combine the varibles together with the DeepHHF score. Receiver operating characteristic curves (ROC) were produced for the test set. The area under ROC (AUROC) scores are provided with 95\% confidence intervals (CI), shown as shaded area and evaluated by bootstrapping the test set. \textbf{a,}~DeepHHF taking as input the 24-hour single ECG recording (red) vs. combining the demographic and comorbidity history features listed in Table~
\ref{tab1} to the DeepHHF score (light green). \textbf{b,}~DeepHHF vs. combining PCP-HF features to the DeepHHF score (teal).}}\label{ext_fig_results_model_vars}
\end{figure*}

\newpage
\begin{figure*}[h]
\centering
\includegraphics[width=0.8\textwidth]{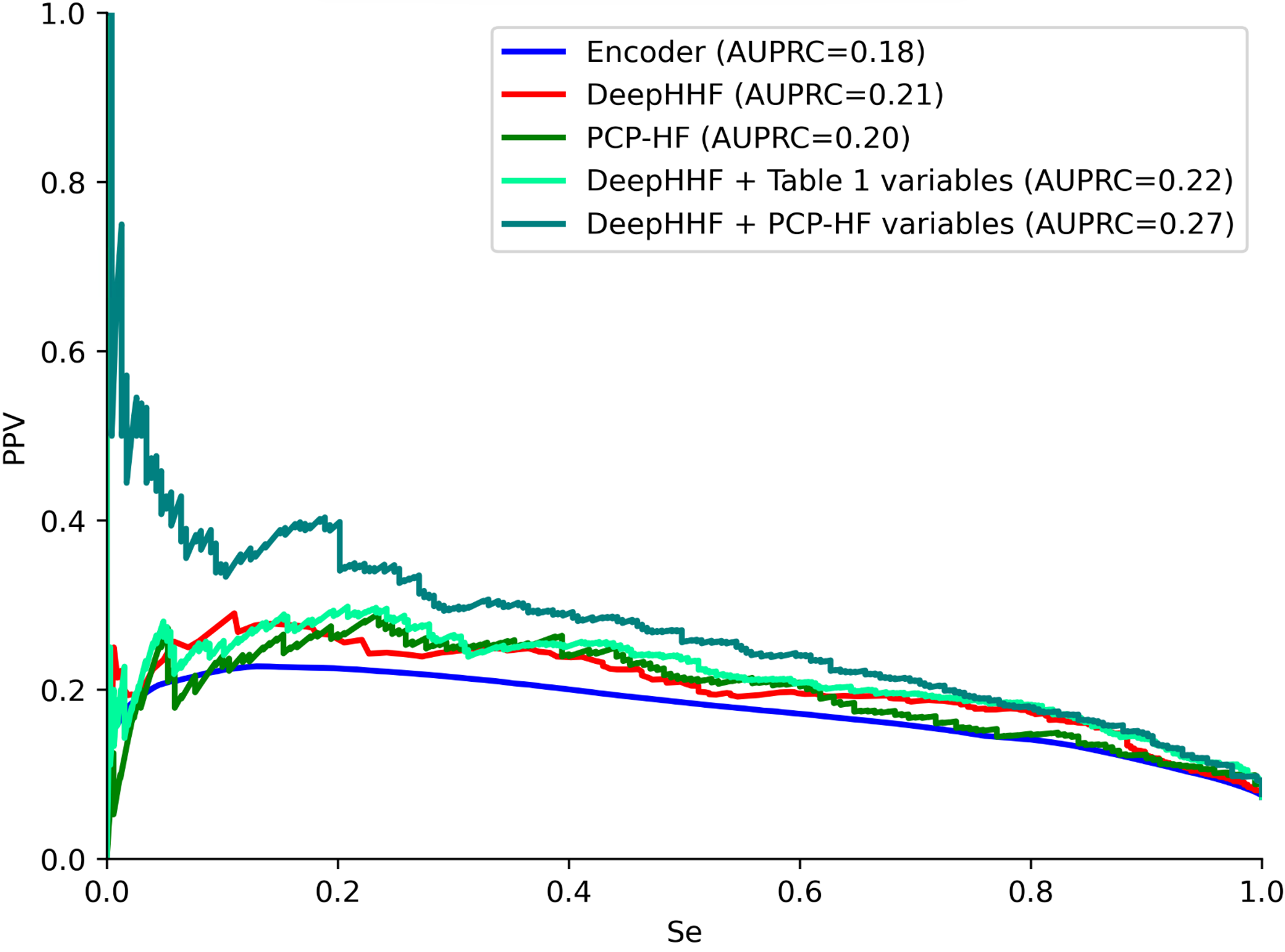}
\caption{\textcolor{black}{\textbf{Precision-recall curves for the different reported models}. The curves of positive predictive value (PPV) versus sensitivity (Se), together with their area under precision-recall curve (AUPRC) scores, are reported for the models shown in Figure~\ref{fig_results_model} and Supplementary Figure~\ref{ext_fig_results_model_vars}.}}\label{ext_fig_pr_curves}
\end{figure*}

\newpage
\begin{figure*}[h]
\centering
\includegraphics[width=\textwidth]{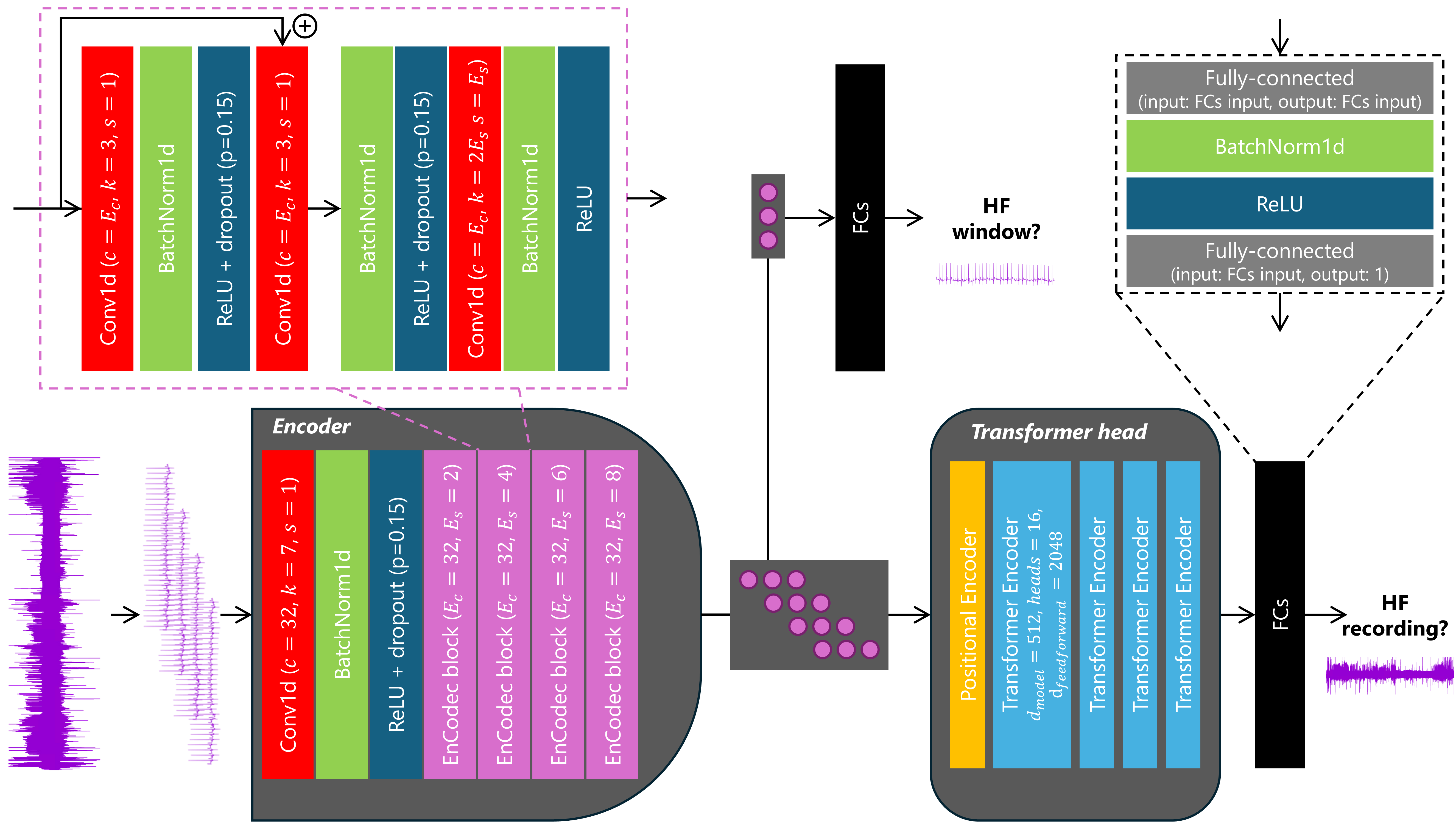}
\caption{\textbf{Model architecture.} The full 24-hour recording is split to 30-second windows, then going into the encoder, which outputs features for each window. The encoder includes EnCodec blocks, adopted from D\'efossez et~al~\cite{Defossez2022HighCompression} (purple frame). These blocks are changing according to their parameters $E_c$ (number of convolution filters) and $E_s$ (convolution stride). Next, in the first training step, each window features are going into fully-connected layers (FCs) block (black frame) for window heart failure (HF) prediction. In the second training step, all windows are used together as input to the transformer head, added with positional encoding, and through three transformer encoders. For the final prediction of the recording, the output of the transformer head is going through a different set of FCs.}
\label{ext_fig_architecture}
\end{figure*}

\newpage
\begin{figure*}[h]
\centering
\includegraphics[width=\textwidth]{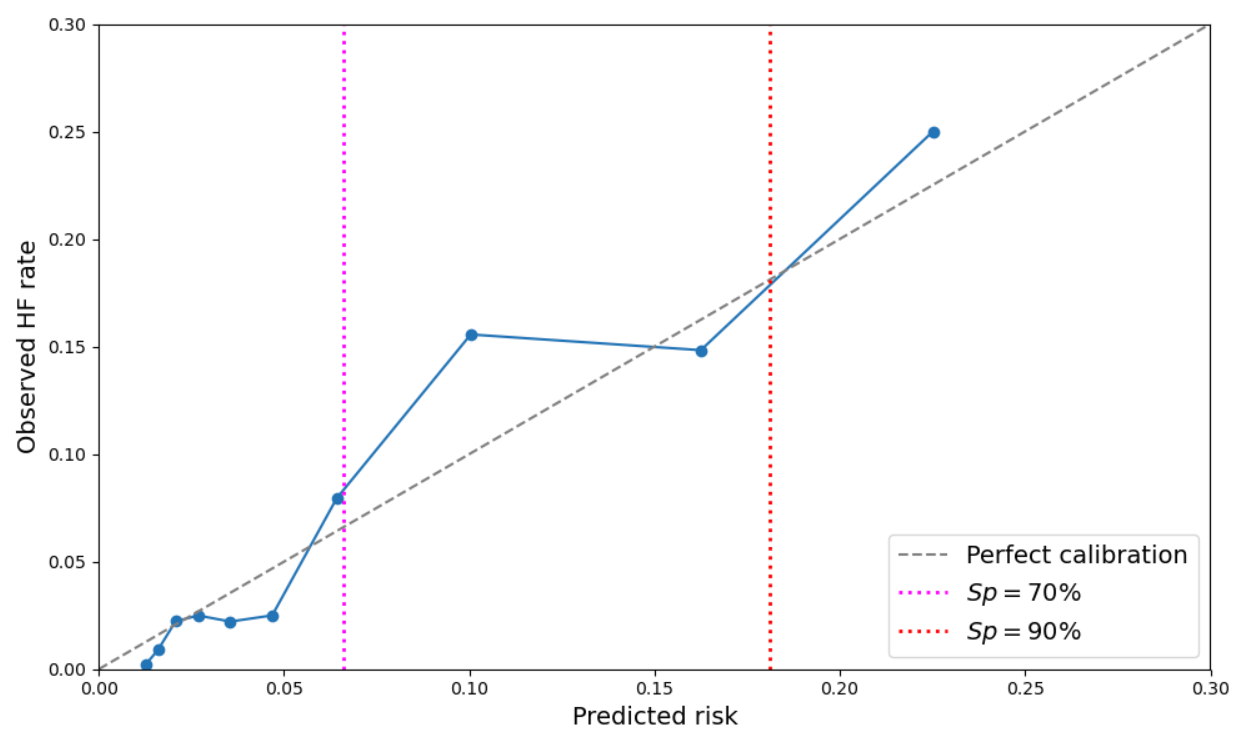}
\caption{\textcolor{black}{\textbf{Calibration curve of DeepHHF predictions on the test set.} The dashed grey diagonal indicates perfect calibration, where predicted probability exactly matches the observed HF rate. The blue curve represents the observed HF rate as a function of the model's post-hoc calibrated predicted risk, \textcolor{black}{evaluated using decile-based binning (10 groups), yielding an ECE of 0.0166.} Vertical dotted lines denote the score thresholds corresponding to 70\% specificity (magenta, $\approx7\%$ predicted risk) and 90\% specificity (red, $\approx18\%$ predicted risk). These operating points illustrate that the model effectively separates patient groups with progressively higher observed HF incidence.}}
\label{ext_fig_calib}
\end{figure*}

\newpage
\begin{figure*}[h]
\centering
\includegraphics[width=0.5\textwidth]{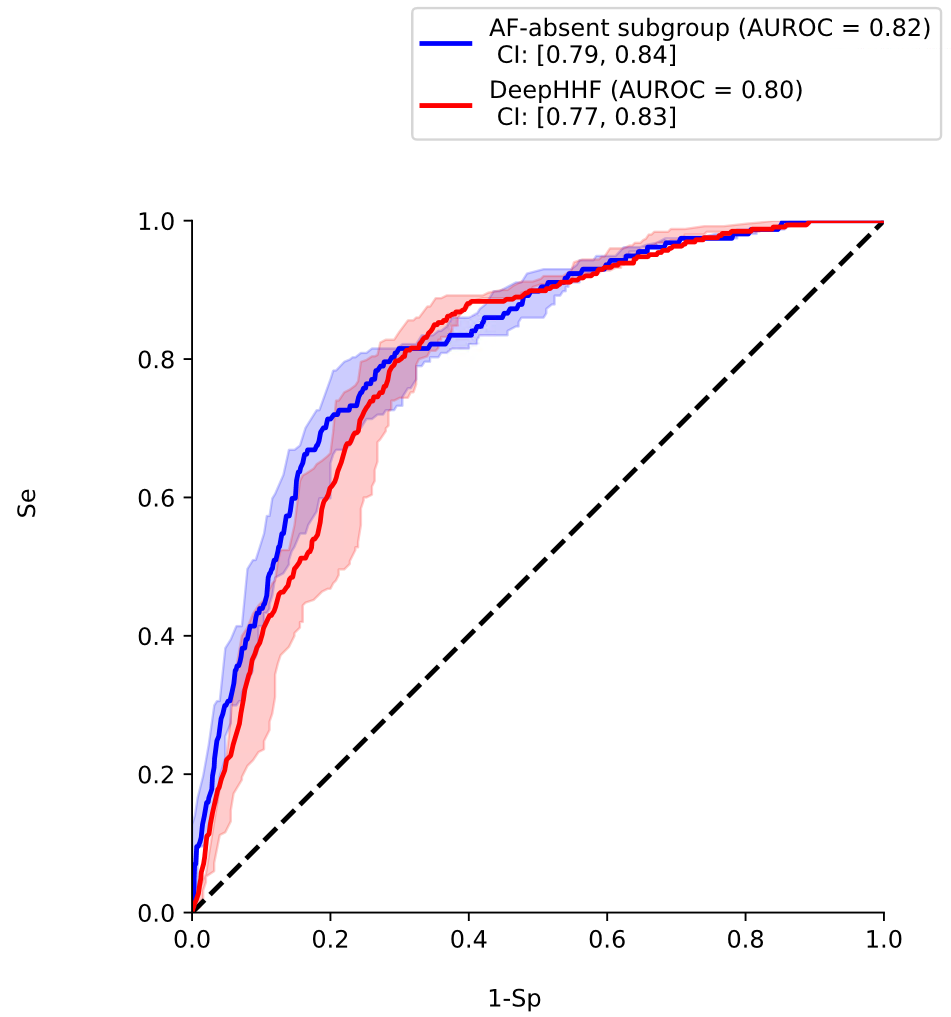}
\caption{\textcolor{black}{\textbf{DeepHHF performance in an atrial fibrillation (AF)-absent subgroup.} Receiver operating characteristic (ROC) curves for the test set comparing the overall DeepHHF model (red, $AUROC=0.80$) with its performance on a strictly AF-absent subgroup (blue, $AUROC=0.82$). The AF-absent subgroup (n=3,139) was defined by excluding all patients with a prior documented EMR diagnosis of AF or atrial flutter. The area under the ROC (AUROC) scores are provided with 95\% confidence intervals (CI), shown as shaded areas, evaluated by bootstrapping. The sustained performance indicates the model identifies pre-HF electrophysiological patterns independent of an AF diagnosis.}}
\label{ext_fig_AFabsent}
\end{figure*}

\newpage
\subsection*{Supplementary tables}

\begin{table}[h]
\caption{International Classification of Diseases - Ninth Revision (ICD-9) codes used for the HF end point and comorbidities.}
\label{ext_tab_icd9}
\begin{tabular}{|p{\dimexpr 0.1\textwidth}p{\dimexpr 0.2\textwidth}p{\dimexpr 0.2\textwidth}p{\dimexpr 0.5\textwidth}|}
\hline
\multicolumn{1}{|c|}{\textbf{\#}} & \multicolumn{2}{c|}{\textbf{Diagnosis}}                                                                               & \textbf{ICD-9 codes}                                                                                              \\ \hline
\multicolumn{4}{|l|}{\textbf{End point:}}                                                                                                                                                                                                                                     \\ \hline
\multicolumn{1}{|c|}{\textbf{1}}  & \multicolumn{2}{l|}{Heart failure (HF)}                                                                               & 428.X,   402.01X, 402.11X, 402.91X, 404.01X, 404.03X, 404.11X, 404.13, 404.93X,   416.11X, 514.2X, 514.3X, 518.4X \\ \hline
\multicolumn{4}{|l|}{\textbf{Comorbidities:}}                                                                                                                                                                                                                                 \\ \hline
\multicolumn{1}{|c|}{\textbf{2}}  & \multicolumn{2}{l|}{Diabetes}                                                                                         & 249-250.X                                                                                                         \\ \hline
\multicolumn{1}{|c|}{\textbf{3}}  & \multicolumn{1}{l|}{\multirow{3}{*}{Ischemic heart disease}}  & \multicolumn{1}{l|}{Any}                              & 410-414.X                                                                                                         \\
\multicolumn{1}{|c|}{\textbf{4}}  & \multicolumn{1}{l|}{}                                         & \multicolumn{1}{l|}{MI or acute coronary   event}     & 410-412.X                                                                                                         \\
\multicolumn{1}{|c|}{\textbf{5}}  & \multicolumn{1}{l|}{}                                         & \multicolumn{1}{l|}{Other}                            & 413-414.X                                                                                                         \\ \hline
\multicolumn{1}{|c|}{\textbf{6}}  & \multicolumn{1}{l|}{\multirow{2}{*}{Cerebrovascular disesse}} & \multicolumn{1}{l|}{Any}                              & 430-438.X                                                                                                         \\ 
\multicolumn{1}{|c|}{\textbf{7}}  & \multicolumn{1}{l|}{}                                         & \multicolumn{1}{l|}{Stroke}                           & 431.X,   433.X1, 434.X1, 435.X, 436.X,438.X                                                                       \\ \hline
\multicolumn{1}{|c|}{\textbf{8}}  & \multicolumn{2}{l|}{Chronic renal failure}                                                                            & 585.6X, 586.X                                                                                                     \\ \hline
\multicolumn{1}{|c|}{\textbf{9}}  & \multicolumn{2}{l|}{Acute renal failure}                                                                              & 584.X                                                                                                             \\ \hline
\multicolumn{1}{|c|}{\textbf{10}} & \multicolumn{2}{l|}{Cardiac  conduction disorders}                                                                    & 426.X                                                                                                             \\ \hline
\multicolumn{1}{|c|}{\textbf{11}} & \multicolumn{1}{l|}{\multirow{2}{*}{Cardiac dysrythmias}}     & \multicolumn{1}{l|}{Any}                              & 427.X                                                                                                             \\ 
\multicolumn{1}{|c|}{\textbf{12}} & \multicolumn{1}{l|}{}                                         & \multicolumn{1}{l|}{Atrial fibrillation   or flutter} & 427.3X                                                                                                            \\ \hline
\multicolumn{1}{|c|}{\textbf{13}} & \multicolumn{2}{l|}{Hypertension}                                                                                     & 401-405.X                                                                                                         \\ \hline
\multicolumn{1}{|c|}{\textbf{14}} & \multicolumn{2}{l|}{Valvular heart disease}                                                                           & 394.X,   385.X, 397.X, 398.X, 424.X                                                                               \\ \hline
\multicolumn{1}{|c|}{\textbf{15}} & \multicolumn{2}{l|}{Chronic obstructive pulmonary disease}                                                            & 490-496.X                                                                                                         \\ \hline
\multicolumn{1}{|c|}{\textbf{16}} & \multicolumn{2}{l|}{Cancer}                                                                                           & 140-159.9X,   160-165.9X, 170-176.9X, 179-189.9X. 190-199.2X, 200-208.92X, 209-209.79X                            \\ \hline
\end{tabular}%

\end{table}

\newpage
\begin{table}[h]
\centering
\caption{Medication group for heart failure label verification and their anatomical therapeutic chemical (ATC) classification codes.}
\label{tab_medications}
\begin{tabular}{|c|l|l|}
\hline
\textbf{\#}                 & \textbf{Heart failure guideline-directed medical therapies groups} & \textbf{Included ATC codes}   \\ \hline
\multirow{2}{*}{\textbf{1}} & \multirow{2}{*}{MRA\tnote{1}}                                               & ATC5Cd= 'C03DA'               \\
                            &                                                                    & ATC7Cd='C03DA01' , 'C03DA04'  \\ \hline
\textbf{2}                  & SGLT2i (DapaEmpa)\tnote{2}                                                   & ATC7Cd='A10BD20'   A10BD15'   \\ \hline
\textbf{3}                  & Agents   acting on the renin-angiotensin system (ACE-I, ARB, ARNI)\tnote{3}         & ATC3Cd='C09'                  \\ \hline
\textbf{4}                  & Low-ceiling Diuretics                                      & ATC4Cd='C03A', 'C03B' \\ \hline
\textbf{5}                  & High-ceiling Diuretics                                       & ATC4Cd='C03C' \\ \hline
\textbf{6}                  & Beta-blockers   (Bblockers)\tnote{4}                                        & ATC3Cd='C07'                  \\ \hline
\end{tabular}%
\footnotesize

\smallskip
\begin{tablenotes} 
\item[1] {Mineralo-corticoid receptor antagonists.}
\item[2] {Sodium-glucose cotransporter 2 inhibitors (including Dapagliflozin and Empagliflozin).}
\item[3] 
{Agents acting on the renin-angiotensin system (including angiotensin-converting enzyme inhibitors (ACE-I), 
angiotensin receptor blockers (ARB), 
angiotensin receptor-neprilysin inhibitors (ARNI)).}
\item[4] {Beta-adrenergic blocking agents.}
\end{tablenotes}
\end{table}


\newpage
\begin{table}[h]
    \centering
    \caption{\textbf{Comparison between patients with different \# of first heart failure (HF) diagnosis repeated in their files.} The table concentrates on the cases of 1, 2 and 3 repeated diagnoses. The table separates between diagnosis given before and after 2018, since after 2018 the data included diagnosis registered in hospitals as well.}
    \begin{tabular}{p{\dimexpr \textwidth}}
         \cr 
    \end{tabular}
    \includegraphics[width=\textwidth]{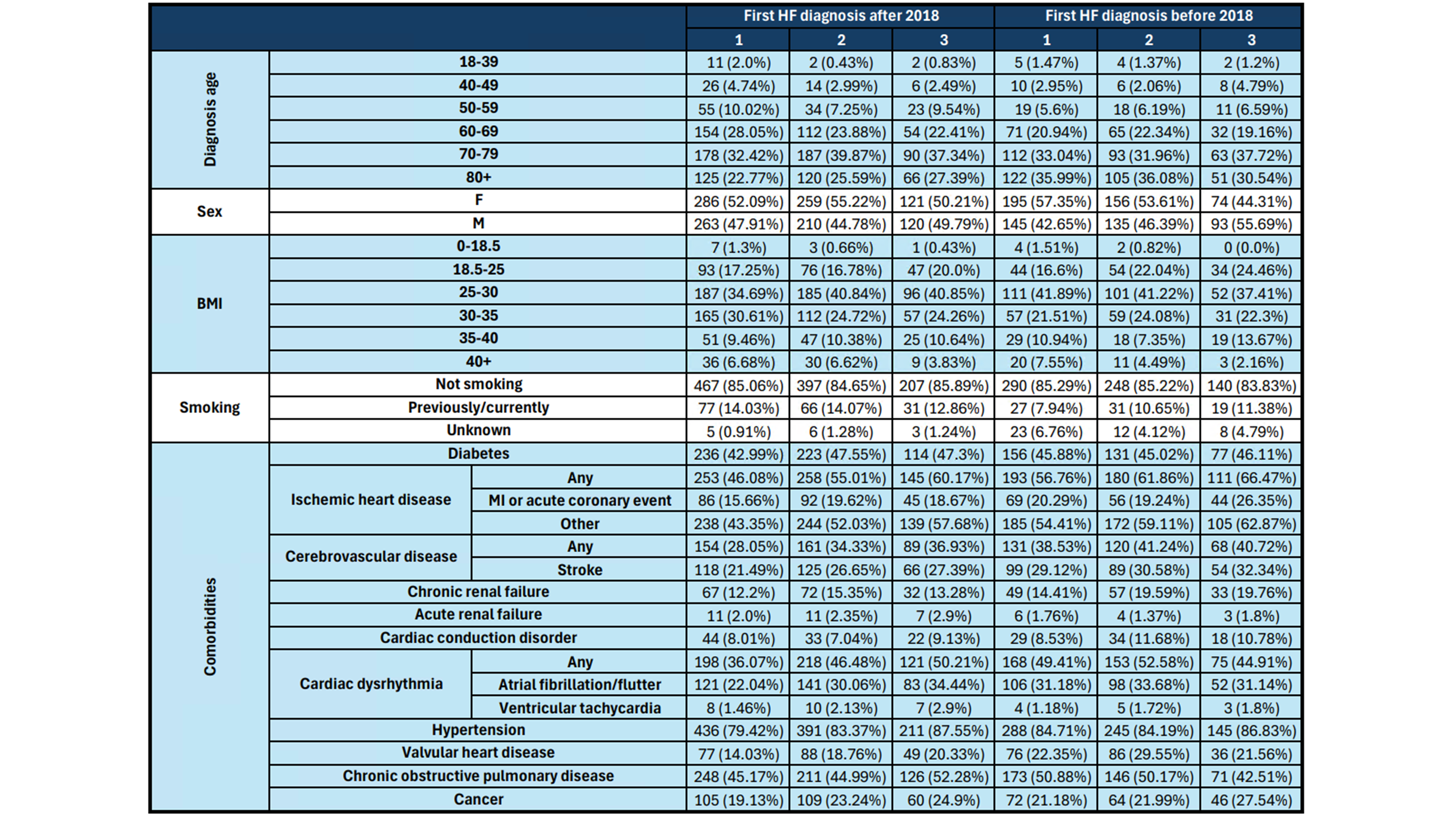}
    \label{ext_fig_tab_diag}
\end{table}

\newpage
\begin{table}[htbp]
\centering
\caption{External Validation Cohort Characteristics}
\label{tab:external_cohort}
\begin{tabular}{lccc}
\toprule
Characteristic & HF+ (n=29) & HF- (n=121) & Total (n=150) \\
\midrule
\textbf{Demographics} &  &  &  \\
Age, years (mean $\pm$ SD) & 70.8 $\pm$ 11.7 & 59.1 $\pm$ 14.2 & 61.4 $\pm$ 14.5 \\
Male sex, n (\%) & 14 (48.3\%) & 65 (53.7\%) & 79 (52.7\%) \\
\addlinespace[0.5em]
\textbf{Comorbidities} &  &  &  \\
Coronary artery disease, n (\%) & 23 (79.3\%) & 4 (3.3\%) & 27 (18.0\%) \\
Chronic kidney disease, n (\%) & 6 (20.7\%) & 4 (3.3\%) & 10 (6.7\%) \\
Diabetes, n (\%) & 11 (37.9\%) & 7 (5.8\%) & 18 (12.0\%) \\
COPD, n (\%) & 0 (0.0\%) & 3 (2.5\%) & 3 (2.0\%) \\
Thyroid disorders, n (\%) & 2 (6.9\%) & 9 (7.4\%) & 11 (7.3\%) \\
Sleep disturbance, n (\%) & 1 (3.4\%) & 0 (0.0\%) & 1 (0.7\%) \\
Prior stroke/TIA, n (\%) & 9 (31.0\%) & 19 (15.7\%) & 28 (18.7\%) \\
\addlinespace[0.5em]
\textbf{Rhythm Profile} &  &  &  \\
AF before/at Holter, n (\%) & 14 (48.3\%) & 1 (0.8\%) & 15 (10.0\%) \\
AF detected after Holter, n (\%) & 12 (41.4\%) & 1 (0.8\%) & 13 (8.7\%) \\
Sick sinus syndrome, n (\%) & 2 (6.9\%) & 0 (0.0\%) & 2 (1.3\%) \\
Conduction disorders, n (\%) & 2 (6.9\%) & 5 (4.1\%) & 7 (4.7\%) \\
\bottomrule
\end{tabular}
\end{table}

\end{document}